\documentclass[lettersize, 1 2pt]{article}

\usepackage{graphics}
\usepackage{float}
\usepackage{theorem}
\usepackage{latexsym}
\usepackage{amsfonts}
\usepackage{amssymb}
\usepackage{amsmath}
\usepackage{graphicx}
\usepackage{natbib}
\usepackage{lscape}
\usepackage{booktabs}
\usepackage{threeparttable}
\usepackage{appendix}
\usepackage{subfig}
\usepackage[colorlinks=true,linkcolor=black,citecolor=black,urlcolor=black,pdfstartview=FitH,bookmarks=false]{hyperref}
\usepackage{setspace}
\usepackage{titlesec}
\usepackage{color}
\usepackage{bm}
\usepackage{multicol}
\usepackage{soul}
\usepackage{tabularx} 

\usepackage[utf8]{inputenc}

\newtheorem{algorithm}{Algorithm}
\newtheorem{theorem}{Theorem}
\newtheorem{corollary}{Corollary}

\newtheorem{lemma}{Lemma}
\newtheorem{assumption}{Assumption}

\theorembodyfont{\rm}
\newtheorem{remark}{Remark}

\usepackage{authblk}

\newcommand{\1}{\mathbf{1}}
\newcommand{\E}{\mathbb{E}}

\newcommand{\p}{\mathbb{P}}

\usepackage{thmtools}
\usepackage{hyperref}

\begin{document}

\title{Lee Bounds with a Continuous Treatment \\
in Sample Selection}
\author{Ying-Ying Lee\footnote{Corresponding author. Department of economics, University of California Irvine, Irvine, CA 92697, U.S.A.  \\
E-mail:\ \href{yingying.lee@uci.edu}{yingying.lee@uci.edu}.
\href{https://sites.google.com/site/yyleelilian}{https://sites.google.com/site/yyleelilian}.
Tel: +1 9498244834. Fax: +1 9498242492.
}
\hspace{1.5cm}
Chu-An Liu\footnote{Institute of Economics, Academia Sinica, Taipei City 115, Taiwan. \\
E-mail:\ \href{caliu@econ.sinica.edu.tw}{caliu@econ.sinica.edu.tw}.
\href{https://chuanliu.weebly.com/}{https://chuanliu.weebly.com/}.
}
}

 \maketitle
\vspace{-25pt}

\begin{center}
\textbf{Abstract}
\end{center}

\vspace{-10pt}
We study causal inference in sample selection models where a continuous or multivalued treat-
ment affects both outcomes and their observability (e.g., employment or survey response). We
generalize the widely used Lee (2009)’s bounds for binary treatment effects. Our key innovation is
a ``sufficient treatment value" assumption that imposes weak restrictions on selection heterogeneity and is implicit in separable threshold-crossing models, including monotone effects on selection. Our double debiased machine learning estimator enables nonparametric and
high-dimensional methods, using covariates to tighten the bounds and capture heterogeneity. Applications to Job Corps and
  Civilian Conservation Corps (CCC) program evaluations reinforce prior findings under weaker assumptions.
\\
\\[6pt]
%

\newpage

\section{Introduction}

Sample selection is a common challenge in studying treatment effects.
A classic question in empirical economics is estimating the effect of training programs on  wages.
There is sample selection because a training program not only affects wages via human capital accumulation, but also
affects the chance that a worker is eventually employed and hence is selected into samples.
We only observe wages of employed workers.
If the estimation is based on a selected sample with observed outcomes,
then one must isolate the effect on employment status (extensive margin) to learn about the  effect on wages (intensive margin).
Nonetheless in survey data, when the treatment affects response behavior, those who respond to the survey (i.e., who are selected into samples) in the treatment group are no longer comparable to respondents from the control group.
Such problems of sample selection, attrition, or missing data arise in fields other than economics; for example,
in medical studies, quality of life after assignment of a new drug is only observed if a patient does not die (truncation by death).
In education, final test score is only observed if a student does not drop out.

Due to the non-random sample selection, the  causal effect on the outcome is not point-identified without imposing further assumptions on functional forms or distributions; e.g., \cite{RePEc:nbr:nberch:10491,Heckman}, \cite{ImbensAngrist}, \cite{ZRM}, \cite{HonoreHu}, \cite{ChenRoth}.
Assuming the treatment variable is randomly assigned (conditional on observables), we build on the seminal work of \cite{HM95} and \cite{LeeBound}, who bound the average causal effect of a binary treatment.
The setup is fully non-parametric and hence allows for general heterogeneity.
Many treatment or policy variables are continuous or discrete multivalued, e.g., hours in a social program, lottery prize, drug dosage, tuition subsidy, cash transfer, air pollution, etc.
As \cite{LeeBound}'s bound has been a common practice in empirical economics, we fill in the important gap to provide a corresponding tool to deal with sample selection in studying the causal effect of a continuous or multivalued treatment.
The replication package of codes and data for our empirical applications is available on the authors' websites for practical implementation.

We provide the worst-case sharp bounds for the average treatment effect, or the average dose-response function,
for {\it always-takers} who are selected into samples, or whose outcomes are observed, regardless of the treatment values they receive.\footnote{
Always-takers are also known as  always-observed, always-responders, always-employed, nonattriters, or survivors.  The concept of always-takers is from the literature on imperfect compliance of treatment \citep{AIR}, where ``taking" is the taking of the treatment affected by an instrument, rather than selection into the sample affected by the treatment, considered in this paper.}
Note that the selected sample given a certain treatment value $d$ consists of always-takers and {\it compliers} who are not selected given other treatment values $d' \neq d$.
The key element of the bounds is the proportion of always-takers in the selected $d$-treated sample, which is used to trim the observed outcomes for the worst-case lower (upper) bound when all always-takes' outcomes are smaller (or larger) than all compliers' outcomes.
Recall that for a binary treatment, \cite{LeeBound} assumes monotone treatment effects on selection, i.e., if a subject is selected in the control group, then it must be selected in the treatment group.
Then only always-takers can be in the selected untreated sample and are identified.
For a continuous or multivalued treatment, we propose a novel {\it sufficient treatment value} assumption on selection: if a subject is selected into samples when it receives the {\it sufficient treatment value}, then it remains selected when receiving {\it any} other treatment values.
So we generalize the monotonicity assumption in \cite{LeeBound} by assuming the sufficient treatment value to be zero for a binary treatment.
Then the selected subjects who are treated with the sufficient treatment value are always-takers.
So the probability of always-takers is identified by the minimum conditional selection probability over the treatment values.

For example, in a standard setting of survey attrition,
if we find the response rate (conditional selection probability) given cash transfer is lowest at \$1,000, then our sufficient treatment value assumption is that everyone who responded to a survey question when receiving a cash transfer of \$1,000 would also have responded if there received any other values (always-takers).  And there are some additional people (compliers) who only responded when they received some other transfer values and did not respond given \$1,000.
We can interpret the sufficient treatment value \$1000 as the least-favored treatment (cash transfer) to induce selection into samples (responding to the survey) in the sense that if a subject is selected under the sufficient treatment value, then it is selected under any other treatment values.




In fact, the sufficient treatment value is implicit under a separable structural error in selection, or in a widely used class of latent variable threshold-crossing models \citep{Vytlacil02}.
This important observation suggests that the sufficient treatment value assumption is not restrictive, and the subpopulation of always-takers is a natural target under minimal assumptions.
The associated always-takers are the largest subpopulation for whom we can partially identify the average effect of switching treatment  over a range of values chosen by researchers for treatment intervention.


Moreover we allow for unconfoundness assumption in observational studies, and subjects with different pretreatment covariates can have different sufficient treatment values.
So the information of covariates could potentially tighten the bounds and  confidence intervals, or capture heterogenous effects that is not revealed without using the covariates, as supported by our empirical illustrations.
Our bounds and asymptotic inference are robust to the extensive margin effect on selection, in the sense that
 when there is no selection bias or no extensive margin, our bounds contain the point-identified causal object.
So we avoid pre-testing the treatment effect on selection.

We note that one might be interested in the partial (or marginal) effect defined as the derivative of the average dose response function.
However, we cannot bound such derivative by the same approach of \cite{LeeBound}.
Instead, we bound the average effects of switching treatment values within a subset of the support of the treatment, or an average derivatives over two treatment values, which could be interesting in practice.

We illustrate our methodology by two applications.
We find significant effects when incorporating covariates, while the bounds estimates without covariates show positive but insignificant effects.
So incorporating covariates is useful to increases precision and captures heterogeneity.
First, we revisit the Job Corps program, one of the largest federally funded job training programs in the U.S.
The evaluation of these programs has been the focus of a substantive methodological literature,
due to the high cost of about \$14,000 on average per participant; see \cite{SBM}, \cite{LeeBound}, \cite{FFGN12ReStat}, among others.
The participants are exposed to different numbers of actual hours of academic and vocational training.
Their labor market outcomes may differ if they accumulate different amounts of human capital acquired through different lengths of exposure.
 We try to understand whether hours of training raise wages by helping them find a job (extensive margin)
 or by increasing human capital that would affect the intensive margin effect on the outcome.
 We find that increasing the training from 1.5 weeks to 9 months increases log weekly earnings by at least 0.224 at  5\% significance level for those always employed ({\it always-takers}).

The second application evaluates the Civilian Conservation Corps (CCC) in \cite{Aizer},
who conduct the first lifetime evaluation of the largest federal youth employment program in U.S. history created to address high youth unemployment during the Great Depression.
The Job Corps is a modern-era job training program that was modeled after the CCC and shares many features.
We bound the effect of service duration on the age at death and
strengthen the findings in \cite{Aizer}.
As differential attrition could bias the OLS estimates, they find that
the effect of duration on longevity is consistently positive and statistically significant under various imputation approaches.
Our bounds suggest that increasing duration from about 3 months to 14 months increases the average death age by at least 1.17 years at 5\% significance level.

Another theoretical contribution of this paper is a weaker sufficiency assumption of a {\it sufficient set} of $M$ treatment values, which is a useful approximation when the error of unobserved heterogeneity is non-separable in the selection equation.
For example of $M=2$, if a program participant is employed under {\it both} one week and fifteen months of training, then this participant is always employed.
Such a weaker identification assumption results in a tradeoff with less informative (wider) bounds.
Furthermore we utilize the well-known Fr\'{e}chet-Hoeffding bounds for the discrete treatment {\it without} imposing any shape restrictions on selection.

We incorporate covariates, following \cite{SGLee} on the generalized Lee bounds for the binary treatment.
Our bound estimator is doubly debiased using an orthogonal moment function and cross-fitting, which enables nonparametric and machine learning methods to handle high-dimensional data, following the recent double debiased machine learning (DML) literature  \citep{CCDDHNR}.
Since the average dose-response function, or the mean potential outcome, and its bounds are functions of the continuous treatment,
such non-regular estimand cannot be estimated at the regular root-$n$ rate.
We use a kernel function for localizing the continuous treatment as in \cite{CL}.

%


The paper is organized as follows.
Section~\ref{SecCT} describes the sample selection model under the potential outcome framework, or equivalently a nonparametric non-separable structural model (e.g., \cite{IN09ETA}).
We discuss related literature.
Section~\ref{SecBound} presents the basic Lee bounds without covariates for a continuous/multivalued treatment under a sufficient value/set assumption on the treatment effect on selection.
We give estimation and inference theory in Section~\ref{SecInf}.
Section~\ref{SecConX} incorporates the covariates and presents the DML inference.
Section~\ref{SecEmCT} and Section~\ref{SecEmCTX} present the empirical illustration on evaluating the Job Corps and the CCC programs.
Appendix contains the main proofs of Theorems and Lemmas.
In the online supplementary appendix, we present the proofs of Corollaries, and supplementary material for the empirical applications.

\section{Sample selection model and related literature}
\label{SecCT}
The researcher chooses a compact subset of the support of the treatment variable $D$, denoted as $\mathcal{D}$, for treatment intervention.
So we aim to learn about the intensive margin effect on the outcome of switching the treatment values over $\mathcal{D}$.
For a continuous $D$, let $\mathcal{D} = [\underline{\mathcal{D}}, \overline{\mathcal{D}}]$;
for a discrete $D$, let $\mathcal{D} = \big\{\underline{\mathcal{D}} =: d_1, d_2,..., d_J:= \overline{\mathcal{D}}\big\}$ with dimension $J$ smaller or equal to the dimension of $D$.

For any treatment value $d\in \mathcal{D}$ that a subject receives, let $Y_d$ be the continuous potential outcome, or the response function of $d$,
and $S_d \in \{0,1\}$ be the potential selection indicator for whether the subject's outcome is observed.
If a subject is treated at $d$, i.e., $D = d$, let the selection status $S = S_d$ and the outcome $Y = Y_d$.
The observed data vector $W = (D,S,S\cdot Y)$, so the outcome is recorded as zero if missing in the sample.

Following the literature and focusing on the sample selection bias, we begin with the independence Assumption~\ref{Aind} on treatment assignment.
After the key results are established, we consider the standard conditional independence assumption given covariates in Section~\ref{SecConX}.
\begin{assumption}[Independence]
$D$ is independent of $\big\{(Y_d, S_d): d\in \mathcal{D}\big\}$.
\label{Aind}
\end{assumption}
Under Assumption~\ref{Aind}, we identify the selection probability at treatment $d$, $\p(S_d=1) = \E[S_d] = \E[S|D=d]$, and hence
the average treatment effect (ATE) on selection $\E[S_d - S_{d'}] = \E[S|D=d] - \E[S|D=d']$, also known as the extensive margin effect of switching treatment from $d'$ to $d$.

Assumption~\ref{Aind} also identifies  the average outcome of the selected population at $d$, $\E[Y_d|S_d = 1] = \E[Y|S=1, D=d]$.
But $\E[Y_d|S_d = 1] - \E[Y_{d'}|S_{d'} = 1]$ is not causal if $\{S_d = 1\}$ and $\{S_{d'} = 1\}$ are different subpopulations.
There are two common assumptions for $\E[Y_d|S_d = 1] - \E[Y_{d'}|S_{d'} = 1]$ to capture the intensive margin:\footnote{
We can decompose $\E[Y_d|S_d = 1] - \E[Y_{d'}|S_{d'} = 1] = InM + ExM$, where $InM := \E[Y_d|S_d = 1] - \E[Y_{d'}|S_{d} = 1]$  from the intensive margin and $ExM = \E[Y_{d'}|S_d = 1] - \E[Y_{d'}|S_{d'} = 1]$  from the extensive margin that cause the selection bias.  Because $\E[Y_{d'}|S_{d} = 1]$ is not identified, we cannot disentangle $InM$ and $ExM$.}
(i) Assume no ATE on selection (no extensive margin), or $\{S_d = 1\} = \{S_{d'}=1\}$ have the same distribution for all $d, d'\in\mathcal{D}$.
(ii) Assume missing at random \citep{Rubin}, so $\E[Y_d|S_d=1] - \E[Y_{d'}|S_{d'}=1]=\E[Y_d - Y_{d'}]$ is the population ATE on the outcome.
But these two assumptions can be restrictive and unrealistic.

To understand the source of selection bias, note that the selected population $\{S_d = 1\}$ is composed of always-takers and $d$-compliers.
Define  \emph{always-takers} AT$:= \{S_{d'}=1\text{ for all } d'\in \mathcal{D}\}$ to be those selected into samples regardless of the treatment value they receive over $\mathcal{D}$.
Define  \emph{$d$-compliers} CP$_d := \{S_d = 1, S_{d'}=0 \text{ for some }d' \in \mathcal{D}\}$ to be those induced to selection due to the treatment value $d$ but are not selected at $d'$.
Recall our goal of recovering the intensive margin effect of switching treatment between any values in $\mathcal{D}$.
Always-takers are the common subpopulation that are selected into samples for {\it all} treatment values in $\mathcal{D}$,
while $d$-compliers are missing in some selected samples with $d'$, $\{D=d', S=1\}$.
As we never observe $d$-compliers in some samples with $d'$, it is not possible to learn about their causal effect of switching treatment from $d$ to $d'$.
So without further assumptions such as functional form for extrapolation, we could only hope to learn about the causal effect for the always-takers.
Therefore our target parameter is  the mean potential outcome $Y_d$ for always-takers,
\begin{align*}
\beta_{d{}}:= \mathbb{E}[Y_d|\{S_{d'}=1 \mbox{ for all } d'\in \mathcal{D}\}]
\end{align*}
 for $d\in\mathcal{D}$.
Note that the definition of always-takers depends on the range of treatment values of interest $\mathcal{D}$.
So the notation $\beta_d$ should depend on $\mathcal{D}$ that is suppressed for simplicity.

When the treatment is continuous, $\beta_{d{}}$ is known as always-takers' {\it average dose-response function}.
When the treatment variable is binary, i.e., $\mathcal{D} = \{0,1\}$, always-takers' ATE is $\beta_{1{}} - \beta_{0{}} = \E[Y_1 - Y_0  \mid S_1=1, S_0=1]$, studied in \cite{LeeBound}.
Nonetheless we cannot determine whether a specific subject is an always-taker or a complier. So we take a bound/partial identification approach following \cite{LeeBound} and \cite{ZR03}.
Our bounds estimation and inference are robust to the extensive margin effect on selection.
\cite{GRR} propose similar worst-case sharp bounds with manipulation-robust inference in regression discontinuity designs.
We also discuss how the sharp bounds might be tightened by our sufficient value/set assumption or the covariates (e.g., \cite{FanPark}).


There are recent developments in sample selection models using the bound/partial identification approach.
\cite{HonoreHu} and \cite{HonoreHuJoE} consider parametric and semiparametric structural models.
In addition to the concern of misspecification, the parametric selection equation often relies on the monotonicity assumption.
\cite{Estrada} studies spillover effects under sample selection, which can be viewed as Lee bound for the multivalued treatment effect.
\cite{HEILER24} and \cite{Olma} study Lee bounds for the conditional average binary treatment effect given a continuous covariate, which is a non-regular estimand as ours.
\cite{KMV} extend Lee bounds for multilayered sample selection to account for training affecting workers sorting to firms.
 \cite{KlineSantos}  assess the sensitivity of empirical conclusions among a continuum of assumptions ordered from strongest (missing at random) to weakest (worst-case bounds).
 See the literature reviews on partial identification in \cite{HoRosen}, \cite{MolinariReview}, \cite{KlineTamer}, and references therein.

Alternatively another literature on sample selection models with {\it exclusion restrictions} or partial randomization assumes a variable $Z$ in the selection equation of $S$ that does not enter $Y$.
In Heckman's classic sample selection model (``Heckit"), the structural equations are linear in $(D,Z)$ and are separable in the normally distributed errors.
\cite{AhnPowell}, \cite{DNV03}, and \cite{EJL16} consider more nonparametric settings.
The standard sample selection model is generally not point-identified without exclusion restrictions.
Nevertheless, it has been noted to be difficult to find a credible instrument $Z$ in practice; see, for example, \cite{HonoreHu}.
\cite{DiNardo} proactively create an instrument by ex-ante randomizing the participants of the Moving to Opportunity experiment to differing intensity of follow-up.
\cite{BCGL} use information on the number of calls made to each individual before responding to the survey to identify the ATE of a binary treatment for a subpopulation of respondents, in the absence of instruments.
 See also \cite{Garlick} for evaluation of various sample selection correction methods and references therein.
\cite{ChenRoth} discuss problems of log-like transformations with zeros and propose some solutions.
We capture general heterogenous causal effects 
without exclusion restrictions and free from misspecification.


\section{Lee Bounds}
 \label{SecBound}
We establish the sharp worst-case bounds for $\beta_d$ with a continuous/multivalued treatment, building on \cite{HM95} and \cite{LeeBound} for a binary treatment.
The upper bound is when all always-takers' wages are larger than all $d$-compliers' wages.
Denote the fraction of always-takers among the selected subjects with treatment $d$ as $p_d$.
Then all always-takers' wages are larger than the $(1-p_d)$-quantile of the observed wage distribution at $d$.
So we can construct the worst-case bound by trimming the upper and lower tails of the observed outcome distribution by $p_d$.
Next we present the well known worst-case bounds based on a given $p_d$, and then we propose identification strategies of $p_d$ for the continuous and multivalued treatment, which is new to the literature.

Independent treatment Assumption~\ref{Aind} identifies the selection probability at $d$ by the conditional selection probability given $d$, $\p(S_d = 1) = \E[S|D=d] =: s(d)$.
If  the proportion of always-takers $\pi_{\text{AT}} := \p(S_{d'} = 1: d'\in\mathcal{D})$ is known, then the fraction of always-takers among the selected subjects with treatment $d$ is $p_d = \p(\text{AT}|S_d = 1) = \pi_{\text{AT}{}}/s(d)$.
Let $Q^d(u)$ be the $u$-quantile of $Y|D=d, S=1$.
Then the bounds of $\beta_d$ are the  trimmed means:
\begin{align}
\rho_{dU}(\pi_{\text{AT}{}}) := \mathbb{E}[Y|Y \geq Q^d(1-\pi_{\text{AT}{}}/s(d)), D=d, S=1]
\label{ELeeUB}
\end{align}
for the upper bound and
$\rho_{dL}(\pi_{\text{AT}{}}) := \mathbb{E}[Y|Y \leq Q^d(\pi_{\text{AT}{}}/s(d)), D=d, S=1]$ for the lower bound.
The key element of the bounds is the proportion of always-takers $\pi_{\text{AT}{}}$.
Once we identify $\pi_{\text{AT}{}}$ and hence $p_d = \pi_{\text{AT}}/s(d)$, we can consistently estimate the bounds.

Note that when there is no complier, the selected sample is composed of always-takers only, so $s(d) = \p(S_{d'}=1: d'\in\mathcal{D})$ is constant
and $p_d = \pi_{\text{AT}{}}/s(d)=1$ for all $d\in\mathcal{D}$.
That is, there is no extensive margin, and $\beta_d = \rho_{dU}(\pi_{\text{AT}{}}) = \rho_{dL}(\pi_{\text{AT}{}}) =  \mathbb{E}[Y|D=d, S=1]$ is point-identified.

Section~\ref{SecIDSV} and Section~\ref{SecIDSS} present novel sufficient assumptions on the treatment effect on selection
to identify the proportion of always-takers $\pi_{\text{AT}}$.
Section~\ref{SecVytlacil} discusses the connection with the structural selection model.
For expositional ease, we focus on the upper bound.

\subsection{Identification of the proportion of always-takers}
\label{SecIDSV}

The key identification Assumption~\ref{Amin} requires one {\it sufficient treatment value} $d_{\text{AT}}$ such that if a subject is selected at $d_{\text{AT}}$ then it will be selected at any treatment values.

\begin{assumption}[Sufficient treatment value]
There exists a treatment value $d_{\text{AT}} \in\mathcal{D}$ such that
 $S_d \geq S_{d_{\text{AT}}}$ almost surely (a.s.) for any $d \in\mathcal{D}$.
\label{Amin}
\end{assumption}
Assumption~\ref{Amin} essentially assumes that always-takers are the selected $d_{\text{AT}}$-receipts, $\{S_{d} =1: d\in\mathcal{D}\} = \{S_{d_{\text{AT}}} = 1\}$.
Together with Assumption~\ref{Aind}, the proportion of always-takers $\pi_{\text{AT}} = \p(S_{d_{\text{AT}}} = 1) = \E[S|D=d_{\text{AT}}] =: s(d_{\text{AT}})$ is identified.

Assuming $s(\cdot)$ to be continuous for a continuous $D$, the extreme value theorem implies that
$d_{\text{AT}} = \arg\min_{d\in\mathcal{D}} s(d)$ exists and $\pi_{\text{AT}} = s(d_{\text{AT}}) = min_{d\in\mathcal{D}} s(d)$ can be estimated from the data.
Notice that $d_{\text{AT}}$ depends on $\mathcal{D}$ that is a range of treatment values chosen by the researcher for policy intervention.
A larger $\mathcal{D}$ results in a smaller $\pi_{{\text{AT}}}$, which trims more observations, and wider bounds.


Importantly Assumption~\ref{Amin} allows any shape of the effect on selection.
Consider an example of $d_{\text{AT}} = 1$.
A 2-complier can have $\{S_{d_{\text{AT}}} = S_1= 0, S_2= 1, S_3 = 0\}$.
In contrast, a stronger monotonicity assumption, which assumes $S_{d'} \geq S_{d}$ a.s.\ for any $d' > d$, rules out this event because it requires $S_3 = 1$ if $S_2 = 1$.

Lemma~\ref{Tmin} formally presents our generalized Lee bounds with a continuous treatment or a discrete multivalued treatment.

\begin{lemma}
Let Assumption~\ref{Aind} hold.  Assuming $s(d) > 0$ for $d\in\mathcal{D}$,
then $\beta_{d{}} \in [\rho_{dL}(\pi_{\text{AT}}), \rho_{dU}(\pi_{\text{AT}})]$ with $\pi_{\text{AT}} = \p(S_{d'} = 1: d' \in\mathcal{D})$ given in equation (\ref{ELeeUB}).
Further let Assumption \ref{Amin} hold.
Then we identify $\pi_{\text{AT}} = s(d_{\text{AT}})$, the sharp bounds for $\beta_d \in [\rho_{dL}(s(d_{\text{AT}})), \rho_{dU}(s(d_{\text{AT}}))]$,
and $\beta_{d_{\text{AT}}} = \mathbb{E}[Y|D=d_{\text{AT}}, S=1]$.
\label{Tmin}
\end{lemma}

\begin{remark}[Binary treatment in \cite{LeeBound}]
Assumption~\ref{Amin} includes the familiar monotonicity assumption for a binary treatment in \cite{LeeBound} that assumes $S_1 \geq S_0$ a.s., i.e., if a subject in the control group $\{D=0\}$ is selected, then it remains selected if it was in the treated group $\{D=1\}$, i.e,  $d_{\text{AT}} = 0$.
So defiers (0-compliers) are excluded \citep{ImbensAngrist}.
Then Lemma~\ref{Tmin} implies Proposition 1a in \cite{LeeBound}: the upper bound of the always-takers' ATE, $\beta_1 - \beta_0 = \E[Y_1 - Y_0| S_0=S_1=1]$, is
$\rho_{1U}(s(0)) - \beta_{d_{\text{AT}}} = \E[Y|Y \geq Q^d(1-s(0)/s(1)), D=1, S=1] - \E[Y|D=0, S=1]$ and the lower bound is $\rho_{1L}(s(0)) - \beta_{d_{\text{AT}}}  = \E[Y|Y \leq Q^d(s(0)/s(1)), D=1, S=1] - \E[Y|D=0, S=1]$.

We remark that if we are only interested in two values of the continuous treatment, then the identification of the bounds for the binary treatment in \cite{LeeBound} can be directly applied to the case of two continuous treatment values.
But policy makers rarely consider only two values, and the always-takers at the two values could be different from the always-takers at another values; for example, $\{S_{d_1} = 1, S_{d_2}=1\} \neq \{S_{d_3} = 1, S_{d_4} = 1\}$.
New challenge in identification arises when we consider many treatment values.
\end{remark}


\subsection{Sufficient set assumption}
\label{SecIDSS}
We introduce a weaker sufficient set
Assumption~\ref{AminJ} that does not assume one sufficient treatment value $d_{\text{AT}}$ but assumes a  set $\mathcal{D}_M$ of $M$ treatment values, which includes Assumption~\ref{Amin} as a special case with $M=1$.
Assumption~\ref{AminJ} essentially assumes that always-takers $\{S_{d} =1: d\in\mathcal{D}\} = \{S_{d} = 1: d\in\mathcal{D}_M\}$.

\begin{assumption}[Sufficient set]
There exists a set of treatment values $\mathcal{D}_M := \big\{d_1, d_2,..., d_M \big\} \subseteq \mathcal{D}$ such that
$S_d \geq \min_{d' \in\mathcal{D}_M} S_{d'}$ a.s.\ for any $d\in\mathcal{D}$.
\label{AminJ}
\end{assumption}

To see how Assumption~\ref{AminJ} is weaker than Assumption~\ref{Amin} or a larger $M$ is weaker, consider an example:
Under Assumption~\ref{AminJ} with $M=3$ and $\mathcal{D}_3 = \{1,2,3\}$ , $\{S_1=0, S_2 = 1, S_3 = 0\}$ and $\{S_1=1, S_2 = 0, S_3 = 1\}$
are both possible values for a complier.
But Assumption~\ref{Amin} with $d_{\text{AT}} = 1$ does not allow a complier to take $\{S_1=1, S_2 = 0, S_3 = 1\}$.
What Assumption~\ref{AminJ} does not allow is this event $\{S_1=1, S_2 = 1, S_{2.5} = 0, S_3 = 1\}$ for example.  But assuming a larger $M=4$ and $\mathcal{D}_4 = \{1,2,2.5,3\}$ would allow a complier to take that.

However, the proportion of always-takers $\pi_{\text{AT}} = \p(S_{d} = 1: d\in\mathcal{D}_M)$ for $M \geq 2$ is not point-identified as we cannot observe the $M$ potential outcomes $\{S_{d}: d \in\mathcal{D}_M\}$ at the same time.
We can use the lower bound of $\pi_{\text{AT}}$ for trimming,
because $\rho_{dU}(\pi_{\text{AT}{}})$ is decreasing in $\pi_{\text{AT}{}}$ and $\rho_{dL}(\pi_{\text{AT}{}})$ is increasing in $\pi_{\text{AT}{}}$.
Theorem~\ref{TminJ} below provides the Lee bounds formally.

Specifically, consider a practical example of $M=2$ and $\mathcal{D}_2 = \{\underline{\mathcal{D}}, \overline{\mathcal{D}}\}$.
It implies that if a subject is selected at the boundary $\underline{\mathcal{D}}$ and $\overline{\mathcal{D}}$,
then this subject must be selected at any treatment value in $\mathcal{D}$.
This example of $M=2$ includes the special case when the selection's response is a concave function of treatment $D$.
The single-peaked pattern may fit well the law of marginal returns.
Therefore $S_{\underline{\mathcal{D}}} = 1$ and $S_{\overline{\mathcal{D}}} = 1$ if and only if  $S_{d} = 1$ for all $d\in\mathcal{D}$.
This observation gives the insight to use the well-known Fr\'{e}chet-Hoeffding bounds
for $\p(S_{\underline{\mathcal{D}}} = 1, S_{\overline{\mathcal{D}}} = 1)$ given in Theorem~\ref{TminJ}.
The bounds resemble the Fr\'{e}chet-Hoeffding bounds for $\p(S_0 = 1, S_1=1)$ for the binary treatment without shape restrictions, as shown in \cite{HSC}.
We require the sufficient set Assumption~\ref{AminJ} due to the continuous treatment variable.
It is important to note that
when the treatment is multivalued discrete with support $\mathcal{D} = \mathcal{D}_M$,
Assumption~\ref{AminJ} holds by construction and is dropped.
So Theorem~\ref{TminJ} provides the sharp bounds without restricting the selection response of the multivalued treatment.


\begin{theorem}
Let $\mathcal{D}_M = \big\{d_1, d_2,..., d_M\big\}\subseteq \mathcal{D}$ with a fixed dimension $M$.
Let Assumption~\ref{Aind} hold.  Then \begin{align*}
\pi_{L}^M := \max\left(
\sum_{d\in \mathcal{D}_M} s(d) - M + 1,
0\right)
\leq
\p(S_{d}=1: d\in\mathcal{D}_M)
\leq \min_{d\in \mathcal{D}_M} s(d)
=: \pi_{U}^M.
\end{align*}
Further let Assumption~\ref{AminJ} hold.
Then
$\pi_{\text{AT}} = \p(S_{d}=1: d\in\mathcal{D}_M) \in \big[\pi_L^M, \pi_U^M\big]$
and
$\beta_{d}  \in \big[\rho_{dL}(\pi_{L}^M), \rho_{dU}(\pi_{L}^M)\big]$.
The bounds are sharp.
\label{TminJ}
\end{theorem}

A final goal is to derive the bounds without restrictions on the selection response of a continuous treatment, i.e., dropping Assumption~\ref{AminJ}.
To make progress, we may assume that the treatment effect is a piecewise constant function $\beta_d = \sum_{m=1}^{M-1} \beta_{d_m} \1\{d \in [d_m, d_{m+1})\}$.
Then treatment can be effectively discretized and
$\beta_{d_m} \in [\rho_{d_mL}(\pi_{L}^M), \rho_{d_mU}(\pi_{L}^M)]$ for $m=1,..,M-1$.
In practice, one might discretize the continuous treatment variable into $M$-multivalued variable for sensitivity analysis.

In general and in theory, we'd like $M$ to be large to allow for a general non-separable nonparametric structural selection model, as discussed in Section~\ref{SecVytlacil}.
However, the bounds can be wide or less informative for a large $M$.
As shown in Theorem~\ref{TminJ}, $\pi_L^M$ can be small, unless $s(d)$ is close to one when there is selection bias.
We illustrate this tradeoff in Section~\ref{SecEmCT} by evaluating the Job Corps program and discuss how to choose $M$.



\begin{remark}[Binary outcome]
{\rm
\cite{KMV} provide the Lee bounds with a binary outcome and a binary treatment.
We can extend their bounds for a binary outcome to a continuous treatment:
$\rho_{dL}(\pi_L^M) = \max\{0, 1- \p(Y = 0|S=1, D=d)/p_d\}$
and $\rho_{dU}(\pi_L^M) = \min\{1, \p(Y=1|S=1, D=d)/p_d\}$, where $p_d = \pi_L^M/s(d)$.\footnote{
$\p(Y=1|S=1, D=d) = \p(Y=1|\text{AT}, D=d) p_d + \p(Y=1|\text{CP}_d, D=d) (1-p_d)$.
The bounds on $ \p(Y=1|\text{AT}, D=d) = \E[Y_d|\text{AT}]$ are obtained by the worst-case bounds of $\p(Y=1|\text{CP}_d, D=d)\in [0,1]$.
}
We focus on the continuous outcome in this paper and  develop the inference for the binary outcome in a separate paper.
}
\label{RBinaryY}
\end{remark}

\subsection{Structural selection equation}
\label{SecVytlacil}
To understand how the sufficient treatment value Assumption~\ref{Amin} and the sufficient set Assumption~\ref{AminJ} impose conditions on the heterogeneity in the structural selection equation, we discuss its relationship with the threshold-crossing model in \cite{Vytlacil02}.
Recall that our potential outcome framework is equivalent to the structural equation $S = \1\{q(D,\eta) \geq 0\}$ with unobserved non-separable and multi-dimensional error $\eta$.
The structural equation $q$ is nonparametric and model-free.
Then we can write the potential variable  $S_d =\1\{q(d, \eta) \geq 0\}$.
\\[5pt]
{\bf Assumption 2$^\prime$ (Latent index selection model)}
{\it (i) Let $S = \1\{q(D) \geq \eta\}$, where $q(d)$ is measurable and nontrivial function of $d$.
(ii) For a continuous treatment with a compact $\mathcal{D}$, there exists $d_{\text{AT}} = \arg\inf_{d\in\mathcal{D}} q(d) \in \mathcal{D}$.}

\medskip

Assumption~2$^\prime$ implies Assumption~\ref{Amin}.
For a multivalued treatment with a finite countable $\mathcal{D}$, $d_{\text{AT}}$ exists under Assumption~2$^\prime$(i).
For a continuous $D$, assuming  $q(\cdot)$ in Assumption~2$^\prime$(i) to be continuous, the extreme value theorem implies (ii).
This important observation suggests our new sufficient treatment value Assumption~\ref{Amin} not restrictive and implied by a common threshold-crossing model with a separable error, or the latent index selection model in \cite{Vytlacil02}.\footnote{
\cite{Vytlacil02} shows that the latent index selection model Assumption~2$^\prime$(i) is equivalent  to the local average treatment effect (LATE) model with Independence (as our Assumption~\ref{Aind}) and Monotonicity assumptions in \cite{ImbensAngrist}.
The LATE Monotonicity assumes the orders of $S_d$ to be the same for everyone, i.e., for all $(d, d')\in\mathcal{D}\times\mathcal{D}$, either $S_d \geq S_{d'}$ $a.s.$, or $S_d \leq S_{d'}$ $a.s$.
Our Assumption~\ref{Amin}  (or Assumption~2$^\prime$(ii)) is weaker than such Monotonicity assumption and only requires the sufficient treatment value $d_{\text{AT}}$ (or  a minimizer of $q(d)$) exists.
}

In the selected sample at $d$, $\{S=1, D=d\} = \{S_d=1, D=d\}$, the subpopulation $\{S_d=1\} = \{\eta \leq q(d)\} = \text{AT} \cup \text{CP}_d$, where  always-takers $\text{AT} = \{\eta \leq q(d_{\text{AT}})\}$ and  $d$-compliers $\text{CP}_d = \{q(d_{\text{AT}}) < \eta \leq q(d)\}$.
The sufficient treatment value has meaningful economic interpretation.
For example, \cite{BCGL} interpret $\eta$ as the individual reluctance to respond to surveys and call the compliers as the marginal respondents.
 Then the sufficient treatment value can be the least-favored treatment value to induce responding to surveys.
 So if  subjects are willing to respond to surveys when receiving $d_{\text{AT}}$, then they continue responding to surveys when receiving any other treatment values.

To further appreciate  always-takers as the target population, we discuss the gettable ATE (GATE) $\E[Y_1 - Y_0|\eta \leq k]$ for some constant $k$, defined by \cite{DiNardo}.
As $k$ increases, GATE converges to the population ATE $\E[Y_1 - Y_0]$.
As subpopulation-specific ATEs are commonplace,
\cite{DiNardo} show how different GATE parameters may be identified under weaker assumptions than in the traditional parametric framework.
We choose $k = q(d_{\text{AT}})$ to characterize always-takers that are the largest subpopulation for whom we can partially identify the ATE of switching treatment values over $\mathcal{D}$, without imposing further assumptions on the functional forms or distributions.

Now we consider a more general non-separable nonparametric model in Assumption 3$^\prime$ that implies  our sufficient set Assumption~\ref{AminJ}.
\\[5pt]
{\bf Assumption 3$^\prime$ (Latent index selection model with non-separable errors)}
{\it Let $S = \1\{q(D,\eta) \geq 0\}$. There exists $d_{\text{AT}}(\eta) = \arg\inf_{d\in\mathcal{D}} q(d, \eta) \in \mathcal{D}_M$ for each $\eta$.}

The unobserved heterogeneity is captured by $\eta$, so there could be an infinite number of types and  the corresponding sufficient value $d_{\text{AT}}(\eta)$ in the most general structural model.
Our sufficient set Assumption~\ref{AminJ} restricts there to be $M$ types of unobserved heterogeneity $\eta$,
in the sense that $d_{\text{AT}}(\eta)$ belongs to $\mathcal{D}_M$.

Under Assumption~\ref{AminJ}$^\prime$,
Assumption~\ref{Amin} can be implied by further assuming $d_{\text{AT}}(\eta) = d_{\text{AT}}$ to be a constant for all individual with $\eta$, and $M=1$.
Therefore we argue that the sufficient value Assumption~\ref{Amin} is reasonable with a separable structural error in selection
as in Assumption 2$^\prime$,
and the sufficient set Assumption~\ref{AminJ} is a useful approximation under more general non-separable errors.

\section{Estimation and inference}
\label{SecInf}
We estimate bounds over an equally spaced grid $\mathcal{D}_J = \{d_1,..,d_J\} \subset \mathcal{D}$ for a continuous treatment.
We can view $\mathcal{D}_J$ as the set of treatment values where the policy maker considers treatment intervention.
The estimation procedure is easy to implement, as the bounds estimates  are sample analogs to the parameters defined in Theorem~\ref{TminJ}.
When $D$ is continuous, we use a kernel function $K_h(D-d) = k((D-d)/h)/h$, where the kernel function $k$ includes a sub-population whose treatment is around $d$, and the size of the sub-population is controlled by the bandwidth $h$ shrinking to zero as the sample size grows.
We provide inference on the causal effect of switching treatment between any two values in $\mathcal{D}_J$.
We present the asymptotic theory for a fixed $J$ and also for $J \rightarrow \infty$ so $\mathcal{D}_J \rightarrow \mathcal{D}$.

For a multivalued discrete treatment,
it is straightforward to use the treatment indicator $\1\{D=d\}$ in place of the binary treatment indicator $D$ in the estimator in \cite{LeeBound} and let $\mathcal{D}_J$ be the (sub)support of $D$.
We develop the inference theory for a discrete treatment in a separate paper.

We implement the estimation procedure using leave-out estimators for $s(d)$ and $Q^d$ in Step 1 and Step 2.
The leave-out estimation is similar to the cross-fitting in the recent double debiased machine learning literature \citep{CCDDHNR}.
The leave-out preliminary estimation achieves stochastic equicontinuity without strong entropy conditions using empirical process theory.
Specifically, for some fixed $L \in \{2,...,n\}$, randomly partition the observation indices into $L$ distinct groups $I_\ell, \ell = 1,...,L$, such that the sample size of each group is the largest integer smaller than $n/L$.
The number of folds $L$ is not random and typically small, such as five
or ten in practice; see, e.g., \cite{CCDDHNR}, \cite{Velez}.
When there is no sample splitting ($L = 1$), $\hat s_1(d)$ and $\hat Q^d_1$ use all observations in the full sample.\footnote{
When $L = n$, $\hat s_\ell$ uses all observations except for the $\ell^{th}$ observation, and is well-known as the leave-one-out estimator (i.e., leave the $\ell^{th}$ observation out), e.g., \cite{PSS89ETA}.
A large $L$ is computationally costly.}

The estimation procedure under Assumption~\ref{Amin} follows four steps:

\paragraph{Step 0.} Estimate the sufficient treatment value $\hat d_{\text{AT}_J} = \arg\min_{d \in \mathcal{D}_J} \hat s(d)$, where
a kernel estimator  $\hat s(d) = \sum_{i=1}^n S_i K_h(D_i - d)/\sum_{i=1}^n K_h(D_i - d)$.
For $d = \hat d_{\text{AT}_{J}}$,
$\hat\beta_{d} =  \sum_{i=1}^n Y_i S_i K_h(D_i - d)/\sum_{i=1}^n S_i K_h(D_i - d)$.
Estimate the proportion of always-takers $\pi_{\text{AT}}$ by $\hat\pi = \min_{d\in\mathcal{D}_J}\hat s(d) =  \hat s(\hat d_{\text{AT}_J})$. 

\bigskip

For $\ell = 1,..., L$, the estimators in Step 1 and in Step 2
use observations not in $I_\ell$, denoted as $I_\ell^c := \{1,...,n\} \setminus I_\ell$.
\paragraph{Step 1.} 
Compute the leave-out kernel estimator  $\hat s_\ell(d) = \sum_{i\in I_\ell^c} S_i K_h(D_i - d)/\sum_{i\in I_\ell^c} K_h(D_i - d)$.
Estimate the trimming probability $p_d$ by $\hat p_\ell = \min\{\hat s_\ell(\hat d_{\text{AT}_J})/\hat s_\ell(d), 1\} -\nu$ for some small positive constant $\nu$ used for robust inference that we explain in the following.

\paragraph{Step 2.}

For $d \neq \hat d_{\text{AT}_{J}}$, estimate the $(1-\hat p_\ell)$-quantile of $Y|D=d, S=1$ by $\hat Q_\ell^d(1-\hat p_\ell)$.
A nonparametric estimator can be the generalized inverse function of the CDF estimate
$\hat F_{Y|D=d,S=1_\ell}(y) = \sum_{i\in I_\ell^c}\1\{Y_i \leq y\} S_i K_h(D_i - d)/\sum_{i\in I_\ell^c} S_i K_h(D_i - d)$.

\paragraph{Step 3.} 

Compute the kernel estimator of $\E[Y{\bf 1}\{Y \geq Q^d(1-p)\}| D=d, S=1]$ and obtain
\begin{align*}
\widehat{\rho_{dU}(\pi)} = \frac{\sum_{\ell=1}^L\sum_{i \in I_\ell}  Y_i{\bf 1}\{Y_i \geq \hat Q_\ell^d(1-\hat p_\ell)\} S_i K_h(D_i - d)}{\sum_{i=1}^n S_i K_h(D_i - d)} \frac{1}{\hat p},
\end{align*}
where $\hat p = \hat\pi/\hat s(d) - \nu$ using the full sample from Step 0.

\medskip



Our inference procedure is robust to extensive margin, allowing the selection probability to be of any unknown functional form with respect to treatment.
The challenge for the continuous or multivalued treatment relative to the well studied binary treatment case comes from an infinite or multiple number of treatment values ($J$) and the corresponding potential selections and outcomes.
Importantly, we allow the sufficient treatment value to be not unique in the sense that there exists a subset  $\mathcal{D}_c \subseteq\mathcal{D}$ such that
$d= \arg\min_{d'\in\mathcal{D}} s(d')$ for any $d \in \mathcal{D}_c$ and $\p(D \in \mathcal{D}_c) > 0$.
So $s(d)$ is constant over $d\in\mathcal{D}_c$, which is implied by no treatment effect on selection (extensive margin) when changing $d$ within the subset $\mathcal{D}_c$.
Then we could use any $d\in\mathcal{D}_c$ as a sufficient treatment value  $d_{\text{AT}}$.
So for any $d \in \mathcal{D}_c$, $s(d_{\text{AT}})/s(d)=1$ and $\beta_d = \rho_{dU}(\pi_{\text{AT}}) = \rho_{dL}(\pi_{\text{AT}})= \E[Y|D=d, S=1] $ is point-identified.
However the asymptotic distributions of the bound-estimators $\big[\widehat{\rho_{dL}(\pi_{\text{AT}})}, \widehat{\rho_{dU}(\pi_{\text{AT}})}\big]$
do not converge to the asymptotic distribution of the point-estimator $\hat\beta_d$ as $s(d_{\text{AT}})/s(d)\rightarrow 1$.
We avoid the complication in testing $s(d) = s(d_{\text{AT}})$, e.g., if the hypothesis $s(d_{\text{AT}})/s(d)=1$ is not rejected, then compute the point-estimator $\hat \beta_d$.
Instead, we estimate the tight bounds using the trimming probability $\hat p_\ell = \min\{\hat s_\ell(\hat d_{\text{AT}_{J}})/\hat s_\ell(d), 1\} - \nu \stackrel{p}{\rightarrow} p_d = s(d_{\text{AT}})/s(d) - \nu \leq 1-\nu < 1$, which contain the untrimmed point-estimator $\hat\beta_{d} =  \sum_{i=1}^n Y_i S_i K_h(D_i - d)/\sum_{i=1}^n S_i K_h(D_i - d)$.\footnote{
In the proof of Theorem~\ref{TEstLB}, we show that $\hat d_{\text{AT}_{J\ell}} = \arg\min_{d\in\mathcal{D}_J}\hat s_\ell(d) = \hat d_{\text{AT}_J}$ when $n$ large enough.
So $\hat p_\ell = \hat s_\ell( \hat d_{\text{AT}_J}) /\hat s_\ell(d) - \nu$.
However, in finite samples, it is possible that $\hat s_\ell(\hat d_{\text{AT}_J})/\hat s_\ell(d) > 1$ for some $d\in\mathcal{D}_J$.
In such case, if $\hat d_{\text{AT}_{J}}$ and $\hat d_{\text{AT}_{J\ell}}$ are in $\mathcal{D}_c$, then
we let $\hat p_\ell \leq 1-\nu$ and estimate bounds of $\beta_{ \hat d_{\text{AT}_{J\ell}}}$ that contain its point-estimate.
}

Therefore we estimate bounds  that are non-sharp with the trimming probability $p_d = s(d_{\text{AT}})/s(d) - \nu$ but still tight with small $\nu$.
We choose this practical and conservative strategy so that our asymptotic theorem and inference are valid regardless of the extensive margin effect on selection, and are easy to implement and interpret.

We show in Theorem~\ref{TEstLB} that the estimation errors of $\hat d_{\text{AT}_{J}}$ and grid approximation are asymptotically ignorable.
That is, for any given $J$ and for $n$ large enough, $\hat d_{\text{AT}_{J}} = d_{\text{AT}_J} := \arg\min_{d \in \mathcal{D}_J} s(d)$.
So for a any fixed set of treatment values $\mathcal{D}_J$, we can find the sufficient treatment value $d_{\text{AT}_J}$ given a large enough sample.
As $\mathcal{D}$ contains an infinite number of values for a continuous treatment, we give conditions on the grid size $J$ going to infinity to approximate $\mathcal{D}$.
We show that as $J, n\rightarrow \infty$, $d_{\text{AT}_J}  \rightarrow d_{\text{AT}} := \arg\min_{d \in\mathcal{D}} s(d)$.
Assumption~\ref{Agrid} below gives conditions on $J$ that depends on the accuracy of $\hat s_\ell(d)$ and the shape of $s(d)$ characterized by $\bar M$.

\begin{assumption}
Let $s^{(m)}(d)$ be the $m^{th}$ derivative of $s(d)$ for $m \in \{1,2,...\}$.
Let $\mathcal{D} = \mathcal{D}_s \cup \mathcal{D}_c$, where $\mathcal{D}_c := \{d: s^{(m)}(d) = 0, \forall m \geq 1\}$
and $\mathcal{D}_s := \{d: s^{(m)}(d) \neq 0, \exists m < \infty\}$.
If $\mathcal{D}_s \neq \emptyset$, let $\bar M = \min\{m: s^{(m)}(d) \neq 0, m=1,2,..., \forall d \in \mathcal{D}_s\} < \infty$.
If $\mathcal{D}_s = \emptyset$, let $\bar M = 0$.
Let an equally spaced grid $\mathcal{D}_J = \{d_1,..,d_J\} \subseteq \mathcal{D}$ with
$J = O({\mathsf{s}_n}^{-1/\bar M})$ and $\mathsf{s}_n := \sup_{d\in\mathcal{D}}|\hat s(d)- s(d)| = o_\p(1)$.
\label{Agrid}
\end{assumption}

$\mathcal{D}_c$ is the set of treatment values not affecting the selection.
For $d \in \mathcal{D}_c$,  $s(d) = c$ for some generic constant $c$, so $s'(d) = 0$.
For $\bar M = 0$ ($\mathcal{D} = \mathcal{D}_c$), there is no extensive margin, so $\beta_d = \E[Y|D=d, S=1]$ is point-identified, and there is no restriction on $J \rightarrow \infty$.
When $s(d)$ is strictly monotone over $\mathcal{D}$, $s'(d) \neq 0$, so $\bar M = 1$ and $\mathcal{D}_c = \emptyset$.
When $s(d)$ is strictly concave, i.e., $s'(d) = 0$ for a $d\in\mathcal{D}$ and $\mathcal{D}_c = \emptyset$, we have $\bar M = 2$.
A larger $\bar M$ implies that it is harder to compare $s(d_j)$ and $s(d_{j+1})$, so we need to estimate $s(d)$ more accurately, i.e., $\mathsf{s}_n$ needs to go to zero faster for a larger $\bar M$ on a given grid.

The kernel estimation is well-studied, and we use the results in \cite{DHB} and \cite{HansenBook}.
\begin{assumption}
\begin{enumerate}
\item[(i)]
The kernel function $k$ is non-negative symmetric bounded kernel with a compact support such that
$\int k(u) du = 1$, $\int u k(u) du = 0$, and $\kappa := \int u^2 k(u) du < \infty$.
Let the roughness of the kernel be $R_k := \int k(u)^2du$.


\item[(ii)] $h\rightarrow 0$, $nh\rightarrow\infty$, $nh^5 \rightarrow c\in [0,\infty)$.


\item[(iii)] 
For $d \in \mathcal{D}$ and $y\in\mathcal{Y}$, $s(d) < 1$; $f_{Y|DS}(y|d,1)$ is continuous and bounded away from $0$;
$F_{Y|DS}(y|d,1) s(d) f_D(d)$ is bounded and has bounded continuous second derivative with respect to $d$;
$var(Y|D=d, S=1)$, $\E\big[|Y|^3\big|D=d, S=1\big]$, and  the second derivative of $\E[Y|D=d, S=1]$
are continuous in $d$.

\end{enumerate}
\label{AEstLB}
\end{assumption}

\begin{theorem}
Let Assumptions~\ref{Aind},  \ref{Agrid}, and \ref{AEstLB} hold.
\begin{enumerate}
\item Under Assumption~\ref{Amin}, $\pi = s(d_{\text{AT}_J})$.
Then for $d\in\mathcal{D}_J$ and $d \neq \hat d_{\text{AT}_J}$, as $n\rightarrow \infty$,
\begin{align}
\sqrt{nh}\Big(\widehat{\rho_{dU}(\pi)} - \rho_{dU}(\pi) - h^2{B}_{dU}\Big) &\stackrel{d}{\rightarrow}\mathcal{N}(0,V_{dU})
\notag \\
\sqrt{nh}\Big(\widehat{\rho_{dL}(\pi)} - \rho_{dL}(\pi) - h^2{B}_{dL}\Big) &\stackrel{d}{\rightarrow}\mathcal{N}(0,V_{dL}),
\label{EAsyD}
\end{align}

where $p = p_d = s(d_{\text{AT}_J})/s(d) - \nu$,
$V_{dU} := p^{-2} \big(V_1 + V_2 + V_{3} + V_{23}\big) R_k /(s(d)f_D(d))$,
$V_{3} := var(Y{\bf 1}\{Y \geq Q^d(1-p)\}|D=d, S=1),
V_2 := p(1-p)Q^d(1-p)^2,
V_{23} := -2p(1-p) \rho_{dU}(\pi) Q^d(1-p),
V_1:=
\big(V_\pi+ p^2 V_{s(d)}\big)s(d)^{-1}
\big(
Q^d(1-p)  - \rho_{dU}(\pi)
\big)^2,$
with $V_{s(d)} := s(d)(1-s(d))$ and $V_\pi = V_{s(d_{\text{AT}_J})}f_D(d)/f_D(d_{\text{AT}_J})$.
For the lower bound, $V_{dL} := p^{-2} \big(V_1 + V_2 + V_{3}^L + V_{23}\big) R_k /(s(d)f_D(d))$,
where $V_{3}^L := var(Y{\bf 1}\{Y \leq Q^d(p)\}|D=d, S=1)$, and
 $V_1$, $V_2$, $V_{23}$ are defined as above with $Q^d(p)$ in place of $Q^d(1-p)$ and with $\rho_{dL}(\pi)$ in place of $\rho_{dU}(\pi)$.
 ${B}_{dU}$ and ${B}_{dL}$ are given explicitly in the proof in the Appendix.

For $d = \hat d_{\text{AT}_J}$,
$\sqrt{nh}\big(\hat\beta_{d_{\text{AT}_J}} - \beta_{d_{\text{AT}_J}} - h^2 B_3\big) \stackrel{d}{\rightarrow} \mathcal{N}\big(0, V_{d_{\text{AT}_J}}\big)$ as $n \rightarrow \infty$,  where $V_{d_{\text{AT}_J}} =  var(Y|D=d_{\text{AT}_J}, S=1) R_k /(s(d_{\text{AT}_J})f_D(d_{\text{AT}_J}))$.

As $J \rightarrow\infty$, $d_{\text{AT}_J} \rightarrow d_{\text{AT}}$ and the above statements hold with $d_{\text{AT}}$ in place of $d_{\text{AT}_J}$.

\item Under Assumption~\ref{AminJ}, choose $\mathcal{D}_M$ to contain $\hat d_{\text{AT}_J}$.
Let $\hat p_\ell = \hat\pi_{L\ell}^M/\hat s_\ell(d)$ in Step 1, and follow Step 2 and Step 3 to obtain the bounds $[\widehat{\rho_{dL}(\pi)}, \widehat{\rho_{dU}(\pi)}]$, where  $\pi = \sum_{d\in\mathcal{D}_M} s(d) - M + 1 > 0$.
Let $V_\pi =
 \sum_{m=1}^M V_{s(d_m)} f_D(d)/f_D(d_m)$.
For $d \in \mathcal{D}_M^c \cap\mathcal{D}_J$, the above asymptotic distributions (\ref{EAsyD}) hold.
For $d \in \mathcal{D}_M \cap\mathcal{D}_J$, the above asymptotic distributions (\ref{EAsyD}) hold with $V_1:=\left(V_\pi + (p^2-2p) V_{s(d)}\right)s(d)^{-1}\left(Q^d(1-p)  - \rho_{dU}(\pi)\right)^2$.

\end{enumerate}

\label{TEstLB}
\end{theorem}

Notice that we allow for no extensive margin, e.g., 
there exists $d \neq d_{\text{AT}_J}$ and $s(d) = s(d_{\text{AT}_J})$.
So we use $p_d = 1 - \nu$ to estimate tight bounds of $\beta_d$, rather than a point-estimand.

The asymptotic variance $V_{dU}$ can be decomposed to the three steps of the estimation procedure:
$V_1$ is from estimating the effect on selection and the trimming probability in Step 1.
$V_2$ is from the quantile regression in Step 2.
$V_3$ comes from the Step 3 trimmed regression.
$V_{23}$ is from the covariance of the Step 2 and Step 3 estimation errors.

Estimation of the variances is easily carried out by replacing all of the above quantities with their sample analogs
or by the sample variances of the influence functions given in equation (\ref{IF}) in the proof of Theorem~\ref{TEstLB} in the appendix.
No additional preliminary estimators are needed.
The confidence interval of at least 95\% coverage can be computed by
$\Big[\widehat{\rho_{dL}(\pi)}-1.96 \times \widehat{\sigma_{dL}}/\sqrt{nh},
\widehat{\rho_{dU}(\pi)} + 1.96 \times \widehat{\sigma_{dU}}/\sqrt{nh} \Big]$ with $\widehat{\sigma_{dL}} := \sqrt{\widehat{V_{dL}}}$
and $\widehat{\sigma_{dU}} := \sqrt{\widehat{V_{dU}}}$.
This interval will asymptotically contain the region $\left[\rho_{dL}(\pi), \rho_{dU}(\pi)\right]$ with at least 95\% probability.




We can bound the ATE of increasing the treatment from $d_1$ to $d_2$,
$\underline{\Delta}_{d_1d_2} := \rho_{d_2L}(\pi) - \rho_{d_1U}(\pi) \leq \beta_{d_2} - \beta_{d_1} \leq
\rho_{d_2U}(\pi) - \rho_{d_1L}(\pi) =: \bar\Delta_{d_1d_2}$.
We note that one might be interested in the partial (or marginal) effect defined as the derivative of $\beta_d$ with respect to $d$, i.e., $\theta_d =  \frac{\partial}{\partial d}\beta_d$.
However, we could not bound such derivative by the same approach of \cite{LeeBound}.
We can view $\Delta_{d_1d_2} := \beta_{d_2} - \beta_{d_1} = \int_{d_1}^{d_2}\theta_s ds$ as an average derivate over $d_1$ to $d_2$.
Corollary~\ref{CATE} provides the asymptotic distribution of the bounds estimators for the ATE $\hat{\underline{\Delta}}_{d_1d_2} := \widehat{\rho_{d_2L}(\pi)} - \widehat{\rho_{d_1U}(\pi)}$ and
$\hat{\bar\Delta}_{d_1d_2} := \widehat{\rho_{d_2U}(\pi)} - \widehat{\rho_{d_1L}(\pi)}$.
Denote the bandwidth in $ \widehat{\rho_{dL}(\pi)}$ and $\widehat{\rho_{dU}(\pi)}$ be $h_{dL}$ and $h_{dU}$, respectively.
\begin{corollary}[ATE]
Let the conditions in Theorem~\ref{TEstLB} hold.
Then $\sqrt{n}V_{Un}^{-1/2}\big(\hat{\bar\Delta}_{d_1d_2} - \bar\Delta_{d_1d_2} - (h_{d_2U}^2{B}_{d_2U} -h_{d_1L}^2{B}_{d_1L})\big) \stackrel{d}{\longrightarrow} \mathcal{N}(0, 1)$
and $\sqrt{n} V_{Ln}^{-1/2}\big(\hat{\underline{\Delta}}_{d_1d_2} - \underline{\Delta}_{d_1d_2}- (h_{d_2L}^2{B}_{d_2L} -h_{d_1U}^2{B}_{d_1U})\big) \stackrel{d}{\longrightarrow} \mathcal{N}(0, 1)$,
where $V_{Un} :=  \E[(\phi_{d_2U} - \phi_{d_1L})^2]$
and $V_{Ln}:=   \E[(\phi_{d_2L} - \phi_{d_1U})^2]$ with the influence functions~$\phi_{dL}, \phi_{dU}$ given explicitly in equation (\ref{IF})  in the Appendix.

Assume Assumption~\ref{Amin}.
When $h_{d2U} = h_{d_1L} = h$, $\lim_{n\rightarrow \infty, h\rightarrow 0} h V_{Un} =
\mathsf{V}_{d_2U} + \mathsf{V}_{d_1L} - 2\mathsf{C}_{d_1d_2U}$.
When $h_{d_2L}=h_{d_1U} = h$, $\lim_{n\rightarrow \infty, h\rightarrow 0} h V_{Ln} = \mathsf{V}_{d_2L} + \mathsf{V}_{d_1U} -2 \mathsf{C}_{d_1d_2L}$,
where $p_{d} = \pi/s(d)$, $\mathsf{C}_{d_1d_2U}:= R_kV_\pi (Q^{d_2}(1-p_{d_2}) - \rho_{d_2U}(\pi))(Q^{d_1}(p_{d_1}) - \rho_{d_1L}(\pi))/(\pi^2f_D(d_{\text{AT}}))$, and $\mathsf{C}_{d_1d_2L}:= R_kV_\pi (Q^{d_1}(1-p_{d_1}) - \rho_{d_1U}(\pi))(Q^{d_2}(p_{d_2}) - \rho_{d_2L}(\pi))/(\pi^2f_D(d_{\text{AT}}))$.
\label{CATE}
\end{corollary}
The variance can be estimated by the plug-in sample analogues  $\hat{\mathsf{V}}_{Un} = n^{-1}\sum_{i=1}^n(\hat\phi_{d_2Ui} - \hat\phi_{d_1Li})^2$
and
$\hat{\mathsf{V}}_{{Ln}} = n^{-1}\sum_{i=1}^n(\hat\phi_{d_2Li} - \hat\phi_{d_1Ui})^2$.
And the 95\% confidence interval $\big[\hat{\underline{\Delta}}_{d_1d_2} - 1.96\times \sqrt{\hat{\mathsf{V}}_{Ln}}/\sqrt{n},
\hat{\bar\Delta}_{d_1d_2} + 1.96\times  \sqrt{\hat{\mathsf{V}}_{Un}}/\sqrt{n} \big]$.

Next we briefly discuss how to choose an undersmoothing bandwidth $h$ smaller than
the optimal bandwidth that minimizes the asymptotic mean squared error (AMSE) such that the bias is first-order asymptotically negligible, i.e., $h^2\sqrt{nh}\rightarrow 0$, so the above confidence interval is valid.

We could estimate the leading bias ${B}_{dU}$ by the method in \cite{PS96} and \cite{CL}.
Let the notation $\widehat{\rho_{dU,b}(\pi)}$ be explicit on the bandwidth $b$ and
$\widehat{{B}}_{dU}:= \big(\widehat{\rho_{dU,b}(\pi)}
- \widehat{\rho_{dU,ab}(\pi)}\big)\big/\big(b^2(1-a^2)\big)$
with a pre-specified fixed scaling parameter $a\in(0,1)$.
From the proof of Theorem 3.2 in \cite{CL}, we can choose $a$ by minimizing the leading term of $var(\hat{B}_{dU})$, i.e., minimizing $(1-a^2)^{-2}a^{-d_T}$ for a $d_T$-dimensional continuous treatment. By deriving the first-order and second-order conditions, we obtain the minimizer $a^*= \sqrt{d_T/(d_T+4)} = \sqrt{1/5}$, for $d_T = 1$ in our case.

Then we propose a data-driven bandwidth $\hat h_{dU}:= (\widehat{V_{dU}}/(4\widehat{{B}}_{dU}^2))^{1/5}n^{-1/5}$
to consistently estimate the AMSE optimal bandwidth $h_{dU}^\ast$ by Theorem 3.2 in \cite{CL}.
For the lower bound, the same estimation applies to the leading bias $\widehat{{B}}_{dL}$ and the AMSE optimal bandwidth $\hat h_{dL}$.

\section{Empirical illustration: Job Corps}
\label{SecEmCT}
We illustrate our method by evaluating the Job Corps program.
We use the Job Corps dataset in \cite{HHLLdata}.
The continuous treatment variable ($D$) is the total hours spent in academic and vocational training.
The outcome variable ($Y$) is  the weekly earnings in the fourth year.
Our sample consists of 4,024 subjects who completed at least 40 hours (one week) of training.
In the online appendix, Figure~\ref{HisT} shows the distribution of $D$, and Table~\ref{TSS_JC} provides brief descriptive statistics.
As our analysis builds on \cite{FFGN12ReStat}, \cite{HHLP}, \cite{HHLL}, we refer the readers to the reference therein for further details of Job Corps.

We use $J=100$ grid points over $\mathcal{D}= [40, 2400]$ that ranges from one week to fifteen months.
We use the Epanechnikov kernel with a undersmoothing bandwidth
 and ten-fold cross-fitting.
The undersmoothing bandwidth at each $d$ is chosen by $0.8\times \min(\hat h_{dL},\hat h_{dU})$, given in Section~\ref{SecInf}, where
we use the rule-of-thumb bandwidth $h_1 = 1.05\times\hat\sigma_D\times n_\ell^{-1/5}\times c_1$ with a constant $c_1=1$ and the sample standard deviation of $D$, $\hat\sigma_D$, to estimate the initial variance and bias with $a = \sqrt{0.2}$ and $b = h_1/a$.
The resulting undersmoothing bandwidth ranges from 349.17 to 626.91. 
The estimated trimming probability is capped by $\hat p_\ell \leq 1 - \nu$ with $\nu = 0.01$.



\begin{figure}[!htp]
\centering
\caption{\small (Job Corps) Estimated selection probability and bounds without covariates}
\includegraphics[width=0.45\textwidth]{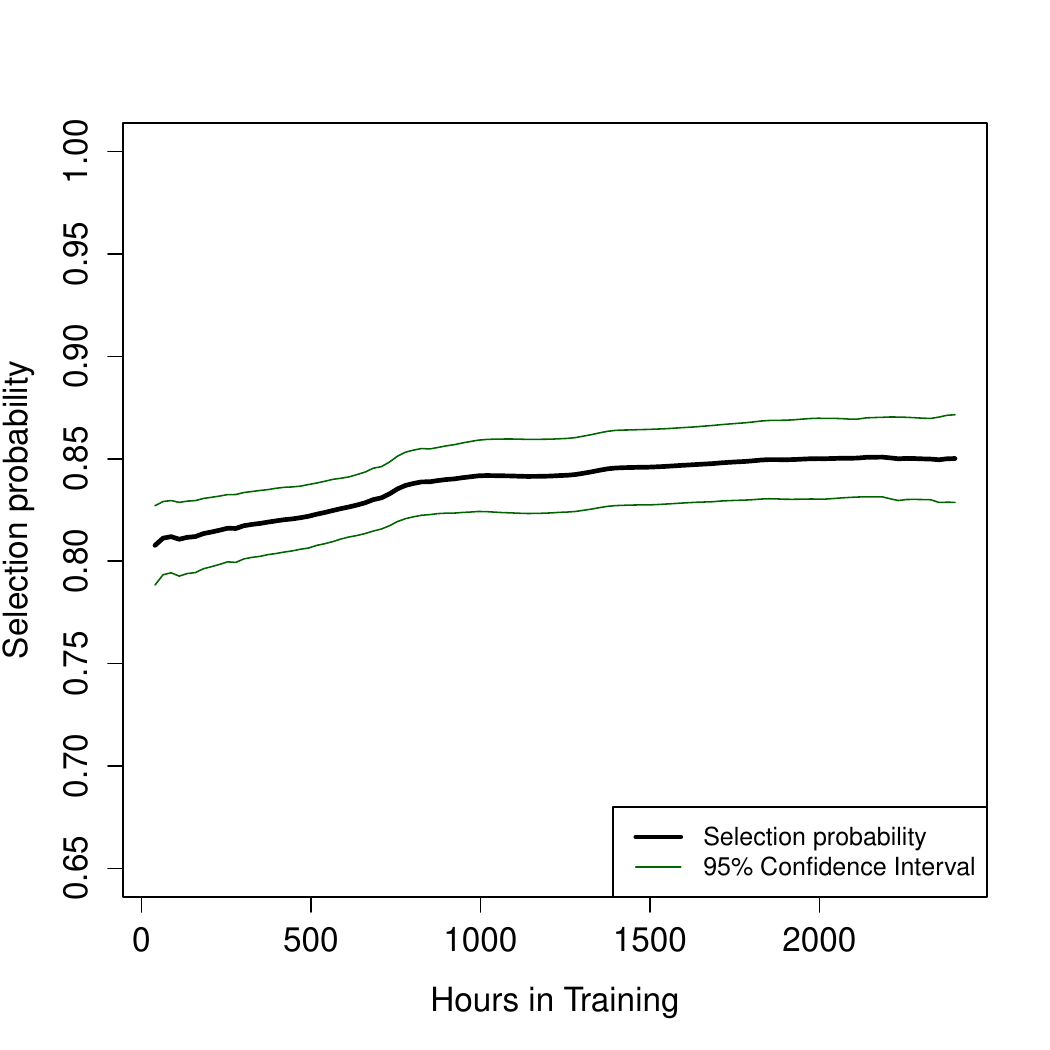}
\includegraphics[width=0.45\textwidth]{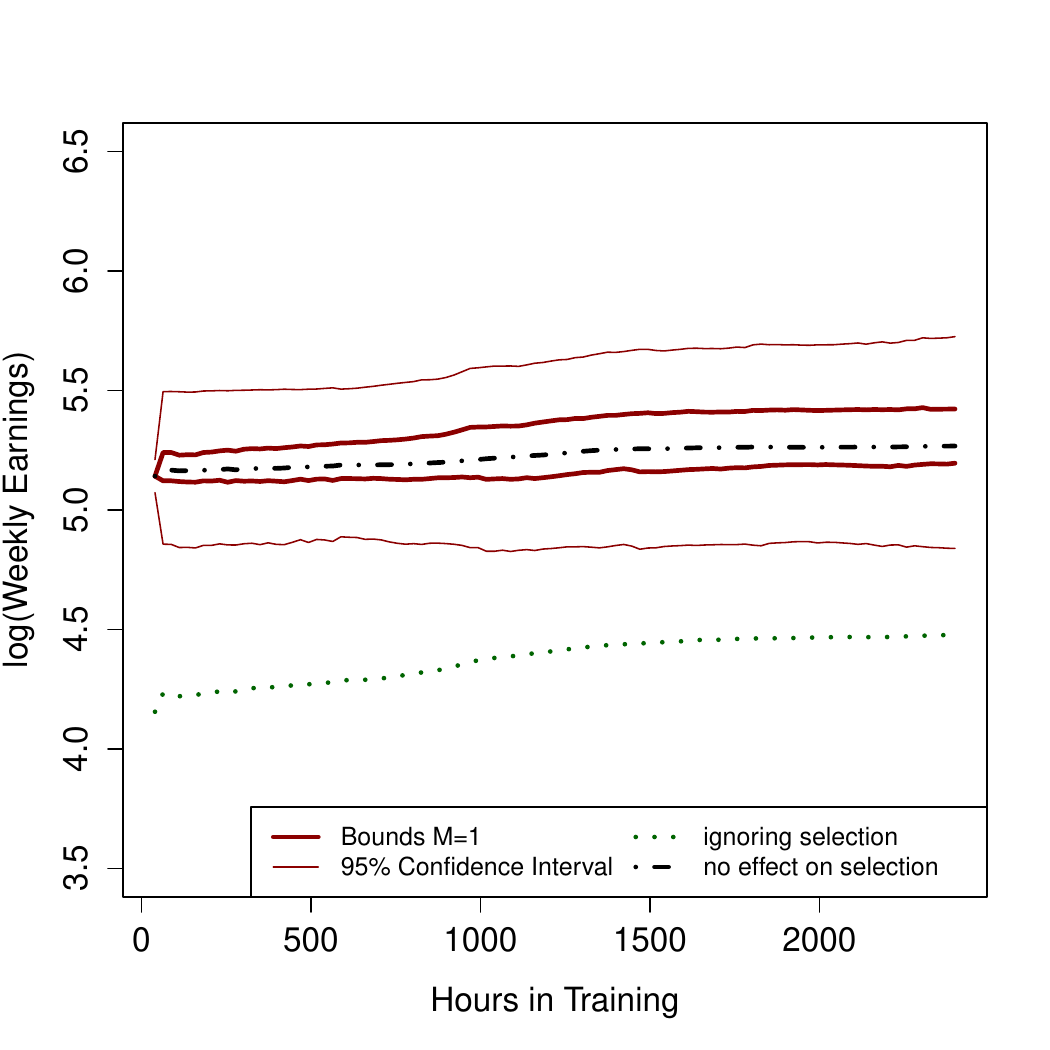}
\label{Fbetad}
\end{figure}

To learn about the effects on selection (or the extensive margin effect), we estimate the average dose-response function of selection $\mathbb{E}[S_d] = s(d)$.
The left panel of Figure~\ref{Fbetad} presents the estimates of the conditional selection probability given $d$, $\hat s(d)$.
More training seems to increase the extensive margin.
The minimum selection probability estimate is $\hat\pi_{\text{AT}} = \min_{d\in\mathcal{D}_J} \hat s(d) =\hat s(40) \approx  0.8082$.
So we use the least treatment value $40$ as the sufficient treatment value, i.e.,
if a participant is employed with one-week training, then this participant will remain employed when receiving more training between one week to fifteen months.

In Figure~\ref{FJCHY} in Section~\ref{ASecJC} in the online appendix, the histograms of $Y$ and $\log(Y)$ in the selected sample $\{Y_i > 0$, or $S_i=1, i=1,...,n\}$ show that weekly earnings has a skewed distribution, and  $\log$(weekly earnings) is closer to normal.
It is well-known (e.g., \cite{ChenRoth}) that averages can be heavily influenced by observations in the tail, especially when the outcome has a skewed distribution, as the weekly earnings in the Job Corps data in Figure~\ref{FJCHY}.
So we estimate the ATE in log, i.e., let $\beta_d = \mathbb{E}[\log(Y_d)|\text{AT}]$, a concave transformation of the outcome that is less heavily influenced by outcomes in the tail of the distribution.
The right panel of Figure~\ref{Fbetad}
shows the estimated bounds  under Assumption~\ref{Amin} with $d_{\text{AT}} = 40$ ($M=1$).
The thinner lines are the 95\% confidence intervals  using the sample variance of the influence function.
We find the largest ATE $\beta_{2400} - \beta_{40} = \E[\log(Y_{2400}) - \log(Y_{40})|\text{AT}]$ bounded by $[0.054, 0.281]$  with a 90\% confidence $[-0.251, 0.543]$.
That is, increasing the training hours from one week to fifteen months would increase the weekly earnings by at least 5.4\%, but this effect is not significant at 10\%  level.
We also estimate the ATE in level and do not find any effects of switching hours within $[40,2400]$; see Figure~\ref{FbetaY} in Section~\ref{ASecJC} in the online appendix.

For comparison, the black dash-dotted line is the estimate of $\mathbb{E}[Y|D=d, S=1]$ assuming that there is no effect on selection, i.e., the selected sample over $\mathcal{D}$ contains only the always-takers, $\{S_{d'}=1 : d'\in\mathcal{D}\} = \{S_d =1\}$ for all $d\in\mathcal{D}$.
That is, $\beta_d = \E[Y|D=d, S=1]$ is point-identified.
Or under the missing at random assumption $\E[Y|D=d, S=1] = \E[Y_d]$.
We also plot the estimates of $\mathbb{E}[Y|D=d]$ ignoring selection using all observations including $\{i: S_i=0\}$.

\begin{figure}[!htp]
\centering
\caption{\small  (Job Corps) Estimated bounds with sufficient sets}
\includegraphics[width=0.4\textwidth]{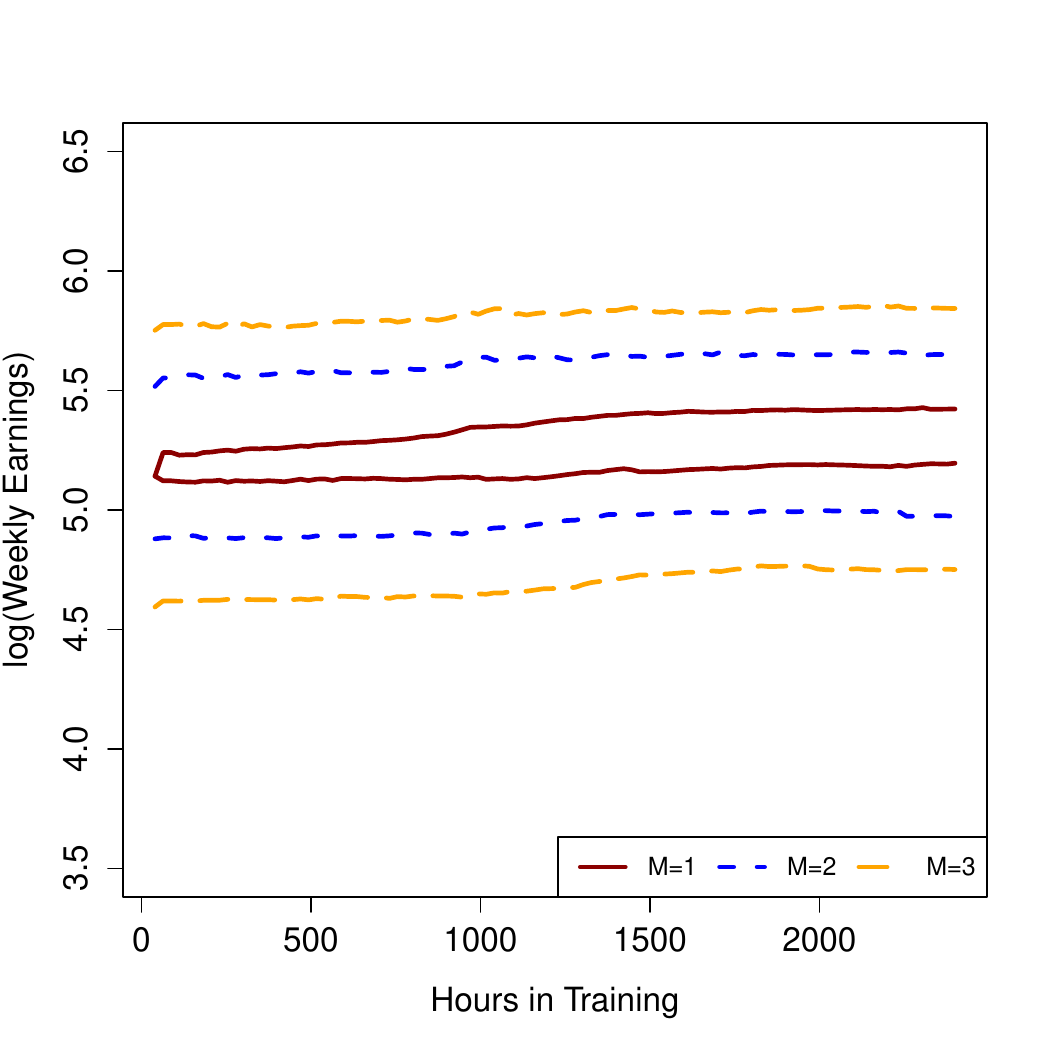}
\label{FBound3}
\end{figure}

\paragraph{Sufficient set}
For sensitivity analysis and illustrating the sufficient set Assumption~\ref{AminJ}, Figure~\ref{FBound3} presents the estimated bounds with $M=1,2,3$ given in Theorem~\ref{TminJ}.
Since $d_{\text{AT}} = \arg\min_{d' \in \mathcal{D}} s(d')$ is estimated as $\hat d_{\text{AT}_J} = 40$, we choose $\mathcal{D}_M$ to contain 40 such that the trimming probability $\pi_L^M/s(d) < 1$.
 It is natural to include the boundary points $\underline{\mathcal{D}} =40$ and $\overline{\mathcal{D}}=2400$ in $\mathcal{D}_M$.
Following the discussion of the non-separable structural selection equation in Assumption~3$^\prime$, for a subject with $d_{\text{AT}}(\eta) = \underline{\mathcal{D}}$, more training hours helps employment.
On the other hand, for a subject with $d_{\text{AT}}(\eta) = \underline{\mathcal{D}}$, the largest treatment value hurts selection probability (employment).
So for $M=2$, we let $\mathcal{D}_2 = \{40, 2400\}$.
Assumption~\ref{AminJ} allows a complier to be employed with $40$ training hours but unemployed with 2400 hours.
And if a participant is employed with both the smallest and largest hours ($d=40, 2400$), then this participant must be employed at any hours between one week to fifteen months, as an always-taker.
The blue dashed line in Figure~\ref{FBound3} uses the lower bound $\hat\pi_L^2 = \hat s(40) + \hat s(2400) - 1\approx 0.658$.

For $M > 2$, we suggest choosing $\mathcal{D}_M$ as a equally spaced sub-grid over $\mathcal{D}_J$ and including $\hat d_{\text{AT}_J}$, if there is no further information on the structural selection model.
So for $M=3$, we let $\mathcal{D}_3 = \{40, 1208, 2400\}$, the orange long-dashed line uses the lower bound $\hat\pi_L^3 = \hat s(40) + \hat s(1208) + \hat s(2400) - 1 \approx 0.5$.
As expected, the bounds are less informative when we assume more treatment values, i.e., a larger $M$.
So we note that the sufficient set assumption is more of theoretical interest than practical.
We focus on sufficient value Assumption~\ref{Amin} in empirical analysis.
Next we incorporate covariates to potentially tighten the bounds and confidence intervals.



\section{Estimation and inference conditional on covariates}
\label{SecConX}
We weaken the independent treatment assumption and the sufficient value/set assumption by conditioning on the covariates.
So the covariates $X$ can serve two purposes:
First, in observational data, it is more plausible and standard to assume conditional independence, also known as  unconfoundedness, selection on observables, or ignorability.
Second, we allow subjects with different pretreatment covariates to
have different sufficient treatment values $d_{\text{AT}x}\in\mathcal{D}$.
It might be reasonable that participants with some particular covariates benefit from more training, but more training could hurt the employment of some participants with different covariates.
So the bounds may be tightened by incorporating the covariates.


After conditional on the covariates, we follow  \cite{SGLee} to represent the generalized Lee bounds as a moment equation and derive an orthogonal moment for it.
The advantage of using the orthogonality moment is that the first-stage estimation has no contribution to the asymptotic variance of the bounds.
We utilize cross-fitting to remove overfitting bias without strong entropy conditions, following the recent double debiased machine learning (DML) literature \citep{CCDDHNR}.
Then we show the asymptotic theory that accommodates low-dimensional smooth and high-dimensional sparse designs.

Notice that $\beta_{d{}}$ and its bounds are functions of $d$ over $\mathcal{D}$ and hence are of infinite dimension for a continuous treatment.
Such non-regular nonparametric estimands cannot be estimated at a regular root-$n$ rate without further assumptions.
The asymptotic theory is more involved with the kernel function and bandwidth~$h$, compared with the semiparametric inference in \cite{CCDDHNR}.
\cite{CL} provide DML inference for $\beta_{d{}}$ that is point-identified when there is no selection bias.
In particular, if researchers assume ``no effect on selection" by using the selected sample excluding those with zero outcomes, then they estimate $\E[\E[Y|S=1, D=d, X]]$.
Or if researchers ``ignore selection" by including all observations with zero outcomes, then they estimate  $\E[\E[Y|D=d, X]]$.
See also \cite{Kennedy}, \cite{SUZ}, and \cite{CNS21EJ} for non-regular estimands and machine learning.

The conditional independence Assumption~\ref{AindX} means that conditional on observables, the treatment variable is as good as randomly assigned, or conditionally exogenous.

\begin{assumption}[Conditional independence]
$D$ is independent of $\big\{(Y_d, S_d): d\in \mathcal{D}\big\}$ conditional on~$X$.
\label{AindX}
\end{assumption}

Assumption~\ref{ACmin} relaxes Assumption~\ref{Amin} to allow subjects with different values of pretreatment covariates $x$ to have different sufficient values $d_{\text{AT}x}\in\mathcal{D}$.


\begin{assumption}[Conditional sufficiency]
For $x \in \mathcal{X}$, there exists $d_{\text{AT}x} \in \mathcal{D}$ such that $\p(S_d \geq S_{d_{\text{AT}x}}, \forall d \in\mathcal{D}|X=x)=1$.
\label{ACmin}
\end{assumption}
As discussed in Assumption~2$^\prime$,
a sufficient condition of Assumption~\ref{ACmin} is to assume a separable structural error $\eta$ in selection $S=\1\{q(D,X) \geq \eta\}$, and there exists $d_{\text{AT}x} = \arg\min_{d\in\mathcal{D}} q(d,x)$.

Let the conditional average dose-response function of always-takers be
$\beta_d(x) := \mathbb{E}[Y_d|\{S_{d'}=1: d' \in \mathcal{D}\}, X=x]$, so $\beta_d = \int_{\mathcal{X}} \beta_d(x) f_X(x|S_{d'}= 1: d' \in \mathcal{D}) dx.$
Let the conditional probability of always-takers be $\pi_{\text{AT}}(x) := \p(S_{d'} = 1: d' \in\mathcal{D}|X=x)$.
Let  the corresponding conditional upper bound given in (\ref{ELeeUB}) be
$\bar\beta_d(x):= \rho_{dU}(\pi_{\text{AT}}(x),x) := \mathbb{E}[Y|Y \geq Q^d(1-\pi_{\text{AT}}(x)/s(d,x),x), D=d, S=1, X=x]$,
where the conditional selection probability $s(d,x) := \p(S=1|D=d, X=x)$ and
$Q^d(u,x)$ is the $u$-quantile of $Y|D=d, S=1, X=x$.
Define the aggregate upper bound for $\beta_{d}$ as
$\bar\beta_{d}:=\int_\mathcal{X}\bar\beta_{d}(x) f_X(x|\text{AT}) dx$.

Lemma~\ref{Lmom} below extends Lemma 1 in \cite{SGLee} to show that the upper bound $\bar\beta_{d}$ is a ratio of two moments,
by replacing the binary treatment indicator $D$ with a kernel function
$K_h(D-d)$.
The moment function for the lower bound is defined analogously in the proof of Lemma~\ref{Lmom} in the Appendix.



\begin{lemma}[Moment-based representation]
\label{Emote}
Assuming Assumption~\ref{AindX}, $s(d,x) = \mathbb{E}[S_d|X=x]$.
Further assuming Assumption~\ref{ACmin}, $\pi_{\text{AT}}(x) = \min_{d\in\mathcal{D}} s(d,x) = s(d_{\text{AT}x},x)$.
Assume $\pi_{AT}=\mathbb{E}\big[\pi_{\text{AT}}(X)\big] > 0$ and
the generalized propensity score $\mu_d(x) := f_{D|X}(d|x) > 0$  with probability one, for $d\in\mathcal{D}$.
Then the sharp upper bound of $\beta_d$ is
\begin{align*}
\bar\beta_{d} = {\mathbb{E}[\bar\beta_{d}(X) \pi_{\text{AT}}(X)]}/{\pi_{\text{AT}}} = \lim_{h\rightarrow 0} {\mathbb{E}\big[m_{dU}(W,\xi)}\big]/{\pi_{AT}}, where \\
m_{dU{}}(W, \xi) := \frac{K_h(D-d)}{\mu_d(X)} S Y\cdot\1\{Y \geq Q^d(1-
\pi_{\text{AT}}(X)/s(d,X),
X)\},
\end{align*}
the nuisance parameter
$\xi(d, x) = \big\{
s(d, x), \mu_d(x), Q^d(1-\pi_{\text{AT}}(x)/s(d,x), x)\big\}$.

\label{Lmom}
\end{lemma}

\begin{remark}[Conditional sufficient set]{\rm
We discuss incorporating the covariates under the sufficient set Assumption~\ref{AminJ}.
We provide a non-sharp bound that can be implemented in practice.
The sharp bound is out of the scope of this paper.
Under Assumption~\ref{AminJ}, Theorem~\ref{TminJ} directly implies the conditional version of the bounds on $\pi_{\text{AT}}(x)$:
\begin{align*}
\pi^M_{L}(x) := \max\left(
\sum_{d\in \mathcal{D}_M} s(d, x) - M + 1,
0\right)
&\leq
\pi_{\text{AT}}(x) = \p(S_{d}=1: d\in\mathcal{D}_M|X=x) \\
&\leq
\min_{d\in \mathcal{D}_M} s(d,x) =: \pi^M_{U{}}(x).
\end{align*}
Then together with the proof of Lemma~\ref{Lmom}, the sharp upper bound $
\mathbb{E}[\rho_{dU}(\pi_{\text{AT}}(X), X) \pi_{\text{AT}}(X)]/\pi_{\text{AT}}$ is not point-identified and
is smaller than $\mathbb{E}[\rho_{dU}(\pi_{L}^M(X), X) \pi_{U}^M(X)]/\pi_{L}^M = \lim_{h\rightarrow 0}\mathbb{E}[m_{dU}(W, \xi)\cdot  {\pi_{U}^M(X)/ \pi_{L}^M(X)}]/\pi_L^J$ with $\pi_{\text{AT}}(X) = \pi_{L}^M(X)$ in $m_{dU}(W,\xi)$, which likely is a non-sharp bound.
Similar, the sharp lower bound
$\mathbb{E}[\rho_{dL}(\pi_{\text{AT}}(X), X) \cdot \pi_{\text{AT}}(X)]/\pi_{\text{AT}}
\geq
\mathbb{E}[\rho_{dL}({\color{black}\pi_{L}^M(X)}, X) \cdot {\color{black}\pi_{L}^M(X)}]/{\color{black}\pi_{U}^M}
= \lim_{h\rightarrow 0}\mathbb{E}[m_{dL}(W,\xi)\cdot  {{\color{black}\pi_{L}^M(X)}/ {\color{black}\pi_{L}^M(X)}}]/{\color{black}\pi_U^M}$,  with $\pi_{\text{AT}}(X) = \pi_{L}^M(X)$ in $m_{dL}(W,\xi)$.
}\end{remark}

\subsection{Estimation and inference}
\label{SecEstX}
We estimate the bounds over an evenly spaced grid $\mathcal{D}_J := \{d_1,...,d_J\} \subseteq \mathcal{D}$.
Let the sets $\mathcal{X}_j := \{x: s(d_j,x)/s(d, x) \leq 1, \forall d \in \mathcal{D}_J\}$ for $j=1,...,J$.
So for $x\in\mathcal{X}_j$, $d_{\text{AT}x_J} = d_j = \arg\min_{d\in\mathcal{D}_J} s(d, x)$.
We can classify the subjects into $J$ groups $\mathcal{X} = \cup_{j=1,...,J}\mathcal{X}_{j}$.
By consistently estimating $s(d,x)$ over the grid $\mathcal{D}_J$,
we show that the mis-classification error is asymptotically first-order ignorable.

We allow the subsets $\mathcal{X}_j$ to overlap, i.e., the sufficient treatment value can be not unique, so there could be no treatment effect on selection over some range in $\mathcal{D}$.
For example, if there exists $x \in \mathcal{X}_{j} \cap \mathcal{X}_{j+1}$ so that $s(d_j,x)/s(d_{j+1},x)=1$, then the bounds degenerate to points at $d_j$ and $d_{j+1}$, i.e., $\bar\beta_{d_j}(x) = \underline{\beta}_{d_j}(x) = \E[Y|D=d_j, S=1, X=x]$ and $\bar\beta_{d_{j+1}}(x) = \underline{\beta}_{d_{j+1}}(x)  = \E[Y|D=d_{j+1}, S=1, X=x]$.
We estimate the robust bounds using the trimming probability $\hat p_{d_{j+1}d_j}(x) = \hat s(d_j,x)/\hat s(d_{j+1},x)-\nu$ for some small positive $\nu$, rather than the untrimmed point-estimator, as discussed in Section \ref{SecInf}.


For $X_i \in \mathcal{X}_j$, define the moment function to be
$m_{dU}^{j}(W_i,\xi) := m_{dU}(W_i,\xi)$ with $\pi_{\text{AT}}(X_i) = s(d_j, X_i)$.
To construct orthogonality as $h\rightarrow 0$,
we derive the correction terms for $s(d_j, x)$, $s(d, x)$, $Q^d(u,x)$, $\mu_d(x)$, respectively, collected in
\begin{align*}
cor_{dU}^{j}(W,\xi) &= Q^d(1-p_{dd_j}(X), X) \cdot \bigg( \frac{K_h(D - d_j)}{\mu_{d_j}(X)} (S-s(d_j, X))
-\frac{K_h(D-d)}{\mu_d(X)} p_{dd_j}(X)  (S-s(d,X)) \notag \\
&\ \ \ + \frac{K_h(D-d) S}{\mu_d(X)}\big(-\1\{Y \geq Q^d(1-p_{dd_j}(X),X)\}  + p_{dd_j}(X)\big)\bigg)
\notag
\\
&\ \ \ +\big(\mu_d(X) - K_h(D-d)\big) \cdot\E\big[Y\big|Y \geq Q^d(1-p_{dd_j}(X), X), D=d, S=1, X\big]\frac{s(d_j, X)}{\mu_d(X)}
\end{align*}
for $p_{dd_j}(X) \leq 1 - \nu < 1$ with $\nu > 0$.
For $d=d_j$, let $\nu = 0$, $p_{dd_j}(X) = 1$, and
\begin{align}
m^j_{dU}(W,\xi) = m^j_{dL}(W,\xi) &= \frac{K_h(D-d)}{\mu_d(X)} S Y,
\notag\\
cor_{dU}^j(W,\xi) = cor_{dL}^j(W,\xi) &= \big(\mu_{d}(X) - K_h(D-d)\big) \E\big[Y\big|D=d, S=1, X\big]\frac{s(d, X)}{\mu_{d}(X)}.
\label{EcorCL}
\end{align}
Then the orthogonal moment function is defined as $g_{dU}^{j} := m_{dU}^{j} + cor_{dU}^{j}$.
Let $g_{dU}(W, \xi) = \sum_{j=1}^J g^j_{dU}(W, \xi)\1\{X \in \mathcal{X}_j\}/\sum_{j=1}^J \1\{X \in \mathcal{X}_j\}$ that allows overlapping $\mathcal{X}_j$s.

These correction terms are derived based on equation (4.7) in \cite{SGLee}.
Specifically let the true nuisance parameter be $\xi_0$ and $\xi_r := \xi_0 + r(\xi - \xi_0)$ for some $\xi$ close to $\xi_0$ and $r \in (0,1)$.
Then we verify that the partial derivative of the moment function $g_{dU}^{j}$ with respect to $r$ is zero.
When $p_{dd_j}(x) =1$, $\bar\beta_d(x)=\underline{\beta}_d(x)$ is point-identified. So the correction term (\ref{EcorCL}) is only for the nuisance function $\mu_{d}(x)$ as derived in \cite{CL}.

The estimation procedure follows four steps:
\paragraph{Step 1.} ($L$-fold Cross-fitting)
As defined in Section~\ref{SecInf},
$L$-fold cross-fitting
randomly partitions the observation indices into $L$ distinct groups $I_\ell, \ell = 1,...,L$.
For $\ell = 1,..., L$, the estimator $\hat \xi_\ell(W)$ for the nuisance function $\xi(W) = \big(s(d,X),  Q^d(1- p_{dd_j}(X), X),\mu_d(X), \E[Y|Y \geq Q^d(1- p_{dd_j}(X), X), D=d, S=1, X], d\in\mathcal{D}\big)$ uses observations not in $I_\ell$, satisfying Assumption \ref{ADMLX} below.

\paragraph{Step 2.} (Double robustness)
For $i\in I_\ell$ and $X_i \in \hat{\mathcal{X}}_{j\ell} = \{x: \hat s_\ell(d_j,x)/\hat s_\ell(d, x) \leq 1, \forall d \in \mathcal{D}_J\}$, estimate the orthogonal moment function by
 $g_{dU}(W_i, \hat \xi_\ell) = \sum_{j=1}^J g^j_{dU}(W_i,\hat \xi_\ell) \1\{X_i \in \hat{\mathcal{X}}_{j\ell}\}/\sum_{j=1}^J \1\{X_i \in \hat{\mathcal{X}}_{j\ell}\}$.

\paragraph{Step 3.}
The DML estimator in \cite{CL} for $\pi_{\text{AT}}$:
$\hat\pi_{\text{AT}} = n^{-1}\sum_{\ell=1}^L\sum_{i\in I_\ell}  \psi(W_i, \hat\xi_\ell)$, where $\psi(W_i, \xi) := K_h(D_i - d_j)(S_i -  s(d_j, X_i))/ \mu_{d_j}(X_i) +  s(d_j, X_i)$ if $X_i \in \hat{\mathcal{X}}_{j\ell}$.

\paragraph{Step 4.}
The DML estimator $\hat{\overline{\beta}}_d = n^{-1}\sum_{\ell = 1}^L \sum_{i \in I_\ell} g_{dU}(W_i, \hat \xi_\ell)/\hat \pi_{\text{AT}}$.

Denote the $L_2$-norm $\big\|\hat \xi- \xi\big\|_2 = \|\Delta\hat\xi(W)\|_2=  \left(\int (\hat \xi(w) -\xi(w))^2 f_W(w) dw\right)^{1/2}$.
\begin{assumption}
\begin{enumerate}
\item[(i)] (Strict overlap) $s(d, x) \in (c, 1-c)$ for some constant $c \in (0, 1/2)$, for all $(d, x) \in \mathcal{D} \times \mathcal{X}$.
 $\inf_{d\in\mathcal{D}}{\rm ess}\inf_{x\in\mathcal{X}} \mu_d(x) \geq c$ for some positive constant $c$.

\item[(ii)]
$s(d,x)$, $f_{YSDX}(y, 1, d, x)$ and $\E[Y|D=d, S=1, X=x]$ are two-times differentiable with respect to $d$ with all two derivatives being bounded uniformly over $\mathcal{Y}\times\mathcal{D}\times\mathcal{X}$.


The derivative of $Q^d(u, x)$ with respect to $u$ and $var(Y|D=d, S=1, X=x)$ is bounded uniformly over $\mathcal{D}\times\mathcal{X}$.

$\E[|Y|^3|S=1, D=d, X=x]$ is continuous in $d$ uniformly over $\mathcal{X}$.


\item[(iii)]
The following terms are $o_\p(1)$ for $d\in\mathcal{D}$:
$\|\Delta\hat\mu_d(X)\|_2$,
$\sup_{y\in\mathcal{Y}_0}\| \Delta\hat{\E}[Y|Y\geq y, S=1, D=d, X]\|_2$,
$\|\Delta\hat s(d, X)\|_2$, $\sup_p\|\Delta\hat Q^d(p,X)\|_2$, where $\mathcal{Y}_0$ be a compact subset of the support of $Y$.


\item[(iv)] The following terms are $o_\p(1/\sqrt{nh})$ for $d\in\mathcal{D}$: \\
$\sup_{y\in\mathcal{Y}_0}\| \Delta\hat{\E}[Y|Y\geq y, S=1, D=d, X]\|_2 \|\Delta\hat\mu_d(X)\|_2$, \\
 $\sup_{p\in(0,1)}\|\Delta\hat Q^d(p,X)\|_2\Big( \sup_{p\in(0,1)}\|\Delta\hat Q^d(p,X)\|_2 + \|\Delta\hat\mu_d(X)\|_2 + \|\Delta\hat s(d, X)\|_2\Big)$,\\
$\|\Delta\hat s(d, X)\|_2 \Big( \|\Delta\hat s(d, X)\|_2 + \sup_{y\in\mathcal{Y}_0}\| \Delta\hat{\E}[Y|Y\geq y, S=1, D=d, X]\|_2 +   \|\Delta\hat\mu_d(X)\|_2\Big)$.

\item[(v)] $h\rightarrow 0$, $nh\rightarrow \infty$, $\sqrt{nh}h^2 \rightarrow c\in[0,\infty)$.

\end{enumerate}
\label{ADMLX}
\end{assumption}

Assumption~\ref{ADMLX}(i) requires the generalized propensity score (GPS) $\mu_d(X)$ to be bounded away from zero, which is the standard overlap assumption  \citep{SGLee}.
We note that such common support assumption should be made with care in practice and is strong especially with many control variables \citep{DAmour}.
In our empirical application, we find a common support by trimming away observations whose estimated GPSs are smaller than some fixed trimming parameter. 
Details are in Section~\ref{SecEmCTX} and Section~\ref{ASecStep0} in the online appendix.

Assumption~\ref{ADMLX}(iii)(iv) gives tractable high-level rate conditions on the nuisance function estimators.
These are standard conditions similarly assumed in the DML literature \citep{CCDDHNR, CL}.
The rate conditions use a ``partial $L_2$" norm in the sense that the regressor $D$ is fixed at $d$ and the expectation is based on the marginal distribution of $X$,
e.g., $\|\Delta \hat s(d,X)\|_2 = \big(\int_{\mathcal{X}} (\hat s(d,x) - s(d,x))^2 f_X(x) dx\big)^{1/2}$.
See \cite{CL} for detailed discussion on the low-level conditions of the nuisance function estimators that satisfy such high-level conditions, such as kernel, series, and neural network in low-dimensional settings with fixed dimension of $X$.
In high-dimensional settings where the dimension of X grows with the sample size, our inference theory is valid as long as the nuisance function estimators satisfy the high-level rate conditions, e.g., Lasso, as we illustrate in Section~\ref{Sec1stEst}.

Similar to Assumption~\ref{Agrid}, Assumption~\ref{Agridx} imposes conditions on the grid.
\begin{assumption}
Let $s^{(m)}(d,x)$ be the $m^{th}$ derivative of $s(d,x)$ w.r.t.\ $d$ for $m \in \{1,2,...\}$.
Let $\mathcal{D} = \mathcal{D}_{sx} \cup \mathcal{D}_{cx}$ for any $x \in \mathcal{X}$, where $\mathcal{D}_{cx} := \{d: s^{(m)}(d,x) = 0, \forall m \geq 1\}$ and $\mathcal{D}_{sx} := \{d: s^{(m)}(d, x) \neq 0, \exists m < \infty\}$.
If $\mathcal{D}_{sx} \neq \emptyset$, let $\bar M_x = \min\{m: s^{(m)}(d,x) \neq 0, m=1,2,..., \forall d \in \mathcal{D}_{sx}\} < \infty$.
If $\mathcal{D}_{sx} = \emptyset$, let $\bar M_x = 0$.
Let $\bar M = \max_{x\in\mathcal{X}} \bar M_x$.
Let an equally spaced grid $\mathcal{D}_J = \{d_1,..,d_J\} \subseteq \mathcal{D}$ with
$J = O({\mathsf{s}_n}^{-1/\bar M})$ and $\mathsf{s}_n := \sup_{d\in\mathcal{D}, x\in\mathcal{X}}|\hat s(d,x) - s(d,x)| = o_\p(1)$.
\label{Agridx}
\end{assumption}

\begin{theorem}
Let Assumptions~\ref{AEstLB}, \ref{AindX}, \ref{ACmin}, \ref{ADMLX}, and \ref{Agridx} hold.
Then for $d\in\mathcal{D}$,
$\hat{\overline{\beta}}_d - \overline{\beta}_d =
 n^{-1}\sum_{i=1}^n
\phi_{dU}(W_i, \xi)
  - \overline{\beta}_d
 + o_\p(1/\sqrt{nh})$, where $\phi_{dU}(W_i, \xi):= g_{dU}(W_i, \xi)/\pi_{\text{AT}} -  \psi(W_i, \xi) \overline{\beta}_d/\pi_{\text{AT}}$.
And $\sqrt{nh}\big(\hat{\bar\beta}_d - \bar\beta_d - h^2\mathsf{B}_{dU}\big)\stackrel{d}{\rightarrow} \mathcal{N}(0,\mathsf{V}_{dU})$. 
Similarly for the lower bounds,
$\hat{\underline{\beta_d}} - \underline{\beta_d} =
 n^{-1}\sum_{i=1}^n
\phi_{dL}(W_i, \xi)
  - \underline{\beta_d}
 + o_\p(1/\sqrt{nh})$, where $\phi_{dL}(W_i, \xi):= g_{dL}(W_i, \xi)/\pi_{\text{AT}} -  \psi(W_i, \xi) \underline{\beta_d}/\pi_{\text{AT}}$.
And $\sqrt{nh}\big(\hat{\underline{\beta_d}} - \underline{\beta_d} - h^2\mathsf{B}_{dL}\big)\stackrel{d}{\rightarrow} \mathcal{N}(0,\mathsf{V}_{dL})$
, where $\mathsf{B}_{dU}$, $\mathsf{V}_{dU}$, $\mathsf{B}_{dL}$, and $\mathsf{V}_{dL}$ are given explicitly in the proof in the Appendix.
\label{TDMLX}
\end{theorem}

We can estimate the asymptotic variance $\mathsf{V}_{dU}$ by the sample variance of the estimated influence function $\hat{\mathsf{V}}_{dU} := hn^{-1}\sum_{i=1}^n\phi_{dU}(W_i,\hat\xi)^2$.
As described in Section~\ref{SecInf}, we could estimate the leading bias $\mathsf{B}_{dU}$ by the method in \cite{PS96} and \cite{CL}.
Let the notation $\hat{\bar \beta}_{d,b}$ be explicit on the bandwidth $b$ and
$\widehat{\mathsf{B}}_{dU}:= \big(\hat{\bar \beta}_{d,b} - \hat{\bar \beta}_{d,ab}\big)\big/\big(b^2(1-a^2)\big)$
with a pre-specified fixed scaling parameter $a\in(0,1)$.
Then a data-driven bandwidth $\hat h_{dU}:= (\widehat{V_{dU}}/(4\widehat{\mathsf{B}}_{dU}^2))^{1/5}n^{-1/5}$ consistently estimates the optimal bandwidth that minimizes the AMSE.
For the lower bound, the same estimation applies to $\hat{\mathsf{V}}_{dL}$, $\widehat{\mathsf{B}}_{dL}$, and $\hat h_{dL}$.
Then we can choose an undersmoothing bandwidth $h$ that is smaller than $\hat h_{dU}$ and $\hat h_{dL}$ to construct the 95\% confidence interval
$\Big[\hat{\underline{\beta_d}}-1.96 \times \widehat{\sigma_{dL}}/\sqrt{nh},
\hat{\bar{\beta_{d}}} + 1.96 \times \widehat{\sigma_{dU}}/\sqrt{nh} \Big]$ with $\widehat{\sigma_{dL}} := \sqrt{\widehat{V_{dL}}}$
and $\widehat{\sigma_{dU}} := \sqrt{\widehat{V_{dU}}}$.
Under additional assumptions, we can straightforwardly show consistency of $\hat{\mathsf{V}}_{dU}$, $\widehat{\mathsf{B}}_{dU}$, and $\hat h_{dU}$ following Theorem 3.2 in \cite{CL}, so we omit the repetition.

We can bound the ATE of increasing the treatment from $d_1$ to $d_2$,
$\underline{\Delta}_{d_1d_2} := \underline{\beta_{d_2}} - \overline{\beta}_{d_1} \leq \beta_{d_2} - \beta_{d_1} \leq \overline{\beta}_{d_2} - \underline{\beta_{d_1}} =: \bar\Delta_{d_1d_2}$.
Denote the bandwidth in $\hat{\underline{\beta}}_{d}$ and $\hat{\bar{\beta}}_{d}$ be $h_{dL}$ and $h_{dU}$, respectively.

\begin{corollary}[ATE]
Let the conditions in Theorem~\ref{TDMLX} hold.
Let $\mathsf{V}_{Un} :=  \E[(\phi_{d_2U} - \phi_{d_1L})^2]$
and
$\mathsf{V}_{Ln}:=   \E[(\phi_{d_2L} - \phi_{d_1U})^2]$. 
Then $\sqrt{n}{\mathsf{V}_{Un}}^{-1/2}\big(\hat{\bar\Delta}_{d_1d_2} - \bar\Delta_{d_1d_2} - (h_{d_2U}^2\mathsf{B}_{d_2U} -h_{d_1L}^2\mathsf{B}_{d_1L})\big) \stackrel{d}{\longrightarrow} \mathcal{N}(0, 1)$
and $\sqrt{n}\mathsf{V}_{Ln}^{-1/2}\big(\hat{\underline{\Delta}}_{d_1d_2} - \underline{\Delta}_{d_1d_2}- (h_{d_2L}^2\mathsf{B}_{d_2L} -h_{d_1U}^2\mathsf{B}_{d_1U})\big)\stackrel{d}{\longrightarrow} \mathcal{N}(0, 1)$.
\label{CATEX}
\end{corollary}

The variance can be estimated by $\hat{\mathsf{V}}_{Un} = n^{-1}\sum_{i=1}^n(\phi_{d_2U}(W_i,\hat\xi) - \phi_{d_1L}(W_i,\hat\xi))^2$
and
$\hat{\mathsf{V}}_{{Ln}} = n^{-1}\sum_{i=1}^n(\phi_{d_2L}(W_i,\hat\xi) - \phi_{d_1U}(W_i,\hat\xi))^2$.
The 95\% confidence interval $\big[\hat{\underline{\Delta}}_{d_1d_2} - 1.96\times \sqrt{\hat{\mathsf{V}}_{Ln}}/\sqrt{n},
\hat{\bar\Delta}_{d_1d_2} + 1.96\times  \sqrt{\hat{\mathsf{V}}_{Un}}/\sqrt{n} \big]$.

\subsection{First-step nuisance parameter estimation}
\label{Sec1stEst}
We illustrate our estimation procedure by applying Lasso methods in Step 1 to estimate the nuisance functions, when $X$ is potentially high-dimensional.
We provide sufficient conditions to verify the high-level Assumption~\ref{ADMLX}.
We follow \cite{SUZ} (SUZ, hereafter) to approximate the outcome, selection, and  treatment models by varying coefficient linear regressions and logistic regressions.
Particularly, the penalized kernel-smoothing least square and maximum likelihood estimations select covariates for each value of the continuous treatment.
The approximation errors satisfy Assumption~\ref{ASUZ} below that imposes sparsity structures
so that the number of effective covariates that can affect them is small.
See \cite{Farrell15}, \cite{CNS21ADML}, for example, for in-depth discussions on the specification of high-dimensional sparse models.

To estimate the $Q^d(p, x)$, we first estimate the conditional CDF $F_{Y|SDX}(y|1,d,x)$ and then compute the generalized inverse function.
To estimate the conditional density $\mu_d(x) = f_{D|X}(d|x)$, we first estimate the conditional CDF $F_{D|X}$ and then take the numerical derivative.
We estimate the $s(d,x)$, $F_{D|X}$, and $F_{Y|SDX}$ by the logistic distributional Lasso regression in \cite{BCFH17} and SUZ.
We modify the penalized local least squares estimators and use the conditional density estimator in SUZ.
For completeness, we present the estimators and asymptotic theory in SUZ in the Appendix and refer readers to SUZ for details.

Let $b(X)$ be a $p\times 1$ vector of basis functions and $\Lambda$ be the logistic CDF.
Define the approximation errors $r_{d}(x;F_D) = F_{D|X}(d|x) - \Lambda(b(x)'\beta_{d})$, $r_{dy}(x;F_Y) = F_{Y|SDX}(y|1,d,x) - \Lambda(b(x)'\alpha_{dy})$, $r_d(x;s) = s(d,x) - \Lambda(b(x)'\theta_d)$, and $r_d(x;\rho) = \rho_{dU}(\pi, x)- b(x)'\gamma_d$.

Denote the usual $1$-norm $\|\alpha\|_1 = \sum_{j=1}^p |\alpha_j|$ for a vector $\alpha= (\alpha_1,..., \alpha_p)$.
Denote $\|W\|_{\p,\infty} = \sup_{w\in\mathcal{W}}|w|$ and
$\|W\|_{\p_{N_\ell}, 2} = (N_\ell^{-1}\sum_{i\notin I_\ell}W_i^2)^{1/2}$ for a generic random variable $W$ with support $\mathcal{W}$.
Assumption~\ref{ASUZ} collects the conditions in Theorems 3.1 and 3.2 in SUZ.
\begin{assumption}[Lasso]
Let $\mathcal{D}_0$ be a compact subset of the support of $D$, $\mathcal{Y}_0$ be a compact subset of the support of $Y$, and
$\mathcal{X}$ be the support of $X$.

\begin{enumerate}
\item[(i)]
(a)~$\|\max_{j \leq p} |b_j(X)|\|_{{\p}, \infty} \leq \zeta_n$
and
$\underline{C} \leq {E}\left[b_j(X)^2\right] \leq 1/\underline{C}$, for some positive constant $\underline{C}$, $j=1,...,p$.
(b)~$\sup_{d \in \mathcal{D}_0, y\in \mathcal{Y}_0} \max(\|\beta_d\|_0, \|\alpha_{dy}\|_0, \|\theta_d\|_0, \|\gamma_d\|_0) \leq \mathfrak{s}$ for some $\mathfrak{s}$ which possibly depends on $n$,
where $\|\theta\|_0$ denotes the number of nonzero coordinates of $\theta$.

(c)~$\sup_{d\in\mathcal{D}_0} \| r_d(X;F_D)\|_{{\p}_n, 2} =O_\p((\mathfrak{s}\log(p \vee n)/n)^{1/2})$
and
$\sup_{d\in \mathcal{D}_0, y \in \mathcal{Y}_0}\Big( \big\| r_{dy}(X;F_Y)k((D-d)/h_1)^{1/2} \big\|_{{\p_n},2} + \big\| r_d(X;s)k((D-d)/h_1)^{1/2} \big\|_{{\p_n},2} + \big\| r_d(X;\rho) k((D-d)/h_1)^{1/2} \big\|_{{\p_n},2}\Big)= O_\p\big(\big(\mathfrak{s}\log(p\vee n)/n\big)^{1/2}\big)$.

(d)~$\sup_{d\in\mathcal{D}_0} \| r_d(X;F_D)\|_{{\p}, \infty} = O((\mathfrak{s}^2\zeta_n^2\log(p \vee n)/n)^{1/2})$
and
$\sup_{d\in \mathcal{D}_0, y \in \mathcal{Y}_0}\big( \big\| r_{dy}(X;F_Y) \big\|_{{\p},\infty} + $\\
$\big\| r_d(X;s) \big\|_{{\p},\infty} + \big\| r_d(X;\rho) \big\|_{{\p},\infty}\big)= O\left(\left(\mathfrak{s}^2\zeta_n^2\log(p\vee n)/(nh_1)\right)^{1/2}\right)$.

(e)~$f_{D|X}(d,x)$ is second-order differentiable w.r.t.\ $d$ with bounded derivatives uniformly over $(d,x)\in\mathcal{D}_0\times\mathcal{X}$.
(f)~$\zeta_n^2\mathfrak{s}^2\iota_n^2\log(p\vee n)/(nh_1) \rightarrow 0$, $nh_1^5/(\log(p\vee n))\rightarrow 0$.

\item[(ii)] Uniformly over $(d,x,y)\in\mathcal{D}_0\times\mathcal{X}\times\mathcal{Y}_0$,
(a)~there exists some positive constant $\underline{C} < 1$ such that $\underline{C}  \leq f_{D|X}(d|x) \leq 1/\underline{C}$,
$\underline{C} \leq F_{Y|SDX}(y|1,d,x) \leq 1-\underline{C}$, and $\underline{C}\leq s(d,x) \leq 1-\underline{C}$;
(b)~$s(d,x)$ and $F_{Y|SDX}(y|1,d,x)$ are three times differentiable w.r.t.\ $d$ with all three derivatives being bounded.
\item[(iii)]
There exists a sequence $\iota_n\rightarrow \infty$ such that $0 < \kappa' \leq
\inf_{\delta\neq 0, \|\delta\|_0 \leq \mathfrak{s}\iota_n} \|b(X)'\delta\|_{{\p}_n,2}\big/\|\delta\|_2
\leq
\sup_{\delta\neq 0, \|\delta\|_0 \leq \mathfrak{s}\iota_n} \|b(X)'\delta\|_{{\p}_n,2}\big/\|\delta\|_2
\leq \kappa'' < \infty$ $w.p.a.1$.


\end{enumerate}

\label{ASUZ}
\end{assumption}

Let Assumption~\ref{ASUZ} hold.
Then Theorems 3.1 and 3.2 in SUZ imply
$\sup_{(d, x)\in\mathcal{D}_0\times\mathcal{X}}|\hat s_\ell(d,x) - s(d,x) |= O_\p(A_n)$,
$\sup_{(d, x)\in\mathcal{D}_0\times\mathcal{X}, u\in (0,1)}|\hat Q^d_\ell(u,x) - Q^d(u,x) |= O_\p(A_n)$,
$\sup_{(d, x,y)\in\mathcal{D}_0\times\mathcal{X}\times\mathcal{Y}_0}|\Delta \hat\E[Y|Y\geq y, S=1, D=d, X=x]|O_\p(A_n)$,
where $A_n = \iota_n(\log(p\vee n) \mathfrak{s}^2\zeta_n^2/(nh_1))^{1/2}$.
And
$\sup_{(d, x)\in\mathcal{D}_0\times\mathcal{X}}|\hat \mu_{d\ell}(x) - \mu_d(x)| = O_\p(R_n)$, where $R_n = h_1^{-1}(\log(p\vee n) \mathfrak{s}^2\zeta_n^2/n)^{1/2}$.
Then we can obtain the $L_2$ rates $\|\hat\xi_\ell - \xi\|_2$ to verify Assumption~\ref{ADMLX}.
In particular a sufficient condition of Assumption~\ref{ADMLX}(iii) is $A_n  \rightarrow 0$ and $R_n \rightarrow 0$.
And a sufficient condition of Assumption~\ref{ADMLX}(iv) is $\sqrt{nh}A_n R_n \rightarrow 0$ and $\sqrt{nh}A_n^2 \rightarrow 0$.

\section{Empirical illustration with covariates}
\label{SecEmCTX}
We illustrate our DML estimator by evaluating the Job Corps and CCC program.
First in Step 0, we prepare a sub-sample that satisfies Assumption~\ref{ADMLX}(i) for estimating the bounds.
We use the full sample to estimate the GPS by $\tilde\mu_d(X_i)$ and the selection probability by $\tilde s(d, X_i)$.
We obtain a sub-sample where $\tilde\mu_d(X_i) \geq trim_{GPS}$ and $\tilde s(d, X_i) \geq 5\%$ for {\it all} $i$ in this sub-sample and for {\it all} $d \in \mathcal{D}_J$, where the trimming parameter $trim_{GPS}$ is based on \cite{Imbens04REStat} and detailed in Section~\ref{ASecStep0} in the online appendix.
Then we use this sub-sample to estimate the bounds following the procedure described in Section~\ref{SecEstX}.
For JC, we remove 11 observations whose $\tilde s(d, X_i) < 5\%$ for some $d \in\mathcal{D}_J$, and no observation is removed by $trim_{GPS}$.
For CCC, we remove 94 observation whose $\tilde \mu_d(X_i) < trim_{GPS}$, and all observations have $\tilde s(d, X_i) \geq 5\%$.
So we only trim a small number of observations relative to the sample size.

In Step 1, we use the Lasso estimation given in Section~\ref{Sec1stEst}.
We present the results with a linear basis function $b(X)$ for the regularized varying coefficient linear regression, e.g., $\rho_{dU}(\pi,X) = X'\gamma_d$.
Results with a quadratic basis function for specification robustness and implementation details are  in Section~\ref{ASecEM} in the online appendix.
Same as the setting in Section~\ref{SecEmCT}, we use the Epanechnikov kernel with a undersmoothing bandwidth and  ten-fold cross-fitting.
Let $h_1 = 1.05\times\hat\sigma_D\times n_\ell^{-1/5}\times c_1$, where we use a range of constants $c_1$ for robustness check.

\begin{figure}[!htp]
\centering
\caption{\small Histograms of the sufficient treatment values}
\includegraphics[width=0.45\textwidth]{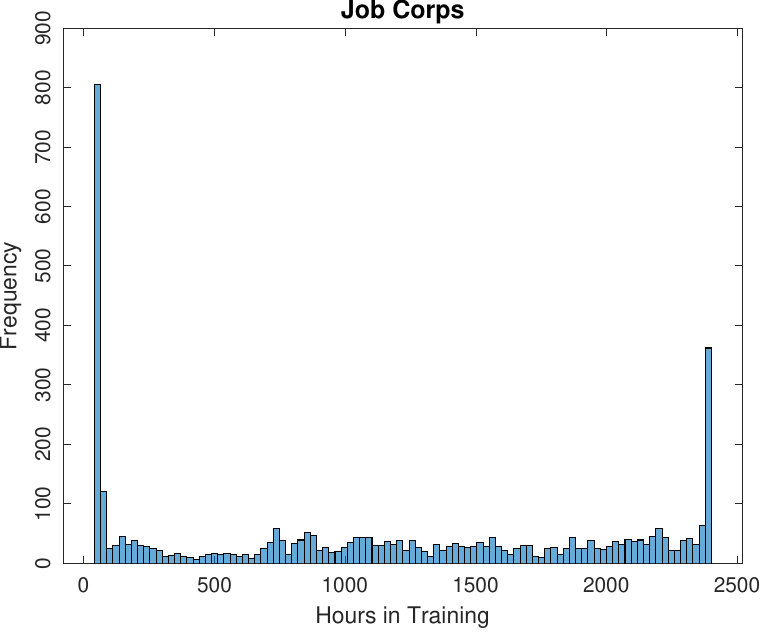}
\includegraphics[width=0.45\textwidth]{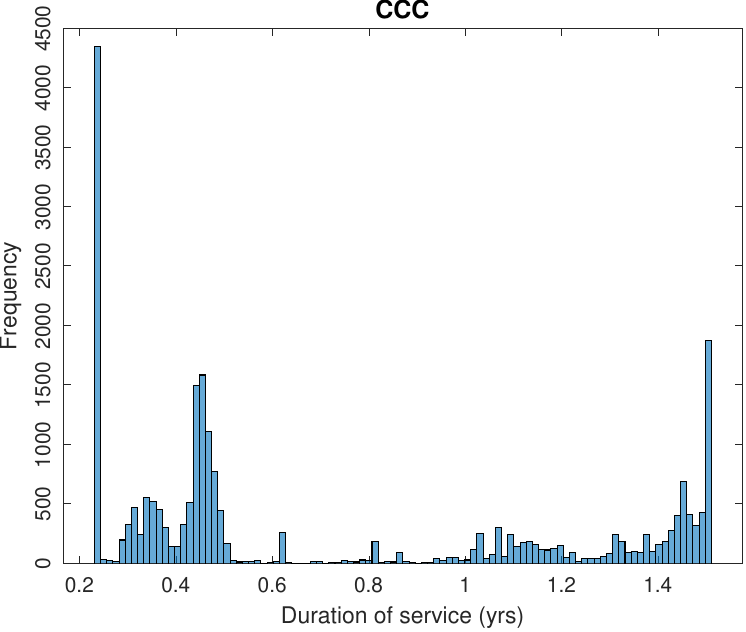}
\label{Fhisdatx}
\end{figure}

We allow subjects to have different sufficient treatment values depending on $X$, i.e., the minimizer of the conditional selection probability given $X$ and $D=d$.
So we might tighten the bounds and the confidence intervals, or capture heterogenous causal effects that are not revealed without the covariates.
Figure~\ref{Fhisdatx} reports histograms of the sufficient treatment values, $\{\hat d_{\text{AT}_xi}, i=1,...n\}$.
For Job Corps in the left panel, most sufficient treatment values are 40 or 2400, but we also see values over $\mathcal{D}$.
In particular, about 20\% of the participants have $d_{\text{AT}x} = 40$, i.e., 40 training hours give them the lowest likelihood of employment.
For about 9\% of the participants who have $d_{\text{AT}x} = 2400$, any hours smaller than 2400 would help employment.
For the CCC data in the right panel,  about 59.2\% of the participants have $d_{\text{AT}x} < 0.5$ years and about 26.19\% of the participants have $d_{\text{AT}x} > 1.2$ years.

\subsection{Job Corps}
\label{SecEmJCX}
We follow the literature to assume the conditional independence Assumption~\ref{AindX}, which is indirectly assessed in \cite{FFGN12ReStat}.
It means that receiving different levels of the treatment is random, conditional on a rich set of observed covariates measured at the baseline survey.
We may further use a control function with an instrumental variable to address the concern that the conditional independence assumption might not hold \citep{IN09ETA, Lee15}.

\begin{figure}[h!]
\centering
\caption{\small (Job Corps) Estimated selection probability and bounds with covariates}
\includegraphics[width=0.46\textwidth]{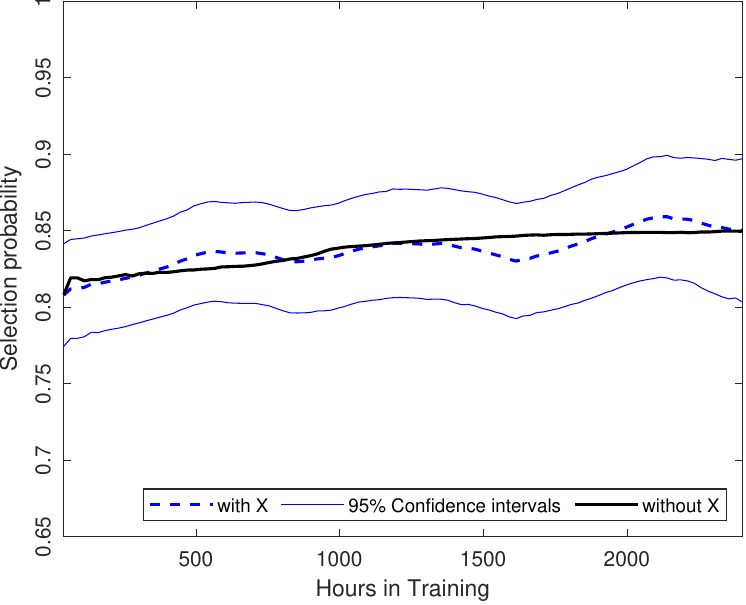}
\includegraphics[width=0.46\textwidth]{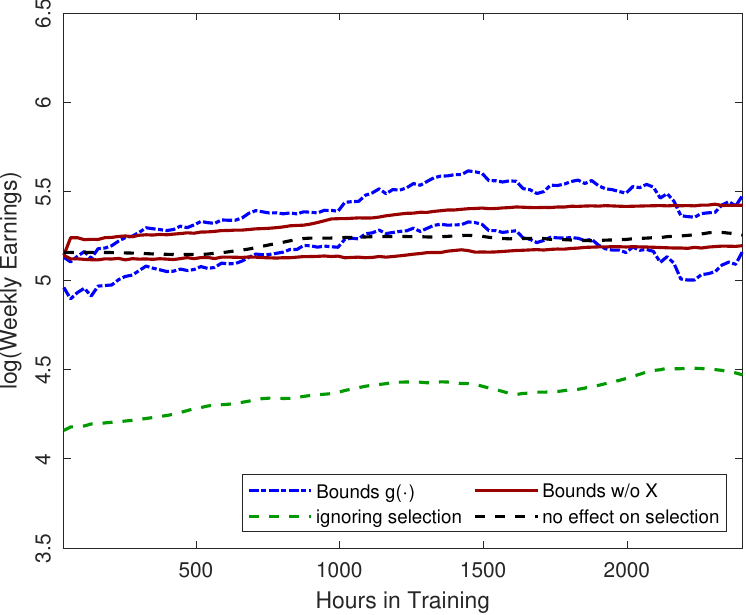}\\
\includegraphics[width=0.46\textwidth]{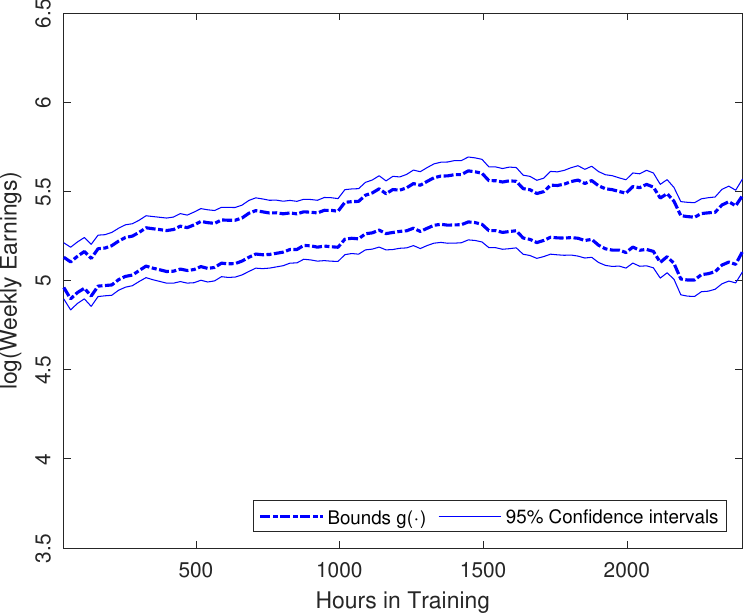}
\includegraphics[width=0.46\textwidth]{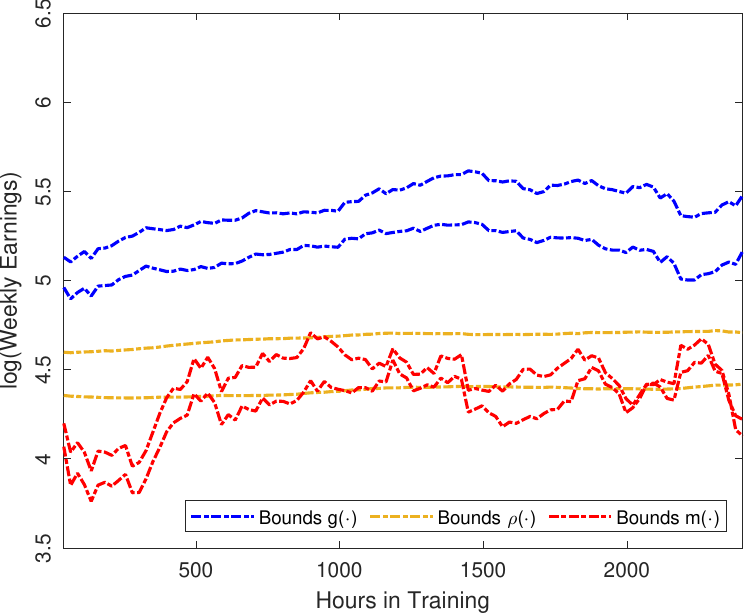}
\label{FbetadXJC}
\end{figure}

The top left panel of Figure~\ref{FbetadXJC} shows the estimated selection probability $\hat\E[S_d]$ with covariates and without covariates.\footnote{We estimate the selection probability $\E[S_d]$ with covariates $X$ by the DML estimator in \cite{CL},
$\hat \E[S_d] = n^{-1}\sum_{\ell = 1}^L\sum_{i\in I_\ell} K_h(D_i-d)(S_i - \hat s_\ell(d, X_i))/\hat\mu_{d\ell}(X_i) + \hat s_\ell(d, X_i)$.}
The bottom left panel of Figure~\ref{FbetadXJC} presents the estimated bounds $[\hat{\underline{\beta_d}}, \hat{\bar\beta}_d]$ and the 95\% confidence intervals.
We find a positive intensive margin effect: increasing the training from 1.5 week to 9 months increases log weekly earnings by at least 0.224, at 5\% significance level.
This is from the largest lower-bound estimate for the ATE  of switching hours over $\mathcal{D}$.
That is, we find the largest ATE $\beta_{1446.465} - \beta_{63.838}$ bounded by $[0.224, 0.718]$ with the 95\%  confidence interval $[0.127, 0.865]$.\footnote{$c_1 = 1$. For robustness check, we find  positive  ATE at $5\%$ significance level over $c_1\in\{0.75, 1, 1.25, 1.5, 1.75\}$ and using a linear or quadratic basis function $b(X)$ in Section~\ref{ASecJC} in the online appendix.
The undersmoothing bandwidths range from 159.008 to 259.490.}
We also not that we do not find significant positive effects on level of weakly earnings.

Consistent with prior empirical research on the Job Corps,
e.g., \cite{FFGN12ReStat} and \cite{HHLL} have found inverted-$U$ shaped average dose-response functions on labor outcomes.
The concave shape is reasonable to consider the optimal treatment intensity in other settings.
Our sufficient treatment values assumption does not assume any shape restrictions and hence includes a concave or monotone selection response.

For comparison, the top right panel presents the bounds without $X$, $[\widehat{\rho_{dL}}, \widehat{\rho_{dU}}]$ (red solid line) given in Figure~\ref{FbetadXJC}.
We also compute two point-estimators of the average dose-response function
``ignoring selection" and ``no effect on selection" by the DML estimator in \cite{CL} incorporating $X$.
The ``ignoring selection" estimates use the full sample including those with zero outcomes,
while the ``no effect on selection" estimates use the selected sample with positive outcomes.
Our bounds are around the ``no effect on selection" estimates, as expected.

To demonstrate the usefulness of the DML method,
the bottom right panel presents two alternative estimators from Lemma~\ref{Lmom}:
the regression-type estimator of
${\mathbb{E}[\bar\beta_{d}(X) \pi_{\text{AT}}(X)]}/{\mathbb{E}\big[\pi_{\text{AT}}(X)\big]}$ (Bounds $\rho(\cdot)$) and
the inverse-probability-weighting estimator of $\lim_{h\rightarrow 0} {\mathbb{E}\big[m_{dU}(W,\xi)}\big]/{\pi_{AT}}$ (Bounds $m(\cdot)$).
These two estimators are not doubly robust, so we see that they are both biased downward, compared with our bounds (Bounds $g(\cdot)$) using a doubly robust moment function.

\subsection{Civilian Conservation Corps (CCC)}
We study the effect of service duration on the age at death, using the data for panel A of Table III in \cite{Aizer}.
They include the 17,639 men (75\% of the original sample size 23,722) who have information on death age and who dies after age 45.
\cite{Aizer} investigate the extent of sample selection and the effects of missing data on their estimates, and conclude modest bias from non-random attrition.
They estimate an accelerated failure time model with added controls for the characteristics of the enrollees and the camps to show that the estimates remain stable.
 They find one more year of training increases the death age by one year. 

From the histograms of the death age and $\log$(death age) in Figure~\ref{FCCCHY} in Section~\ref{ASecCCC} in the online appendix,
we see that the distribution of $\log$(death age) is more skewed.
So we use the level of {\it death age} as our outcome variable $Y$ and estimate the bounds for $\beta_d = \E[death\ age|\text{AT}]$.

\begin{figure}[h!]
\centering
\caption{(CCC) Estimated selection probability and bounds without covariates}
\includegraphics[width=0.45\textwidth]{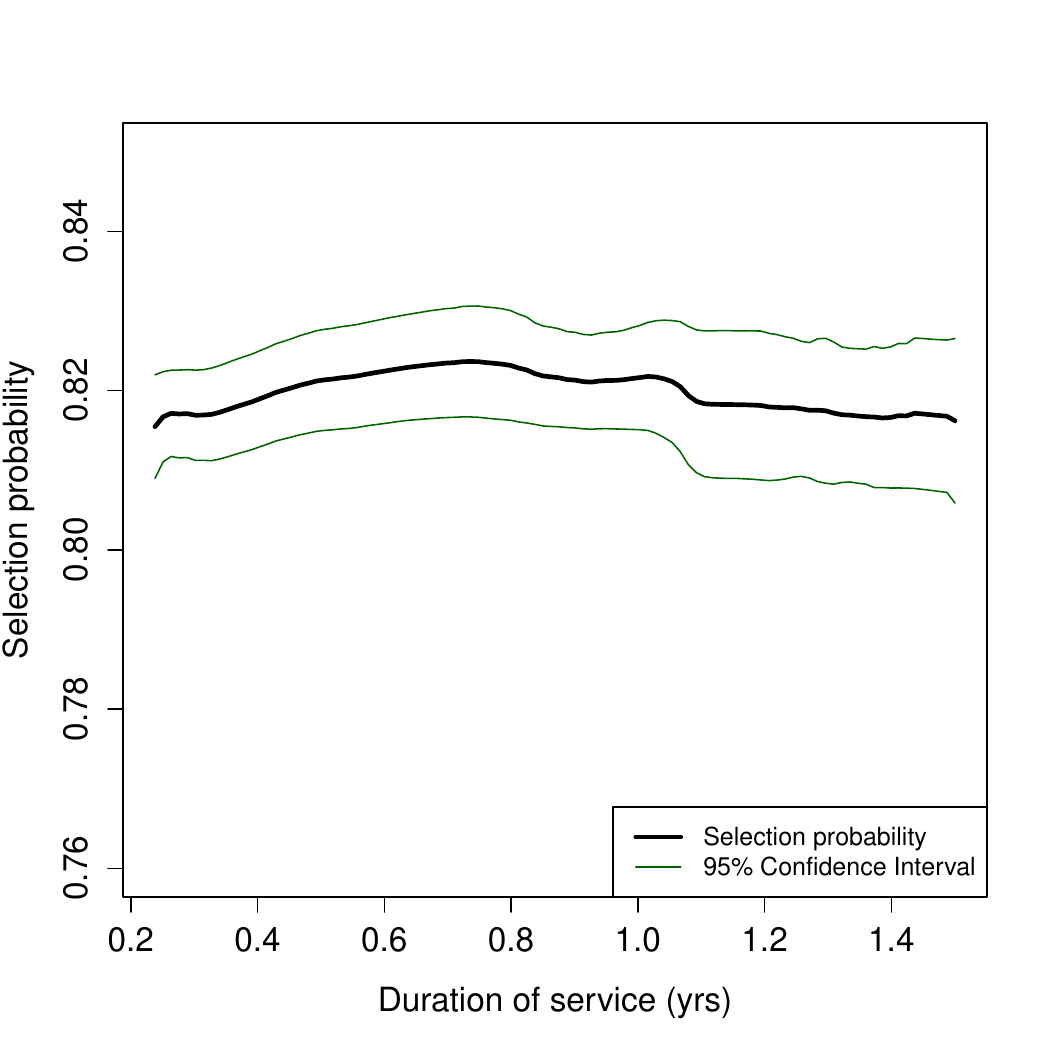}
\includegraphics[width=0.45\textwidth]{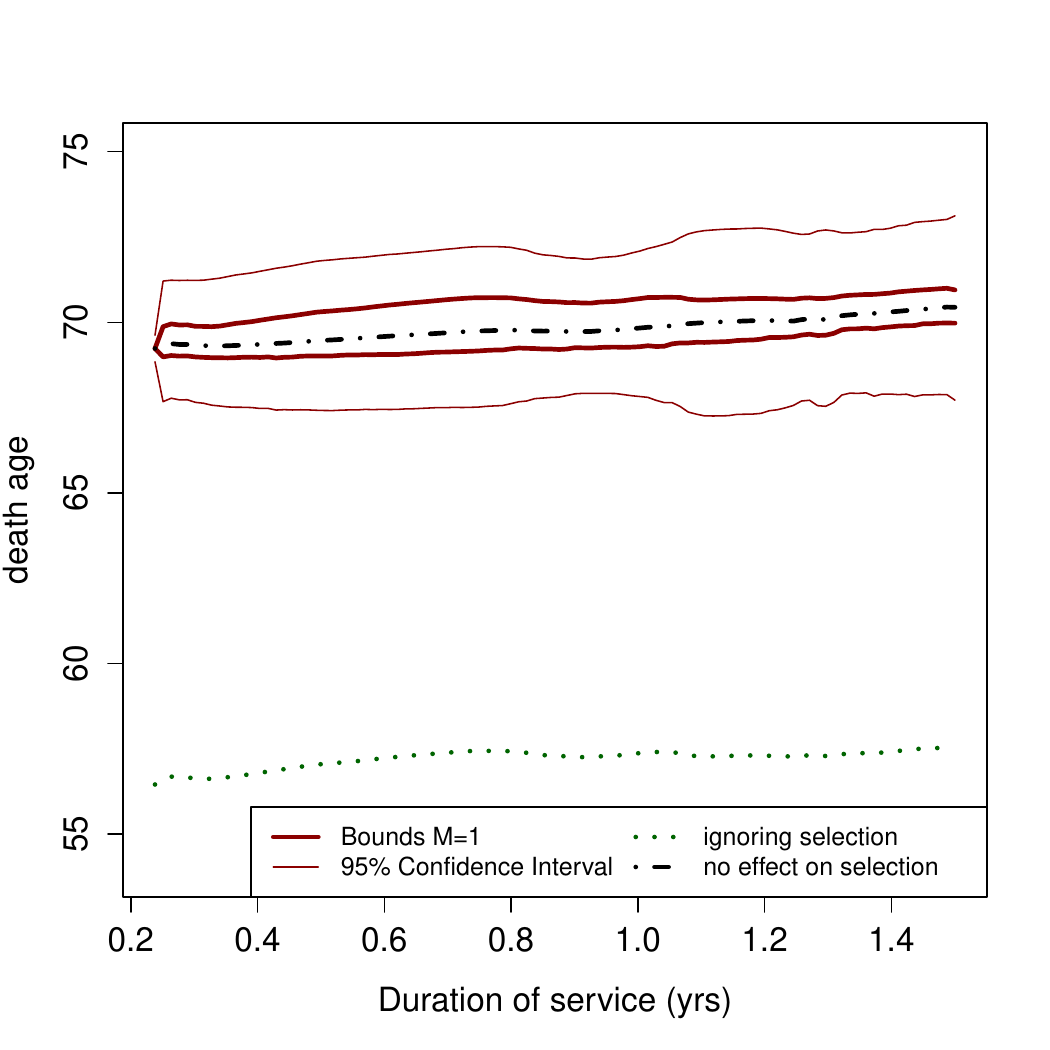}
\label{Fagegrid100}
\end{figure}

We consider one hundred equally spaced grid points $\mathcal{D}_{100} \subset \mathcal{D} = [0.238, 1.5]$, which are from the 15\% to 85\% quantiles of duration in years.
Figure~\ref{Fagegrid100} reports the estimates without using covariates.
The largest ATE is when increasing duration from 0.238 to 1.488 years with the bounds $[0.747, 1.765]$ and 90\% confidence interval $[-1.031, 3.476]$.\footnote{
The undersmoothing bandwidths range from 0.1909 to 0.4409.
$\nu=0.01$ and $c_1=1$.
For robustness check, the lower bound estimates of the largest ATE with $c_1 \in \{0.75, 1,1.25,1.5, 1.75, 2\}$ are positive but insignificant at 10\% level.
}
The proportion of always-takers, or the minimum selection probability, is $\hat\pi_{\text{AT}} = \hat s(0.2382) = 0.8155$.
Our sufficient treatment value Assumption~\ref{Amin} means that if the participants' death ages are observed when they received 0.2382 years of service, then they would remained observed for any duration over these one hundred grid points.

\begin{figure}[h!]
\centering
\caption{(CCC) Estimated selection probability and bounds with covariates over $\mathcal{D}_{100}$}
\includegraphics[width=0.49\textwidth]{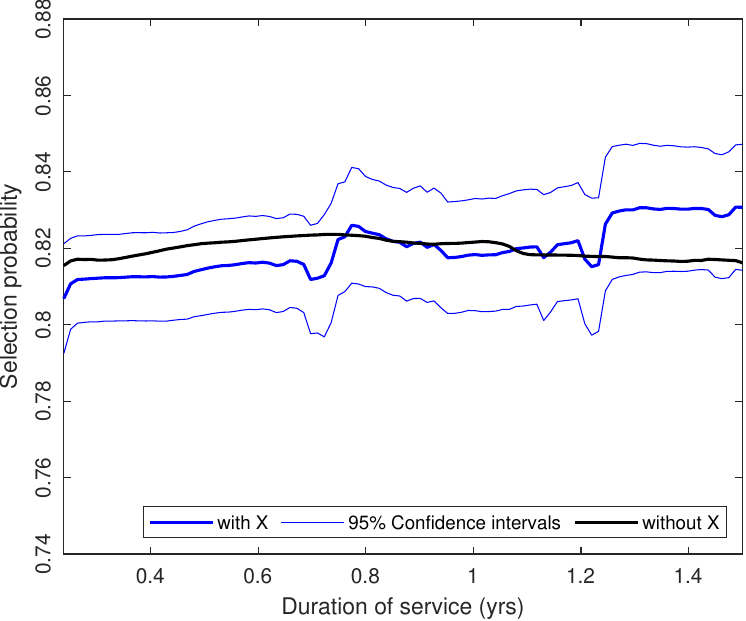}
\includegraphics[width=0.49\textwidth]{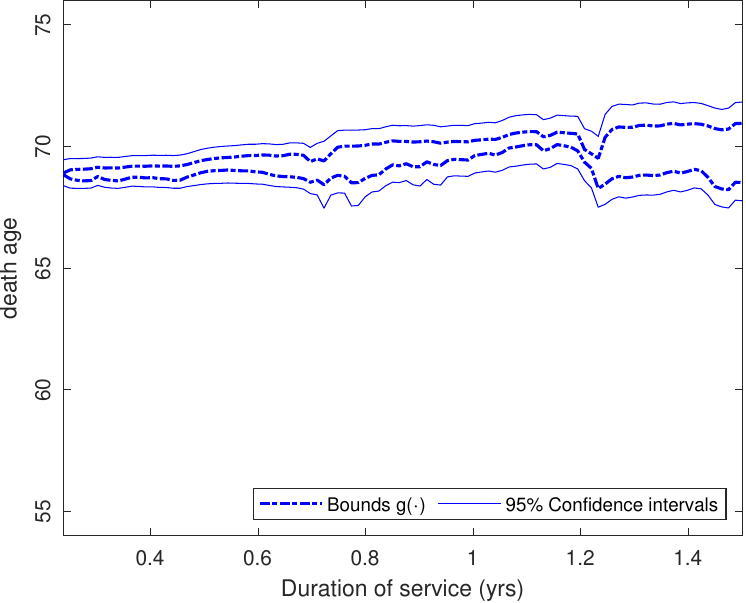}
\label{Fagegrid100X}
\end{figure}

\begin{figure}[h!]
\centering
\caption{(CCC) Estimated selection probability and bounds with covariates over $\mathcal{D}_{30}$}
\includegraphics[width=0.49\textwidth]{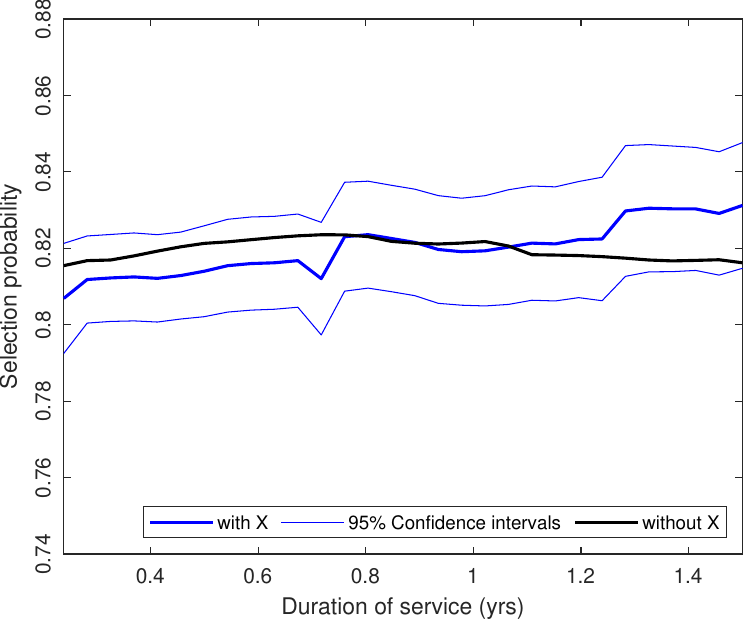}
\includegraphics[width=0.49\textwidth]{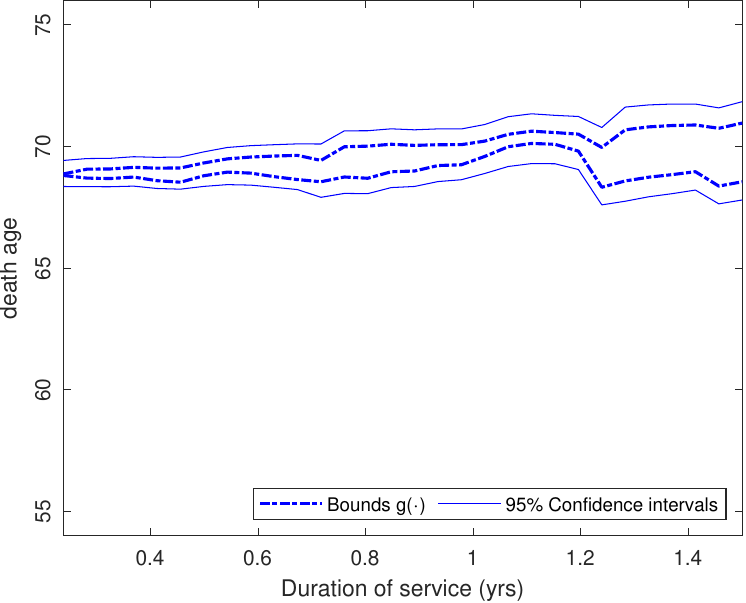}
\label{Fagegrid30X}
\end{figure}

Next we include covariates as in column (3) Add Indiv Controls in Panel A in Table III in \cite{Aizer}.
Figure~\ref{Fagegrid100X} reports the estimates with covariates.
The largest ATE is when increasing duration from  0.238 to 1.156  years with the bounds $[1.169, 1.755]$ and 95\% confidence interval $[0.366, 2.795]$.
That is, increasing duration from  0.238 to 1.156 years would at least increase the average death age by 1.169 years and the effect is significant at 5\% level, which is consistent with the findings in \cite{Aizer}.\footnote{
$\nu=0.01$ and $c_1=1.5$. For robustness check, the lower bounds with $c_1 \in \{1.75, 2\}$ and linear/quadratic basis functions range from $0.288$ to $0.54$ but are not significant at 10\% level.
The undersmoothing bandwidths range from 0.162 to 0.313.
}
However, the effects are not significant for other choices of $c_1$.
By the histogram in Figure~\ref{FhistdCCC} and summary statistics in Table~\ref{SSCCC} in the online appendix, the distribution of duration might have mass points at 0.5, 1, 1.5, 2 years.
So the caveat of implementing our method on this dataset is that  the smoothness conditions could be violated.
This may explain why the estimates in Figure~\ref{Fagegrid100X} are not smooth.
Nevertheless, we are able to implement our method at a coarser grid with 100 grid points and assume the distributions are smooth locally at these grid points.
We also consider thirty equally spaced grid points $\mathcal{D}_{30} \subset \mathcal{D} = [0.238, 1.5]$ in Figure~\ref{Fagegrid30X}
and obtain similar results as $\mathcal{D}_{100}$.\footnote{The largest  ATE is when increasing the duration from 0.238 to 1.109 years
with the bounds $[1.255, 1.822]$ and 95\% confidence interval $[0.43, 2.918]$.  $c_1=1.5$.  The undersmoothing bandwidths range from 0.158 to 0.294.
}

\section{Conclusion}
We study causal effects of a treatment/policy variable that could be either continuous, multivalued discrete or binary, in a sample selection model where the treatment affects the outcome and also researchers' ability to observe the outcome.
To account for the non-random selection into samples,
we provide sharp bounds for the mean potential outcome of always-takers whose outcomes are observed regardless of their treatment value, generalizing \cite{LeeBound}'s bound.
We propose a novel sufficient treatment value (set) assumption to (partially) identify the share of always-takers in the observed selected $d$-receipts for each treatment value $d$.

By incorporating pretreatment covariates $X$, we allow for unconfoundedness and
allow subjects with different values of $X$ to have different sufficient treatment values, which might tighten the bounds, increase precision (tighten the confidence intervals), and reveal heterogeneity, as we illustrate in our empirical analysis of the Job Corps and CCC programs.
The inference procedure is robust to extensive margin, allowing for any unknown functional form of the selection probability with respect to treatment.


There are many potential applications of our bounds for continuous/multivalued treatments.
For example,
\cite{ACL} study the multivalued treatments of the Workforce Investment Act program on participants’ earnings.
\cite{swedish} use Swedish lotteries data to study the effect of wealth on labor supply.

\appendix
\appendixpage
\singlespacing


\paragraph{Proof of Lemma~\ref{Tmin}:}
The proof is implied by Theorem~\ref{TminJ} with $M=1$ and $d_1 = d_{\text{AT}}$.
\hfill $\square$

\paragraph{Proof of Theorem~\ref{TminJ}:}
$\p(S_{d_1} = S_{d_2} = ... = S_{d_M} = 1) =
\p(S_{d_1} = 1, \{S_{d_2} = ... = S_{d_M} = 1\}) =
\p(S_{d_1} = 1) + \p(\{S_{d_2} = ... = S_{d_M} = 1\}) -
\p(S_{d_1} = 1 \mbox{ or } \{S_{d_2} = ... = S_{d_M} = 1\})
\geq \p(S_{d_1} = 1) + \p(\{S_{d_2} = ... = S_{d_M} = 1\}) - 1$ that is the Fr\'{e}chet-Hoeffding bound.
The same argument gives $\p(\{S_{d_m} = ... = S_{d_M} = 1\}) \geq \p(S_{d_m} = 1) + \p(\{S_{d_{m+1}} = ... = S_{d_M} = 1\}) - 1$
for $m=2,...,M-1$.
Under Assumption~\ref{Aind}, $\p(S_{d_m} = 1) = s(d_m)$.
So we obtain $\pi_L^M$ by induction.

Denote the always-takers $\text{AT} = \{S_d=1: d\in\mathcal{D}\}$
and the $d$-complier $\text{CP}_d =
\{S_d = 1, S_{d'}=0 \text{ for some } d'\in\mathcal{D}\}$.
By the law of iterated expectations,
\begin{align*}
\mathbb{E}[Y|S = 1, D=d] =
\mathbb{E}[Y_d|S_d = 1] &=
\mathbb{E}[Y_d|S_d = 1, \text{AT}]
\cdot \p(S_d = 1, \text{AT}| S_d = 1) \\
&\ \ + \mathbb{E}[Y_d|S_d = 1, \text{CP}_d]\cdot (1-\p(S_d = 1, \text{AT}| S_d = 1))\\
& =\beta_{d} \cdot \pi_{\text{AT}}/s(d)
+ \mathbb{E}[Y_d|S_d = 1, \text{CP}_d] \cdot(1-\pi_{\text{AT}}/s(d)).
 \end{align*}
Then we apply Proposition 1a in \cite{LeeBound}.
Specifically, we replace $\Delta^{UB}_0$ in \cite{LeeBound}
with $\E[Y|D=d, S=1, Y \geq Q^d(1-\pi_{\text{AT}}/s(d))]$,
replace $p_0$ with $1-\pi_{\text{AT}}/s(d)$, replace $D=1$ with $D=d$, replace $Y_1^*$ with $Y_d$,  replace $\{S_0=1, S_1 = 1\}$ with AT , replace $\{S_0=1, S_1=1\}$ with CP$_d$, and remove $\E[Y|D = 0, S = 1]$.
Then the same arguments in the proof of Proposition 1a in \cite{LeeBound} yields
that $\beta_{d} \leq
\mathbb{E}[Y|Y \geq Q^d(1-\pi_{\text{AT}}/s(d)), D=d, S=1] =:
\rho_{dU}(\pi_{\text{AT}{}})$ that is sharp.

Since $\rho_{dU}(\pi_{\text{AT}{}})$ is decreasing in $\pi_{\text{AT}}$,
$\rho_{dU}(\pi_{\text{AT}{}}) \leq \rho_{dU}(\pi_L^M)$ that is sharp.
A similar argument for the sharp lower bound follows.
\hfill $\square$

\paragraph{Proof of Theorem~\ref{TEstLB}:}
We prove the result using a given $d_{\text{AT}}$ and $\mathcal{D}_M$.
In the later section {\bf Misclassification} of the proof, we show that using $\hat d_{\text{AT}_J}$ is equivalent asymptotically.
Let $W_i$ be a generic random variable and $\mathcal{W}_\ell^c$ denote the observations $W_i$ for $i \notin I_\ell$.
Denote the sample average operator over all observations as $\E_n[W]:= n^{-1}\sum_{i=1}^n W_i$,
the sample average operator over the observations in group $I_\ell$ as $\E_{n_\ell}[W] := n_\ell^{-1}\sum_{i\in I_\ell} W_i$,
the sample average operator over the observations in group $I_\ell^c$ as $\E_{n_\ell}^c[W] := (n-n_\ell)^{-1}\sum_{i\in I_\ell^c} W_i$,
and the conditional mean operator given the observations  in group $I_\ell^c$ as $\E_\ell[W] := \E[W| \mathcal{W}_\ell^c]$.
Denote as $\1_i := \1\{Y_i \geq Q^d(1-p)\}$ and $\hat \1_i^\ell := \1\{Y_i \geq \hat Q_\ell^d(1-\hat p_{\ell})\}$ for $i \in I_\ell$.

We first derive the asymptotically linear representation
$\widehat{\rho_{dU}(\pi)} - \rho_{dU}(\pi)
= \E_n \left[\phi_{dU}
\right] + o_{\p}(1/\sqrt{nh})$, where
for $p = p_d < 1$,\footnote{To simplify notation and without loss of clarity, we suppress the subscript of $d$ in $p_d$.}
\begin{align}
\label{IF}
\phi_{dU} &:= (\phi_{12} + \phi_3 - \rho_{dU}(\pi) \phi_1) p^{-1},\\
\phi_{3}&:=\frac{K_h(D - d) S}{s(d)f_D(d)} \left(Y \1 -  \E[Y\1|D=d, S=1]  \right),
\notag\\
\phi_{12} = \phi^U_{12} &:=
 \phi_1 Q^d(1-p) - \phi_2(1-p) Q^d(1-p) f_{Y|DS}(Q^d(1-p)|d, 1),
\notag\\
\phi_1 &:= \left( \phi_\pi
- p\phi_s(d)  \right)/s(d), \mbox{with}\notag\\
&\phi_s(d) := (S - s(d))\frac{K_h(D - d)}{f_D(d)},
\hat \pi_{\ell} - \pi =  \E_{n_\ell}^c[\phi_\pi] + o_{\p}(1/\sqrt{nh}), \notag \\
\hat Q^d_\ell(\tau) - Q^d(\tau) &= \E^c_{n_\ell}[\phi_2(\tau)] + o_{\p}(1/\sqrt{nh}), \mbox{with}\ \phi_2(\tau) := \frac{K_h(D-d)S\left(\tau - {\bf 1}\left\{ Y \leq Q^d(\tau)\right\}\right)}{s(d)f_D(d)f_{Y|DS}(Q^d(\tau)|d,1)}.
\notag
\end{align}
For $\pi = \sum_{d\in\mathcal{D}_M} s(d) - M +1$, $\phi_{\pi} = \sum_{d\in\mathcal{D}_M} \phi_s(d)$.

For the lower bound $\widehat{\rho_{dL}(\pi)} - \rho_{dL}(\pi)
= \E_n \left[\phi_{dL} \right] + o_{\p}(1/\sqrt{nh})$,
let $\1 = \1\{Y \leq Q(p)\}$, $\hat \1_i^\ell = \1\{Y_i \leq \hat Q^d_\ell(\hat p_\ell)\}$, and
$\phi_{12} = \phi^L_{12} := \phi_1 Q^d(p) + \phi_2(p) Q^d(p) f_{Y|DS}(Q^d(p)|d, 1)$.
For $p = 1$, $\phi_{dU} = \phi_{dL} := \frac{K_h(D - d) S}{s(d)f_D(d)} \left(Y -  \E[Y|D=d, S=1]  \right)$.

We focus on the upper bound $\rho_{dU}(\pi)$ in the proof, and similar arguments for the lower bound $\rho_{dL}(\pi)$ are given later.

Let 
$\rho_{dU}(\pi)
=: num/p$,
where $num:=\mathbb{E}[Y\1\{Y \geq Q^d(1-p)\}|D=d, S=1]$.
By a linearization of the estimator,
\begin{align}
\widehat{\rho_{dU}(\pi)} - \rho_{dU}(\pi)
&=  \frac{\widehat{num} - num}{p} - \frac{\rho_{dU}(\pi)}{p}(\hat p - p) + O_{\p}\left(\|\widehat{num} - num\|\|\hat p - p\| + \|\hat p - p\|^2\right).
\label{Elinear}
\end{align}

Decompose the numerator estimator $\widehat{num} = L^{-1}\sum_{\ell=1}^L\widehat{num}^\ell$ in (\ref{Elinear}) to the Step1\&Step2 estimation error and the Step 3 estimation,
\begin{align*}
\widehat{num}^\ell
=
\frac{\E_{n_\ell} \left[Y\hat \1^\ell K_h(D - d)S\right]}{\E_{n} \left[K_h(D -d)S\right]}
= \widehat{num}_{12}^\ell + \widehat{num}_3^\ell, \mbox{ where}
\end{align*}
\begin{align*}
\widehat{num}_{12}^\ell := \frac{\E_{n_\ell} \left[Y (\hat \1^\ell - \1) K_h(D - d)S\right]
}{\widehat{den}}, \
\widehat{num}_{3}^\ell := \frac{
 \E_{n_\ell} \left[Y \1 K_h(D - d)S\right]}{\widehat{den}},\
  \widehat{den} := \E_{n} \left[K_h(D -d)S\right].
\end{align*}

To derive the influence function at each step,
we use the following four Claims, whose proofs are in the online appendix.
\paragraph{Claim-Step1:} For any $d\in\mathcal{D}$,
(i) $\hat s(d) - s(d) =  \E_{n}\left[\phi_s(d)\right] + O_{\p}(1/(nh) + h^4)  = O_{\p}(1/\sqrt{nh}+ h^2).$

(ii)
$\lim_{h\rightarrow 0 }h \E[\phi_s^2(d)] =  V_{s(d)} R_k/f_D(d)$.

(iii) For any $d'\neq d$, $\lim_{h\rightarrow 0}h\E[\phi_{s(d')} \phi_{s(d)}] = 0$.


(iv) Let $\hat p = \hat \pi/\hat s(d) - \nu$. $\hat p - p =  \E_{n}[\phi_1] + O_{\p}(1/(nh) + h^4)$.

\paragraph{Claim-Step2:}
Theorems 3.7 and 4.1 in \cite{DHB} provide that
for any $d\in\mathcal{D}$, $\hat Q^d(\tau) - Q^d(\tau) = \E_n[\phi_2(\tau)] + o_{\p}(1/\sqrt{nh})$, uniformly in $\tau\in[0,1]$.
And $V_{Q}(\tau) := \lim_{h\rightarrow 0} h\E[\phi_2(\tau)^2]= \frac{\tau(1-\tau)R_k}{s(d)f_D(d) f_{Y|DS}(Q^d(\tau)|d,1)^2}$.


\paragraph{Claim-Step3:}
$\widehat{den} - den = O_{\p}(1/\sqrt{nh} + h^2)$, where $den :=  s(d)f_D(d)$.
$\widehat{num_3}^\ell - \E[Y\1|D=d, S=1]  =  \E_{n_\ell}[\phi_3] + o_{\p}(1/\sqrt{nh}).$
$\lim_{h\rightarrow 0} h\E[\phi_3^2]  = V_{3}R_k/(f_D(d)s(d))$.

\paragraph{Claim-SE:}
$\widehat{num_{12}}^\ell = num_{12}^\ell + o_{\p}(1/\sqrt{nh})$,
where
\begin{align*}
num_{12}^\ell :=
 (\hat p_{\ell} - p)Q^d(1-p) - \left(\hat Q_\ell^d(1- p) - Q^d(1-p)\right)Q^d(1-p) f_{Y|DS}(Q^d(1-p)|d,1)
\end{align*}
for $\rho_{UB}(\pi)$, and
$num_{12}^\ell :=
 (\hat p_{\ell} - p)Q^d(p)
+ \left(\hat Q_\ell^d(p) - Q^d(p)
\right)Q^d(p) f_{Y|DS}(Q^d(p)|d,1)$
for $\rho_{LB}(\pi)$.

\begin{align*}
&\widehat{num}^\ell - num
= num_{12}^\ell + \widehat{num}_{3}^\ell - \E[Y\1|D=d, S=1] + o_{\p}((nh)^{-1/2}) \\
&=
(\hat p_\ell - p) Q^d(1-p)  - \left(\hat Q_\ell^d(1-p) - Q^d(1-p)
\right)Q^d(1-p) f_{Y|DS}(Q^d(1-p)|d,1)
\\
&\ \ \ + \E_{n_\ell}[\phi_3]
 + o_{\p}((nh)^{-1/2})
 \\&
 =
\E_{n_\ell}^c[\phi_1] Q^d(1-p)  - \E_{n_\ell}^c[\phi_2]
Q^d(1-p) f_{Y|DS}(Q^d(1-p)|d,1)
+ \E_{n_\ell}[\phi_3]
 +  o_{\p}((nh)^{-1/2}),
 \end{align*}
where
the first and second equalities is by Claim-SE and Claim-Step3,
and the third equality is by Claim-Step1 and Claim-Step2.

Note that $L^{-1}\sum_{\ell = 1}^L \E_{n_\ell}^c[W] = L^{-1}\sum_{\ell = 1}^L (n - n_\ell)^{-1} \sum_{i\in I_\ell^c} W_i
= (L(n-n/L))^{-1}(L-1)\sum_{i=1}^n W_i
= \E_n[W]$.
And $L^{-1}\sum_{\ell = 1}^L \E_{n_\ell}[W] = \E_n[W]$.
Then $\widehat{num} - num = L^{-1}\sum_{\ell=1}^L\widehat{num}^\ell - num = \E_n\big[
\phi_1Q^d(1-p) - \phi_2 Q^d(1-p)f_{Y|DS}(Q^d(1-p)|d,1) + \phi_3
\big]  + o_{\p}(1/\sqrt{nh})
= \E_n[\phi_{12} + \phi_3] + o_{\p}(1/\sqrt{nh})$.
By (\ref{Elinear}), we obtain~$\phi$.

\paragraph{Variance.}
We derive the asymptotic variance $V_{dU}$ by
$\E\left[\left(\phi_{12}^U + \phi_3 - \rho_{dU}(\pi) \phi_1\right)^2\right]
= \E\big[{\phi_{12}^U}^2 + \phi_3^2 + \rho_{dU}(\pi)^2 \phi_1^2
+ 2\phi_{12}^U\phi_3 - 2\rho_{dU}(\pi)\phi_{1}\phi_{12}^U
- 2\rho_{dU}(\pi)\phi_1\phi_3
\big]$.
Since $\phi_1$ does not involve $Y$, the law of iterated expectations yields $\lim_{h\rightarrow 0}h\E[\phi_1 \phi_2] = 0$
and $\lim_{h\rightarrow 0}h\E[\phi_1 \phi_3] = 0$.

$\lim_{h\rightarrow 0}  h\E[\phi_1^2] =\lim_{h\rightarrow 0}  (h\E[\phi_\pi^2] + p^2 \cdot h\E[\phi_s^2(d)])/s(d)^2 = \left(V_\pi + p^2 V_{s(d)}\right)R_k/(f_D(d)s(d)^2)$, by Claim-Step1 (ii)(iii),
where $V_\pi = V_{s(d_{\text{AT}_J})}f_D(d)/f_D(d_{\text{AT}_J})$ if $\pi = s(d_{\text{AT}_J})$.
Under Assumption~\ref{AminJ}, when $d\in\mathcal{D}_M$ so that $\pi$ contains $s(d)$,
there is an additional term in $\lim_{h\rightarrow 0}  h\E[\phi_1^2]$, \\
$-2p\lim_{h\rightarrow 0}h\E[\phi_\pi\phi_s(d)]/s(d)^2 = -2pV_{s(d)}R_k/(f_D(d)s(d)^2)$.

$\lim_{h\rightarrow 0} h\E\left[{\phi_{12}^U}^2\right] =
\lim_{h\rightarrow 0}
h\E\left[\phi_1^2 \right] Q^d(1-p)^2
+ h\E\left[\phi_2(1-p)^2 \right]
Q^d(1-p)^2 f_{Y|DS}(Q^d(1-p)|d, 1)^2.$

By Claim-Step2, $\lim_{h\rightarrow 0}h\E[\phi_2(1-p)^2] = V_Q(1-p) = \frac{p(1-p)R_k}{s(d)f_D(d) f_{Y|DS}(Q^d(1-p)|d,1)^2}$.
We obtain $V_2$.

$\lim_{h\rightarrow 0} h\E[\phi_1\phi_{12}^U] = \lim_{h\rightarrow 0}
h\E[\phi_1^2]  Q^d(1-p).$

For $\E[\phi_{12}^U\phi_{3}]$,
\begin{align}
&h\E[\phi_2(\tau) \phi_3] \times s(d)^2f_D(d)^2 f_{Y|DS}(Q^d(\tau)|d,1)\notag\\
&=h\E\left[SK_h(D-d)^2 \E\left[
\left(Y{\bf 1} - \E[Y{\bf 1}|D=d, S=1]\right)\left(\tau - \1\{Y\leq Q^d(\tau)\}\right)|D,S\right]\right]
\notag\\
&= h\E\Big[SK_h(D-d)^2\Big(
\E[Y{\bf 1}|D,S] \tau - \E[Y{\bf 1}|D=d, S=1]\tau \notag\\
&\ \ \ + \E[Y{\bf 1}|D=d, S=1]\E[{\bf 1}\{Y \leq Q^d(\tau)\}|D,S]\Big)\Big]
\label{E123}
\\
&\rightarrow R_k s(d)f_D(d)\tau\E[Y{\bf 1}|D=d,S=1] \notag
\end{align}
as $h\rightarrow 0$.
So
$\lim_{h\rightarrow 0} h\E[\phi_{12}^U\phi_3]
= \lim_{h\rightarrow 0} -h\E[\phi_2(1-p) \phi_3] Q^d(1-p) f_{Y|DS}(Q^d(1-p)|d, 1)
= V_{23}R_k/(2 f_D(d)s(d)).
$


For $\rho_{LB}(p)$ with $\1 = \1\{Y\leq Q^d(p)\}$,
(\ref{E123}) becomes
$h\E\Big[SK_h(D-d)^2\Big(
\E[Y{\bf 1}|D,S] p - \E[Y{\bf 1}|D=d, S=1]p
+ \E[Y{\bf 1}|D=d, S=1]\E[{\bf 1}\{Y \leq Q^d(p)\}|D,S]\Big)
- \E[Y\1|D,S]\Big]
\rightarrow R_k s(d)f_D(d)(p-1)\E[Y{\bf 1}|D=d,S=1]$.
So
$\lim_{h\rightarrow 0} h\E[\phi_{12}^L\phi_3]
= \lim_{h\rightarrow 0} h\E[\phi_2(p) \phi_3] Q^d(p) f_{Y|DS}(Q^d(p)|d, 1)
= -(1-p)\E[Y{\bf 1}|D=d,S=1] Q^d(p) R_k /(s(d)f_D(d))
= V_{23}R_k/(2 f_D(d)s(d)).
$


Putting together the above results yields
$V_{dU} = \lim_{h\rightarrow 0} p^{-2}\Big\{
h \E[\phi_1^2]\cdot\big(
Q^d(1-p)^2
+ \rho_{dU}(\pi)^2
- 2 \rho_{dU}(\pi) Q^d(1-p)
\big)
+ h\E[\phi_2(1-p)^2] \cdot Q^d(1-p)^2 f_{Y|DS}(Q^{d}(1-p)|d,1)^2 + h\E[\phi_3^2]
+ h2\E[\phi_{12}\phi_3]\Big\}$.

\paragraph{Bias.}
We show  $\E[\phi_{dU}] = h^2 {B}_{dU} + o_\p(h^2)$, where ${B}_{dU} := \big( {B}_\pi - p{B}_s(d)\big) Q^d(1-p)/s(d) - {B}_2(1-p) Q^d(1-p) f_{Y|DS}(Q^d(1-p)|d, 1) + {B}_3 - \rho_{dU}(\pi) {B}_1) p^{-1}$ derived below.
So the bias is first-order asymptotically ignorable by assuming $\sqrt{nh}h^2 =o(1)$.

Assuming the second derivatives of $s(d)$ and $f_D(d)$ are bounded continuous,
\begin{align*}
\E[\phi_s(d)] &= \E[(s(D) - s(d)) K_h(D-d)/f_D(d)] = h^2{B}_s(d)
 + o_\p(h^2), \mbox{where}\\
 {B}_s(d)&:= \frac{\kappa}{2} \frac{d^2}{dv^2} \big\{(s(v) - s(d)) f_D(v) \big\}\big|_{v=d}/f_D(d).
\end{align*}

Under Assumption~\ref{Amin}, ${B}_\pi = {B}_s(d_{\text{AT}})$.
Under Assumption~\ref{AminJ}, $\E[\phi_\pi] = \sum_{d\in\mathcal{D}_J} \E[\phi_s(d)]$, so ${B}_\pi = \sum_{d\in\mathcal{D}_J} {B}_s(d)$.
We obtain $B_1 = \big(B_\pi - pB_s\big)/s(d)$.
By the same argument,
\begin{align*}
\E[\phi_2(1-p)] &= \E\left[(\pi - F_{Y|DS}(Q^d(1-p)|D,1)) \frac{K_h(D-d) s(D)}{s(d) f_D(d) f_{Y|DS}(Q^d(1-p)|d, 1)}  \right] \\
&= h^2 {B}_2(1-p)+ o_\p(h^2), \mbox{where}
\end{align*}
\begin{align*}
{B}_2(1-p)&:= \frac{\kappa}{2} \frac{d^2}{dv^2} \big\{ (\pi -  F_{Y|DS}(Q^d(1-p)| v, 1)) f_D(v) s(v)\big\}\big|_{v=d}/(s(d) f_D(d) f_{Y|DS}(Q^d(1-p)|d, 1));\\
\E[\phi_3] &= \E\left[\frac{K_d(D-d) s(D)}{s(d) f_D(d)}\big(\E[Y\1|D,S=1] - \E[Y\1|D=d, S=1] \big)\right] = h^2{B}_3 + o_\p(h^2), \mbox{where}\\
{B}_3&:=\frac{\kappa}{2} \frac{d^2}{dv^2} \big\{ f_D(v) s(v) \big(\E[Y\1|D=v,S=1] - \E[Y\1|D=d, S=1] \big)
\big\}\big|_{v=d}/(s(d) f_D(d)).
\end{align*}

Similarly for the lower bound, ${B}_{dL} := \big( {B}_\pi - p{B}_s(d)\big) Q^d(p)/s(d) - {B}_2(p) Q^d(p) f_{Y|DS}(Q^d(p)|d, 1) + {B}_3 - \rho_{dU}(\pi) {B}_1) p^{-1}$, where $\1 = \1\{Y \leq Q^d(p)\}$.

\paragraph{Asymptotic normality.}
The asymptotic normality follows from the Lyapunov central limit theorem with the third
absolute moment.
Specifically, the Lyapunov condition holds if $\lim_{n\rightarrow \infty} n^{-1/2} h^{3/2}$\\
$ \E[|\phi|^3] = 0$.
Under the condition that $\E[|Y|^3{\bf 1}|D=d, S=1]$ is continuous in $d \in \mathcal{D}$, we can show that $\E[|\phi_3|^3] = O(h^{-2})$ by a similar algebra as for the variance in Claim-Step 1(ii).

%

\paragraph{Misclassification.}
Let the true $ \tau_{j} =  s(d_{j+1}) -  s(d_j)$ with the estimate $\hat \tau_{j} = \hat s(d_{j+1}) - \hat s(d_j)$.
First consider $\hat d_{\text{AT}_J} = \arg\min_{d\in\mathcal{D}_J} \hat s(d) = d_{j'}\in \mathcal{D}_c$ and $\tau_{j'} = 0$.
As both $d_{j'}$ and $d_{d'+1}$ are minimizers of $s(d)$, $\beta_{d_{j'}}$ and $\beta_{d_{j'+1}}$ are point-identified.
There is no misclassification problem,
as we can use $\hat d_{\text{AT}_J} = d_{j'}$ and $\hat p_\ell = \min\{\hat s_\ell(d_{j'})/\hat s_\ell(d_{j'+1}) ,1\} - \nu$ in finite samples to estimate valid bounds at $d = d_{j'+1}$ that contains the point estimates $\hat \beta_{d_{j' + 1}}$.

We consider misclassification when $\hat \tau_{j} \leq 0 < \tau_{j}$ or $\hat \tau_{j} \geq 0 > \tau_{j}$ for some $j \in\{1,...,J-1\}$.
Suppose $\tau_j \neq 0$, $d_{\text{AT}} =  \arg\min_{d\in\mathcal{D}} s(d) \in \mathcal{D}_s$, and
$d_{\text{AT}_J} =  \arg\min_{d\in\mathcal{D}_J} s(d) \in \mathcal{D}_s$ .
Because $|\hat \tau_{j} - \tau_{j}| \leq |\hat s(d_{j+1}) - s(d_{j+1})| + |\hat s(d_{j}) - s(d_{j})| \leq 2\mathsf{s}_n$,
misclassification implies that $0 < |\tau_{j}| \leq 2\mathsf{s}_n$.
For a discrete multivalued treatment, $|\tau_j| > 2\mathsf{s}_n$ for $n$ large enough, which implies correct classification.
For a continuous treatment, let $\mathsf{D} = \overline{\mathcal{D}} - \underline{\mathcal{D}}$, so $d_{j+1} - d_j = \mathsf{D}/(J-1)$ for all equally spaced grid point $d_j \in\mathcal{D}_J$.
By the Taylor expansion and mean-value theorem,
for $d_j \in \mathcal{D}_s$, for some $\bar d_j$ between $d_j$ and $d_{j+1}$, and for some generic constant $C$,
{\small \begin{align*}
|\tau_j| = \left|\sum_{m=1}^{\bar M-1} s^{(m)}(d_j) \mathsf{D}^m/(m! (J-1)^m) + s^{(\bar M)}(\bar d_j) \mathsf{D}^{\bar M}/({\bar M}! (J-1)^{\bar M})\right| \geq C/J^{\bar M}.
\end{align*}
}
Assuming for $n$ large enough, $2\mathsf{s}_n < CJ^{-\bar M}$, and hence $|\tau_{j}| > 2\mathsf{s}_n$, which implies correct classification, for all $j \in \{1,.., J-1\}$ and $J \geq 2$.
So for $n$ large enough, there is no misclassification
and $\hat d_{\text{AT}_J} = d_{\text{AT}_J} := \arg\min_{d\in\mathcal{D}_J} s(d)$.
Similarly in each leave-out group $\ell$,
 $\hat d_{\text{AT}_{J\ell}} = \arg\min_{d\in\mathcal{D}_J} \hat s_\ell(d) = \hat d_{\text{AT}_J} = d_{\text{AT}_J}$ for $n$ large enough.

Since $s(d)$ is continuous, as $J \rightarrow \infty$, $|d_{\text{AT}_J} - d_{\text{AT}}| \leq \mathsf{D}/(J-1)$.
So $d_{\text{AT}_J}  \rightarrow d_{\text{AT}}$ as $J \rightarrow \infty$.  \hfill$\square$

\paragraph{Proof of Lemma~\ref{Lmom}:}
%

Let $p_d(X) := \pi/s(d,X)$. $\mathbb{E}[m_{dU}(W,\xi)|X=x]$
{\small \begin{align*}
&=  \mathbb{E}[
\mathbb{E}[Y\1\{Y \geq Q^d(1-p_d(x), x)\}| S=1, D, X=x] K_h(D-d)|S=1,X=x] \frac{\p(S=1|X=x)}{\mu_d(x)}\\
&=
 \int
\mathbb{E}[Y\1\{Y \geq Q^d(1-p_d(x), x)\}| S=1, D = d+uh, X=x] k(u) f_{D|X=x, S=1}(d+uh) du \\
&\ \ \ \times\frac{\p(S=1|X=x)}{
\mu_d(x)}
\\
&=
\mathbb{E}[Y\1\{Y \geq Q^d(1-p_d(x), x)\}| S=1, D = d, X=x] \cdot f_{D|X=x, S=1}(d) \frac{\p(S=1|X=x)}{
\mu_d(x)}  + o(h)
\\&=
\mathbb{E}[Y|  Y \geq Q^d(1-p_d(x), x)\}, S=1, D = d, X=x] p_d(x)\cdot
s(d, x) + o(h)
=
\rho_{dU}(\pi, x)\cdot\pi+ o(h).
\end{align*}}
Then the aggregate sharp upper bound is
$\int_\mathcal{X}\rho_{dU}(\pi_{\text{AT}}(x),x) f_X(x|S_{d'}=1: d'\in\mathcal{D})dx
=
$\\$\int_\mathcal{X}\rho_{dU}(\pi_{\text{AT}}(x),x) \pi_{\text{AT}}(x)  f_X(x)\frac{f_X(x|\text{AT})}{\pi_{\text{AT}}(x)f_X(x)}dx
= \lim_{h \rightarrow 0}
\mathbb{E}\big[m_{dU}(W, \xi)\big]/\pi_{\text{AT}}.$


Similarly for the lower bound,
define
\begin{align*}
m_{dL}(W,\xi)&:= \frac{K_h(D-d)}{\mu_d(X)}\cdot S\cdot Y\cdot\1\{Y \leq Q^d(p_{d}(X), X)\}.
\end{align*}
For $X_i \in \mathcal{X}_j$, define the moment function to be
$m_{dL}^{j}(W,\xi) := m_{dL}(W,\xi)$ with $\pi_{\text{AT}}(X) = s(d_j, X)$.
To construct orthogonality as $h\rightarrow 0$,
$cor_{dL}^{j}(W,\xi) = Q^d(p_{dd_j}(X), X) \bigg( \frac{K_h(D - d_j)}{\mu_{d_j}(X)} (S-s(d_j, X))
-\frac{K_h(D-d)}{\mu_d(X)} p_{dd_j}(X)  (S-s(d,X))
 + \frac{K_h(D-d) S}{\mu_d(X)}\big(1\{Y > Q^d(p_{dd_j}(X),X)\} - 1 + p_{dd_j}(X)\big)\bigg)
 +(\mu_d(X) - K_h(D-d)) \E[Y|Y \leq Q^d(p_{dd_j}(X), X), D=d, S=1, X]\frac{s(d_j, X)}{\mu_d(X)}.$
Then the orthogonal moment function is $g_{dL}^{j} := m_{dL}^{j} + cor_{dL}^{j}$.
\hfill $\square$

\paragraph{Proofs of Theorem \ref{TDMLX}:}
For ease of exposition, we collect the notations below.
\paragraph{Notations.}
$\kappa:= \int_{-\infty}^\infty u^2 k(u) du$.
$\mathsf{K}_d := K_h(D-d)$, $\mathsf{s}_d := s(d,X)$, $\lambda_d := 1/\mu_d(X)$.
\\
For the upper bound, $\mathsf{Q} := Q^d(1- p_{dd_j}(X), X)$,  ${\mathsf 1} := \1\{Y \geq Q^d(1- p_{dd_j}(X), X)\}$, $\rho := \E[Y|Y \geq \mathsf{Q}, D=d, S=1, X]$.
\\
$\mathsf{B}_{dU} := \big(\mathsf{B}_{gU} - \mathsf{B}_\psi {\bar\beta}_d\big) /\pi_{\text{AT}}$,
where $\mathsf{B}_\psi:=
\E\Big[
2 \frac{\partial}{\partial d} s(d,X) \frac{\partial}{\partial d} f_{D|X}(d|X)/f_{D|X}(d|X) + \frac{\partial^2}{\partial d^2} s(d,X)
\Big]\kappa/2$,
{\small \begin{align*}
 \mathsf{B}_{gU} &:= \frac{\kappa}{2}\E\Bigg[
 \frac{\partial^2}{\partial d^2} \big(f_{D|X}(d|X) s(d,x) \E[Y|Y\geq \mathsf{Q},S=1, D=d, X]  \big(1-F_{Y|SDX}(\mathsf{Q}|1,d,X)\big) \big)
\\
&\ \ + \mathsf{Q} \lambda_{d_j} \frac{\partial^2}{\partial d_j^2}\big(f_{D|X}(d_j,X) s(d_j,X) \big)
- \mathsf{Q}\lambda_{d_j} \mathsf{s}_{d_j}  \frac{\partial^2}{\partial d_j^2}f_{D|X}(d_j|X)
\\&\ \ +(\mathsf{Q} - \rho) \lambda_{d} \mathsf{s}_{d_j}\frac{\partial^2}{\partial d^2}f_{D|X}(d,X)
- \mathsf{Q} \lambda_{d} \frac{\partial^2}{\partial d^2}\big(f_{D|X}(d|X)s(d,X)(1-F_{Y|SDX}(\mathsf{Q}|S=1,d,X))\big)
  \Bigg].
  \end{align*}
  }
$\mathsf{V}_{dU} := \big(\mathsf{V}_{gU} + \mathsf{V}_\psi \bar\beta_d^2 - 2\bar\beta_d\mathsf{V}_{gU\psi}\big)/\pi_{\text{AT}}^2$,
where $\mathsf{V}_\psi := R_k\E[\mathsf{s}_{d_j}(1-\mathsf{s}_{d_j})\lambda_{d_j}]$,
$\mathsf{V}_{gU} :=
\E\big[var(Y|Y \geq \mathsf{Q}, S=1, D=d, X) \mathsf{s}_{d_j}\lambda_d +
\big(
 \mathsf{Q}^2 \lambda_{d_j}+
 (\rho-\mathsf{Q})^2\lambda_{d}
 \big)
 \mathsf{s}_{d_j}(1-\mathsf{s}_{d_j})
\big] R_k$, and $\mathsf{V}_{gU\psi} = R_k\E[\mathsf{Q}\lambda_{d_j} \mathsf{s}_{d_j}(1-\mathsf{s}_{d_j})]$.
\\[5pt]
$\mathsf{C}_{d_1d_2U} := R_k \E\big[{Q}^{d_1}(p_{d_1d_j}(X),X){Q}^{d_2}(1-p_{d_2d_j}(X),X)\lambda_{d_j} \mathsf{s}_{d_j}(1-\mathsf{s}_{d_j})\big]$.
\\
\\
For the lower bound,
$\mathsf{Q} := Q^d(p_{dd_j}(X),X)$, $\mathsf{1}:=\1\{Y \leq Q^d(p_{dd_j}(X),X)\}$, and $\rho := \E[Y|Y \leq \mathsf{Q}, D=d, S=1, X]$.
\\
$\mathsf{B}_{dL} := \big(\mathsf{B}_{gL} - \mathsf{B}_\psi \underline{\beta_d}\big) /\pi_{\text{AT}}$,
where \begin{align*}
 \mathsf{B}_{gL} &:= \frac{\kappa}{2}\E\Bigg[
 \frac{\partial^2}{\partial d^2} \big(f_{D|X}(d|X) s(d,x) \E[Y|Y\leq \mathsf{Q},S=1, D=d, X]  F_{Y|SDX}(\mathsf{Q}|1,d,X) \big)
\\
&\ \ + \mathsf{Q} \lambda_{d_j} \frac{\partial^2}{\partial d_j^2}\big(f_{D|X}(d_j,X) s(d_j,X) \big)
- \mathsf{Q}\lambda_{d_j} \mathsf{s}_{d_j}  \frac{\partial^2}{\partial d_j^2}f_{D|X}(d_j|X)
\\&\ \ +(\mathsf{Q} - \rho) \lambda_{d} \mathsf{s}_{d_j}\frac{\partial^2}{\partial d^2}f_{D|X}(d,X)
- \mathsf{Q} \lambda_{d} \frac{\partial^2}{\partial d^2}\big(f_{D|X}(d|X)s(d,X)F_{Y|SDX}(\mathsf{Q}|S=1,d,X)\big)
  \Bigg].
  \end{align*}
  $\mathsf{V}_{dL} := \big(\mathsf{V}_{gL} + \mathsf{V}_\psi \underline{\beta_d}^2 - 2\underline{\beta_d}\mathsf{V}_{gL\psi}\big)/\pi_{\text{AT}}^2$,
where $\mathsf{V}_{gL} :=
\E\big[var(Y|Y \leq \mathsf{Q}, S=1, D=d, X) \mathsf{s}_{d_j}\lambda_d +
\big(
 \mathsf{Q}^2 \lambda_{d_j}+
 (\rho-\mathsf{Q})^2\lambda_{d}
 \big)
 \mathsf{s}_{d_j}(1-\mathsf{s}_{d_j})
\big] R_k$,
and $\mathsf{V}_{gL\psi} = R_k\E[\mathsf{Q}\lambda_{d_j} \mathsf{s}_{d_j}(1-\mathsf{s}_{d_j})]$.
\\[5pt]
$\mathsf{C}_{d_1d_2L} := R_k \E\big[{Q}^{d_1}(1-p_{d_1d_j}(X),X){Q}^{d_2}(p_{d_2d_j}(X),X)\lambda_{d_j} \mathsf{s}_{d_j}(1-\mathsf{s}_{d_j})\big]$.

\bigskip

We can write $\hat{\bar\beta}_d = \hat A/\hat B$, where $A = \E[g_{dU}(W, \xi)]$ and $B = \pi_{\text{AT}}$.
By linearization, $\hat A/\hat B = A/B + (\hat A - A)/B - (\hat B-B) A/B^2 + O_\p\big(|(\hat A-A)(\hat B-B) + (\hat B-B)^2|\big)$.

By Theorem 3.1 in \cite{CL},
$\hat\pi_{\text{AT}}= n^{-1}\sum_{i=1}^n \psi(W_i, \hat\xi_\ell) = n^{-1}\sum_{i=1}^n \psi(W_i, \xi) + o_\p(1/\sqrt{nh})$,
the bias $\E[\hat\pi_{\text{AT}}] - \pi_{\text{AT}} = h^2\mathsf{B}_\psi + o_\p(h^2)$, and
$\sqrt{nh}\big(\hat\pi_{\text{AT}} - \pi_{\text{AT}} - h^2\mathsf{B}_\psi\big)\stackrel{d}{\rightarrow} \mathcal{N}(0,\mathsf{V}_\psi)$

The proof focuses on deriving the asymptotically linear representation of the numerator $\hat A$: $n^{-1}\sum_{i=1}^n g_{dU}(W_i, \hat\xi_\ell) = n^{-1}\sum_{i=1}^n g_{dU}(W_i, \xi) + o_\p(1/\sqrt{nh})$.
By linearization, we obtain the asymptotically linear representation $\hat{\overline{\beta}}_d - \overline{\beta}_d =
 n^{-1}\sum_{i=1}^n
 \phi(W_i, \xi) - \overline{\beta}_d  + o_\p(1/\sqrt{nh})$ and
the bias  $\E[\hat{\bar\beta}_d] - {\bar\beta}_d  = h^2\mathsf{B}_d+ o_\p(h^2)$.
Then we will show $\sqrt{nh}\big(\hat{\bar\beta}_d - \bar\beta_d - h^2\mathsf{B}_d\big)\stackrel{d}{\rightarrow} \mathcal{N}(0,\mathsf{V}_d)$.
We first show the special case under Assumption~\ref{ACmin} with $J=1$, i.e., everyone has the same known unique sufficient treatment value $d_{\text{AT}}$.
The results for $J \geq 2$ are implied if there is no classification error.
Finally we show that the classification error is asymptotically ignorable.

The proof follows the arguments in the proof of Theorem 3.1 in \cite{CL} for the DML estimator of $\beta_d$ under unconfoundedness without selection.
The new challenges for $\bar\beta_d$ arise due to more nuisance functions $\xi$ that
enter the orthogonal moment function nonlinearly and complicate the notations.
We re-define the nuisance functions so that the orthogonal moment function is linear in each of the re-defined nuisance functions.

The nuisance function estimator $\hat\xi_\ell$ uses observations $\mathcal{W}_\ell^c$.
We will show $L^{-1}\sum_{\ell=1}^L \E_{n_\ell} g(W,\hat\xi_\ell) = \E_{n} g(W,\xi) + o_\p(1/\sqrt{nh})$ by showing $\E_{n_\ell} g(W,\hat\xi_\ell) - \E_{n_\ell} g(W,\xi) = o_\p(1/\sqrt{nh})$.
Below we suppress the subscript $\ell$ in $\hat\xi_\ell$ to simplify notation as $\hat\xi$.
We focus on the upper bound as the same arguments apply to the lower bound.

The orthogonal moment function $g_{dU}^{j} := m_{dU}^{j} + cor_{dU}^{j}$ is denoted as
\begin{align*}
g(W, \xi)
&=\mathsf{K}_d \lambda_d SY\mathsf{1} + \mathsf{Q} \bigg( \mathsf{K}_{d_j} \lambda_{d_j} (S-\mathsf{s}_{d_j})
+ \mathsf{K}_d \lambda_{d} \big(\mathsf{s}_{d_j}  - S\mathsf{1}\big)\bigg)
 +(1- \mathsf{K}_d \lambda_d) \rho \mathsf{s}_{d_j}, 
\end{align*}
where the nuisance parameter $\xi = (\mathsf{s}_{d_j}, \mathsf{Q}, \lambda_d, \lambda_{d_j}, \rho, \mathsf{1})$.

Let $\xi_{-\iota} = \xi \setminus \xi_\iota$ for $\iota =1,..,6$.
As $g(W, \xi)$ is linear in each element in $\xi$, we can write $g(W, \hat\xi_\iota, \xi_{-\iota}) - g(W, \xi_\iota, \xi_{-\iota}) =  \Delta\hat\xi_\iota \cdot g_{-\iota}(\xi_{-\iota})$.
And $g_{-\iota}(\xi_{-\iota})$ is also linear in each of the remaining elements in $\xi_{-\iota}$.

For example, as $\xi_1 = \mathsf{s}_{d_j}$ and $\xi_{-1} = (\mathsf{Q}, \lambda_d, \lambda_{d_j}, \rho, \mathsf{1})$,
we can write $g(W,  \hat\xi_1, \xi_{-1}) - g(W,  \xi_1, \xi_{-1}) = \Delta\hat{\mathsf{s}}_{d_j} \cdot g_{-1}(\xi_{-1})$, where
$g_{-1}(\xi_{-1}) = \mathsf{Q}\big(-\mathsf{K}_{d_j} \lambda_{d_j} + \mathsf{K}_{d} \lambda_{d}\big) + (1- \mathsf{K}_{d} \lambda_{d}) \rho.$

As $\xi_2 = \mathsf{Q}$ and $\xi_{-12} = \xi \setminus (\xi_1, \xi_2) = \xi_{-1}\setminus\xi_2 = (\lambda_d, \lambda_{d_j}, \rho, \mathsf{1})$, and $g_{-12} (\xi_{-12})= -\mathsf{K}_{d_j} \lambda_{d_j} + \mathsf{K}_{d} \lambda_{d}$,
we can write $g_{-1}(\hat\xi_{2}, \xi_{-12})-g_{-1}(\xi_2, \xi_{-12}) = \Delta\hat\xi_{2} \cdot g_{-12}(\xi_{-12}) = \Delta\hat{\mathsf{Q}}\big(-\mathsf{K}_{d_j} \lambda_{d_j} + \mathsf{K}_{d} \lambda_{d}\big)$.

Further as $\xi_3 = \lambda_d$ and $g_{-123}(\xi_{-123}) = \mathsf{K}_{d}$, we can write $g_{-12} (\hat\xi_3, \xi_{-123}) - g_{-12} (\xi_3, \xi_{-123}) = \Delta\hat\xi_3 \cdot g_{-123}(\xi_{-123}) = \Delta{\lambda}_d \mathsf{K}_{d}$.

We decompose the remainder term $\E_{n_\ell} \big[g(W, \hat \xi) - g(W, \xi)\big]$ for $\ell = 1,..,L$ to the following and control each term to be $o_\p(1/\sqrt{nh})$:
\begin{align}
\E_{n_\ell} \big[g(W, \hat \xi) - g(W, \xi)\big]= \sum_{\iota = 1}^6\E_{n_\ell} \big[g(W, \hat \xi_\iota, \xi_{-\iota}) - g(W, \xi)\big]- \E_\ell \big[g(W, \hat \xi_\iota, \xi_{-\iota}) - g(W, \xi)\big] \tag{SE}
\label{ESE}\\
+ \E_\ell \big[g(W, \hat \xi_\iota, \xi_{-\iota}) - g(W, \xi)\big] \tag{DR} \label{EDR}\\
+ O_\p\left(\sum_{\iota \neq j \in\{1,...,6\}}\|\hat \xi_\iota - \xi_\iota\|_2\|\hat \xi_j - \xi_j\|_2\right) + ...
\tag{Rem} \label{ERem}
\end{align}

\paragraph{(DR)}
We show the sufficient conditions for (\ref{EDR})$=o_p(1/\sqrt{nh})$ are
$\sqrt{nh}h^2 \E[|\Delta\hat\xi_\iota|] = o_\p(1)$ for $\iota = 1, 2, 3, 4, 5$ and $\sqrt{nh}\|\hat{\mathsf{Q}} - \mathsf{Q}\|_2^2 = o_\p(1)$. 

For $\iota = 1,...,5$, $\xi_\iota$ is a function of $X$.
We can show $\E[g_{-\iota}(\xi_{\iota})|X] = O_\p(h^2)$ by Assumption~\ref{ADMLX}.
This is the key benefit of the doubly robust moment function.
Then by the law of iterated expectations,
$\big| \E_\ell\big[g(W, \hat \xi_\iota, \xi_{-\iota}) - g(W, \xi)\big] \big| =
\big| \E_\ell \big[\Delta\hat\xi_\iota \cdot g_{-\iota}(\xi_{-\iota}) \big] \big|
\leq
\E_\ell\big[\big|\Delta\hat\xi_\iota\big| \cdot \big|\E[g_{-\iota}(\xi_{-\iota})|X] \big| \big]
= O_\p(\E_\ell\big[\big|\Delta\hat\xi_\iota\big|\big] h^2).$
So it suffices to assume $\sqrt{nh}h^2 \E[|\Delta\hat\xi_\iota|] = o_\p(1)$.

For example,
by a standard algebra for kernel, we can show that $\E[\mathsf{K}_{d}|X] = \mu_d(X) + O_\p(h^2)$.
For  $\xi_1 =   \mathsf{s}_{d_j}$, $
\big| \E_\ell \big[g(W, \hat \xi_1, \xi_{-1}) - g(W, \xi)\big] \big| \leq
\E_\ell\big[\big|\hat{\mathsf{s}}_{d_j} - \mathsf{s}_{d_j}\big| \cdot \big|\E[ \mathsf{Q} (-\mathsf{K}_{d_j}\lambda_{d_j} + \mathsf{K}_{d}\lambda_{d}) + (1- \mathsf{K}_{d}\lambda_{d})\rho|X]\big|\big]
= O_\p\big(\E_\ell\big[\big|\Delta\hat{\mathsf{s}}_{d_j}\big|\big] h^2\big)$.

The same argument applies to $\xi_\iota = \mathsf{Q},  \lambda_d, \lambda_{d_j}, \rho$.
We can verify $\E[g_{-\iota}(\xi_{-\iota})|X] = O_\p(h^2)$ by Assumption~\ref{ADMLX}(ii), where $g_{-2}(\xi_{-2}) = \mathsf{K}_{d_j} \lambda_{d_j} (S-\mathsf{s}_{d_j})
+ \mathsf{K}_d \lambda_{d} \big(\mathsf{s}_{d_j}  - S\mathsf{1}\big)$,
$g_{-3}(\xi_{-3}) =
\mathsf{K}_d SY\mathsf{1} + \mathsf{Q}  \mathsf{K}_d  \big(\mathsf{s}_{d_j}  - S\mathsf{1}\big)
 - \mathsf{K}_d \rho \mathsf{s}_{d_j}$,
 $g_{-4}(\xi_{-4}) =  \mathsf{Q} \mathsf{K}_{d_j} (S-\mathsf{s}_{d_j})$, $g_{-5}(\xi_{-5}) = (1- \mathsf{K}_d \lambda_d) \mathsf{s}_{d_j}$.

Now consider (\ref{EDR}) with $\xi_6 = \mathsf{1}$.
The following second equality uses $\int^{\infty}_{\hat{\mathsf{Q}}} (y - \mathsf{Q}) f_{Y|SDX} dy
= \int^{\infty}_{{\mathsf{Q}}} (y - \mathsf{Q}) f_{Y|SDX} dy - \frac{1}{2}\mathsf{Q} f_{Y|SDX}(\mathsf{Q})(\hat{\mathsf{Q}} - \mathsf{Q})^2 + o_\p((\Delta \hat{\mathsf{Q}})^2)$, by Taylor expansion, Leibniz rule, and uniformly bounded $f'_{Y|SDX}$.
So  $\E_\ell\big[\Delta\hat\xi_6 \cdot g_{-6}(\xi_{-6})\big] =\E_\ell\big[(\hat{\mathsf{1}} - \mathsf{1})\cdot
\mathsf{K}_d\lambda_d S(Y - \mathsf{Q})
\big]
$
\begin{align*}
&=
\E_\ell\Big[\mathsf{K}_d\lambda_d S\cdot
\Big(
\int^{\infty}_{\hat{\mathsf{Q}}} (y - \mathsf{Q}) f_{Y|SDX} dy -
\int^{\infty}_{{\mathsf{Q}}} (y - \mathsf{Q}) f_{Y|SDX} dy
\Big)\Big]
\\&=
\E_\ell\Big[ \big(s(d,X)+ O_\p(h^2)\big)\cdot
\Big(
- \frac{1}{2}\mathsf{Q} f_{Y|SDX}(\mathsf{Q})(\hat{\mathsf{Q}} - \mathsf{Q})^2 + o_\p((\Delta \hat{\mathsf{Q}})^2)
\Big)
\Big]
\\
&= O_\p(\|\Delta\hat{\mathsf{Q}}\|_2^2).
\end{align*}
So it suffices to assume $\sqrt{nh}\|\hat{\mathsf{Q}} - \mathsf{Q}\|_2^2 = o_\p(1)$.

\paragraph{(SE)}
We show that the sufficient condition  for (\ref{ESE})$=o_p(1/\sqrt{nh})$ is $\|\Delta\hat\xi_\iota\|_2 = o_\p(1)$ for $\iota = 1,...,5$,
due to cross-fitting.

Define $\Delta_{i\ell} := g(W_i, \hat \xi_\iota, \xi_{-\iota}) - g(W_i, \xi) - \E_\ell\big[g(W_i, \hat \xi_\iota, \xi_{-\iota}) - g(W_i, \xi)\big]
= \Delta\xi_\iota g_{-\iota}(\xi_{-\iota}) - \E_\ell\big[\Delta\xi_\iota g_{-\iota}(\xi_{-\iota})\big]$.
By construction and independence of $\mathcal{W}_\ell^c$ and $W_i$ for $i\in I_\ell$, $\E_\ell[\Delta_{i\ell}] =0$ and $\E_\ell[\Delta_{i\ell} \Delta_{j\ell}] = 0$ for $i, j\in I_\ell$.
If $\E_\ell[(\sqrt{h/n_\ell} \sum_{i\in I_\ell}\Delta_{i\ell} )^2] = h \E_\ell[\Delta_{i\ell}^2] = o_p(1)$, then the conditional Markov's inequality implies that $\sqrt{h/n_\ell} \sum_{i\in I_\ell}\Delta_{i\ell} = o_\p(1)$.
So it suffices to show that $h \E_\ell[\Delta_{i\ell}^2] = o_p(1)$.

Because $\E_\ell\big[\Delta\xi_\iota g_{-\iota}(\xi_{-\iota})\big]=o_\p(1/\sqrt{nh})$ as shown above for (\ref{EDR}), $h \E_\ell[\Delta_{i\ell}^2] = O_\p\Big(
h \E_\ell\big[
(\Delta\xi_\iota)^2$\\
$\times g_{-\iota}(\xi_{-\iota})^2
\big]
\Big) = O_\p\Big(
h \E_\ell\big[
(\Delta\xi_\iota)^2 \E\big[g_{-\iota}(\xi_{-\iota})^2|X\big]
\big]
\Big) = O_\p\Big(\E_\ell\big[(\Delta\xi_\iota)^2 \big]\Big)$,
by showing $h\E\big[g_{-\iota}(\xi_{-\iota}^2)|X\big] = O_\p(1)$.

Specifically $h \E[\mathsf{K}_{d}^2|X] = \int h^{-1}k((D-d)/h)^2 f_{D|X}(D|X)dD = \int k(u)^2 f_{D|X}(d+uh|X) du
= R_k f_{D|X}(d|X) + o_\p(1) = O_\p(1)$.
For example of $\xi_1 = \mathsf{s}_{d_j}$,
$h \E_\ell[\Delta_{i\ell}^2] = O_\p\Big( h \E_\ell[(\Delta \hat{\mathsf{s}}_{d_j})^2 \cdot \E[(\mathsf{Q} (-\mathsf{K}_{d_j}\lambda_{d_j} + \mathsf{K}_{d}\lambda_{d}) + (1- \mathsf{K}_{d}\lambda_{d})\rho)^2|X]]\Big)
= O_\p(\|\Delta \hat{\mathsf{s}}_{d_j}\|_2^2).$
The same argument applies to $\xi_\iota = \mathsf{Q}, \rho, \lambda_d, \lambda_{d_j}$.

For $\xi_6 = \mathsf{1}$,
it suffices to show that
$h\E_\ell\big[(\hat{\mathsf{1}} - \mathsf{1})^2\cdot
\mathsf{K}_d^2\lambda_d^2 S(Y - \mathsf{Q})^2
\big] = hE_\ell\big[
\mathsf{K}_d^2\lambda_d^2 S\cdot
\E[\1\{\mathsf{Q} < Y \leq \hat{\mathsf{Q}}\}\cdot
(Y - \mathsf{Q})^2
|S, D, X]\big]= o_\p(1)$, considering $\hat{\mathsf{Q}} > \mathsf{Q}$ without loss of generality.
By integration by parts, $\E[\1\{\mathsf{Q} < Y \leq \hat{\mathsf{Q}}\}\cdot
(Y - \mathsf{Q})^2|S, D, X] =
\int_{\mathsf{Q}}^{\hat{\mathsf{Q}}} (y - \mathsf{Q})^2 f_{Y|SDX}(y) dy =
(\hat{\mathsf{Q}} - \mathsf{Q})^2 F_{Y|SDX}(\hat{\mathsf{Q}} ) -
2\int_{\mathsf{Q}}^{\hat{\mathsf{Q}}} (y - \mathsf{Q}) F_{Y|SDX}(y) dy = O_\p(|\Delta\hat{\mathsf{Q}}|^2) = o_\p(1)$,
because the last term
$\int_{\mathsf{Q}}^{\hat{\mathsf{Q}}} (y - \mathsf{Q}) F_{Y|SDX}(y) dy =
\frac{1}{2}(\hat{\mathsf{Q}} - \mathsf{Q})^2 F_{Y|SDX}(\bar{\mathsf{Q}}) = O_\p(|\Delta\hat{\mathsf{Q}}|^2)$
for some $\bar{\mathsf{Q}}$ between $\mathsf{Q}$ and $\hat{\mathsf{Q}}$, by Taylor expansion and Leibniz rule.
By the same algebra of kernel as above, we can show $hE_\ell[\mathsf{K}_d^2\lambda_d^2 S] = O_\p(1)$.
Then we obtain the result.

\paragraph{(Rem)}
We show that the sufficient condition for (\ref{ERem})$=o_\p(1/\sqrt{nh})$ is $\|\Delta\hat\xi_\iota\|_2\|\Delta\hat\xi_j\|_2 = o_\p(1/\sqrt{nh})$ for $\iota \neq j$.

Consider a simplified case when there are three elements in $\xi = (\xi_1,\xi_2,\xi_3)$.
The same arguments apply to the general result with more complicated notations.
We suppress $W$ in the function $g(W,\xi) = g(\xi_1,\xi_2,\xi_3)$ when there is no confusion.
We show the remainder terms in (\ref{ERem})
{\small \begin{align}
&g(\hat\xi) - g(\xi) - \big(g(\hat \xi_1, \xi_{-1}) - g(\xi)\big) -\big(g(\hat \xi_2, \xi_{-2}) - g(\xi) \big) - \big(g(\hat \xi_3, \xi_{-3}) - g(\xi) \big)
\notag\\
&-g(\xi_1, \hat\xi_2, \hat \xi_3) + g(\xi_1, \hat\xi_2, \hat \xi_3)
\notag\\
&=
\Delta\hat\xi_1\Delta\hat\xi_2 g_{-12}(\xi_3)
+ \Delta\hat\xi_1\Delta\hat\xi_3g_{-13}(\xi_2) + \Delta\hat\xi_2 \Delta\hat\xi_3 g_{-23}(\xi_1)  + \Delta\hat\xi_1\Delta\hat\xi_2 \Delta\hat\xi_3 g_{-123}, \mbox{where}
\label{ERem1} \\
&g(\hat\xi) - g(\xi_1, \hat\xi_2, \hat\xi_3)
 - \big(g(\hat \xi_1, \xi_{-1}) - g(\xi)\big)
= \Delta\hat\xi_1g_{-1}(\hat\xi_2,\hat\xi_3) - \Delta\hat\xi_1g_{-1}(\xi_2,\xi_3)
+  \Delta\hat\xi_1g_{-1}(\xi_2,\hat\xi_3) -  \Delta\hat\xi_1g_{-1}(\xi_2,\hat\xi_3)
\notag\\
&=  \Delta\hat\xi_1 \Delta\hat\xi_2
g_{-12}(\hat\xi_3) + \Delta\hat\xi_1 \Delta\hat\xi_3 g_{-13}(\xi_2)
+ \Delta\hat\xi_1 \Delta\hat\xi_2 g_{-12}(\xi_3)
- \Delta\hat\xi_1 \Delta\hat\xi_2 g_{-12}(\xi_3)
\notag\\
&= \Delta\hat\xi_1\Delta\hat\xi_3g_{-13}(\xi_2) + \Delta\hat\xi_1\Delta\hat\xi_2 g_{-12}(\xi_3)
+ \Delta\hat\xi_1\Delta\hat\xi_2\big(g_{-12}(\hat\xi_3) - g_{-12}(\xi_3)\big) \notag
\\
&= \Delta\hat\xi_1\Delta\hat\xi_3g_{-13}(\xi_2) + \Delta\hat\xi_1\Delta\hat\xi_2 g_{-12}(\xi_3)
+ \Delta\hat\xi_1\Delta\hat\xi_2 \Delta\hat\xi_3 g_{-123},  \mbox{ and}  \notag \\
& -\big(g(\hat \xi_2, \xi_{-2}) - g(\xi) \big) - g(\hat \xi_3, \xi_{-3}) + g(\xi_1, \hat\xi_2, \hat \xi_3)
 = -\Delta\hat\xi_2 g_{-2}(\xi_1,\xi_3) +\Delta\hat\xi_2 g_{-2}(\xi_1, \hat\xi_3)
= \Delta\hat\xi_2 \Delta\hat\xi_3 g_{-23}(\xi_1). \notag
 \end{align}
}
We start with the term $\Delta\hat\xi_1\Delta\hat\xi_2 g_{-12}(\xi_{-12})$ in (\ref{ERem1}).
We show $\E_\ell\left[\left| \sqrt{{h}/{n_\ell}} \sum_{i\in I_\ell} \Delta\hat\xi_1\Delta\hat\xi_2 g_{-12}(\xi_{-12})\right|\right] = o_\p(1)$
by focusing on the second term in $g_{-12} (\xi_{-12})= -\mathsf{K}_{d_j} \lambda_{d_j} + \mathsf{K}_{d} \lambda_{d}$ below, as the same argument applies to the first term.
So (\ref{ERem}) is $o_\p(1/\sqrt{nh})$ follows by the conditional Markov's inequality and triangle inequalities.
{\footnotesize
\begin{align*}
&\E_\ell\left[\left| \sqrt{{h}/{n_\ell}} \sum_{i\in I_\ell} \Delta\hat\xi_1\Delta\hat\xi_2 \mathsf{K}_{d} \lambda_{d}\right|\right]
\leq
\sqrt{n_\ell h} \int_\mathcal{X}\int_\mathcal{D}\left|  \Delta\hat{\mathsf{s}}_{d_j}\right|\left|\Delta\hat{\mathsf{Q}}\right| K_d(D-d) \lambda_{d}f_{DX}(D,X)dDdX \\
&\leq
\sqrt{n_\ell h} \left( \int_\mathcal{X}\int_\mathcal{D}\left(\Delta\hat{\mathsf{s}}_{d_j}\right)^2K_d(D-d) \lambda_{d}f_{DX}(D,X)dDdX\right)^{1/2}
\left( \int_\mathcal{X}\int_\mathcal{D}\left(\Delta\hat{\mathsf{Q}}\right)^2 K_d(D-d) \lambda_{d}f_{DX}(D,X)dDdX \right)^{1/2}\\
&= O_\p\left(\sqrt{n_\ell h} \left( \int_\mathcal{X}\left(\Delta\hat{\mathsf{s}}_{d_j}\right)^2 f_{X}(X)dX\right)^{1/2}\left( \int_\mathcal{X}\left(\Delta\hat{\mathsf{Q}}\right)^2 f_{X}(X)dX \right)^{1/2}\right) = o_\p(1)
\end{align*}
}
by Cauchy-Schwarz inequality and Assumption~\ref{ADMLX}.
The same argument applies to other terms in (\ref{ERem1}).
Specifically, $g_{-13}(\xi_{-13}) = \mathsf{K}_{d} (\mathsf{Q} - \rho)$, $g_{-14}(\xi_{-14}) = -\mathsf{K}_{d_j}\mathsf{Q}$, $g_{-15}(\xi_{-15}) = 1-\mathsf{K}_{d} \lambda_{d}$, $g_{-16}(\xi_{-16})=0$.
By the law of iterated expectations and assuming $\E\big[|g_{-1\iota}(W,\xi_{-1\iota})|\big|X\big]$ is uniformly bounded, for $\iota = 3,4,5$, $\E_\ell\left[\left| \sqrt{{h}/{n_\ell}} \sum_{i\in I_\ell} \Delta\hat\xi_1\Delta\hat\xi_\iota g_{-1\iota}(W,\xi_{-1\iota}) \right|\right]
\leq
\sqrt{n_\ell h} \E\Big[\big| \Delta\hat\xi_1\big|\big|\Delta\hat\xi_\iota\big|
\E\big[|g_{-1\iota}(W,\xi_{-1\iota})|\big|X\big]\Big]
\leq
\sqrt{n_\ell h} \left( \int_\mathcal{X}\big( \Delta\hat\xi_1\big)^2 \E\big[|g_{-1\iota}(W,\xi_{-1\iota})|\big|X\big] f(X)dX\right)^{1/2}
\left( \int_\mathcal{X} \big( \Delta\hat\xi_\iota\big)^2 \E\big[|g_{-1\iota}(W,\xi_{-1\iota})|\big|X\big]f_{X}(X)dX \right)^{1/2}
= O_\p\left(\sqrt{n_\ell h} \left( \int_\mathcal{X}\big( \Delta\hat\xi_1\big)^2 f_{X}(X)dX\right)^{1/2}\left( \int_\mathcal{X}\big( \Delta\hat\xi_\iota\big)^2 f_{X}(X)dX \right)^{1/2}\right) = o_\p(1).
$

The same argument applies to $\xi_\iota$, $\iota = 2,3,4,5$.
Specifically we can show that $g_{-23} = \mathsf{K}_{d} (\mathsf{s}_{d_j} - S\mathsf{1})$, $g_{-24} = \mathsf{K}_{d_j} (S-\mathsf{s}_{d_j})$, $g_{-25} = 0$,
$g_{-34} = 0$, $g_{-35} = -\mathsf{K}_{d_j}\mathsf{s}_{d_j}$,
$g_{-45} = 0$.

Now consider $\xi_6 = \mathsf{1} = \1\{Y\geq \mathsf{Q}\}$.
Then $g_{-6}(\xi_{-6}) = \mathsf{K}_{d} \lambda_{d}S(Y-\mathsf{Q})$, $g_{-26}(\xi_{-26}) = -\mathsf{K}_{d} \lambda_{d}S$,
$g_{-36}(\xi_{-36}) = \mathsf{K}_{d} S(Y-\mathsf{Q})$,
and $g_{-\iota 6} = 0$ for $\iota = 1, 4, 5$.

We first compute $\E_\ell\left[\left(\Delta\hat{\mathsf{1}}\right)^2\Big|S, D, X\right] =
\E_\ell\left[\left(\1\{Y\geq \hat{\mathsf{Q}}\} - \1\{Y\geq \mathsf{Q}\}\right)^2\Big|S, D, X\right]
= \big|F_{Y|SDX}(\hat{\mathsf{Q}}) - F_{Y|SDX}({\mathsf{Q}})\big|
= f_{Y|SDX}(\mathsf{Q})\big|\hat{\mathsf{Q}}-\mathsf{Q}\big|+ O_\p((\Delta\hat{\mathsf{Q}})^2)$.
For $\Delta\hat\xi_2\Delta\hat\xi_6 g_{-26}(\xi_{-26})$ in (\ref{ERem1}),
{\small \begin{align*}
&\E_\ell\left[\left| \sqrt{{h}/{n_\ell}} \sum_{i\in I_\ell} \Delta\hat{\mathsf{Q}} \Delta\hat{\mathsf{1}} \mathsf{K}_{d} \lambda_{d}S\right|\right]
\leq
\sqrt{n_\ell h} \E_\ell\left[\left|   \Delta\hat{\mathsf{Q}} \right|\left|\Delta\hat{\mathsf{1}}\right| K_d(D-d) \lambda_{d}S \right]\\
&\leq
\sqrt{n_\ell h} \left( \int_\mathcal{X}\int_\mathcal{D}\left(\Delta\hat{\mathsf{Q}}\right)^2K_d(D-d) \lambda_{d}s(D,X)f_{DX}(D,X)dDdX\right)^{1/2}
\\ \ \ \
&\times\left( \int_\mathcal{X}\int_\mathcal{D}\E_\ell\left[\left(\Delta\hat{\mathsf{1}}\right)^2|S=1,D,X\right] K_d(D-d) \lambda_{d}s(d,X)f_{DX}(D,X)dDdX \right)^{1/2}
\\
&= O_\p\left(\sqrt{n_\ell h} \left( \int_\mathcal{X}\left(\Delta\hat{\mathsf{Q}}\right)^2 f_{X}(X)dX\right)^{1/2}\left( \int_\mathcal{X}\left|\Delta\hat{\mathsf{Q}}\right| f_{X}(X)dX \right)^{1/2}\right),
\end{align*}
}
which is $o_\p(1)$ by assuming $\sqrt{nh}\|\Delta\hat{\mathsf{Q}}\|_2^2 = o_\p(1)$.

Now we turn to $\Delta\hat\xi_3\Delta\hat\xi_6 g_{-36}(\xi_{-36})$ in (\ref{ERem1}).
We first compute $\E_\ell\left[\left(\Delta\hat{\mathsf{1}}\right)^2 |Y-\mathsf{Q}| |S=1, D,X\right]
= \E_\ell\left[\left(\1\{\mathsf{Q} > Y \geq \hat{\mathsf{Q}}\} + \1\{\mathsf{Q} \leq Y < \hat{\mathsf{Q}}\}\right)
|Y-\mathsf{Q}||S=1, D,X\right]
=
\1\{\mathsf{Q} > \hat{\mathsf{Q}}\}\int_{\hat{\mathsf{Q}}}^{{\mathsf{Q}}} (\mathsf{Q}-Y) f_{Y|SDX}(y)dy
+
\1\{\mathsf{Q} < \hat{\mathsf{Q}}\} \int^{\hat{\mathsf{Q}}}_{{\mathsf{Q}}} (Y - \mathsf{Q}) f_{Y|SDX}(y)dy =
\frac{1}{2} f_{Y|SDX}(\mathsf{Q}) (\Delta\hat{\mathsf{Q}})^2 + o_\p((\Delta\hat{\mathsf{Q}})^2 ).
$
Then{\small
\begin{align*}
&\E_\ell\left[\left| \sqrt{{h}/{n_\ell}} \sum_{i\in I_\ell} \Delta\hat{\lambda}_d \Delta\hat{\mathsf{1}}  \mathsf{K}_{d} S(Y-\mathsf{Q}) \right|\right]
\leq
\sqrt{n_\ell h} \E_\ell\left[\left|   \Delta\hat{\lambda}_d \right|\left|\Delta\hat{\mathsf{1}}\right|  \mathsf{K}_{d} S|Y-\mathsf{Q}| \right]\\
&\leq
\sqrt{n_\ell h} \left( \int_\mathcal{X}\int_\mathcal{D}\left(\Delta\hat{\lambda}_d \right)^2
\mathsf{K}_{d} s(D,X)\E[|Y-\mathsf{Q}||S=1,D,X]
f_{DX}(D,X)dDdX\right)^{1/2}\\
&\ \ \ \times\left( \int_\mathcal{X}\int_\mathcal{D}\E_\ell\left[\left(\Delta\hat{\mathsf{1}}\right)^2
|Y-\mathsf{Q}| \Big|S=1, D,X\right]
 \mathsf{K}_{d} s(D,X) f_{DX}(D,X)dDdX \right)^{1/2}\\
&=
O_\p\bigg(\sqrt{n_\ell h} \Big( \int_\mathcal{X}\left(\Delta\hat{\lambda}_d \right)^2s(d,X)\E[|Y-\mathsf{Q}||S=1,D=d,X]
f_{DX}(d,X)dX\Big)^{1/2} \\
&\ \ \ \times\Big( \frac{1}{2}\int_\mathcal{X}\int_\mathcal{D}
 f_{Y|SDX}(\mathsf{Q}) (\Delta\hat{\mathsf{Q}})^2
 s(d,X) f_{DX}(d,X)dX \Big)^{1/2}\bigg)
+ o_\p(h^2),
\end{align*}
}
which is $o_\p(1)$ by assuming $\sqrt{nh}\|\Delta\hat\lambda_d\|_2\|\Delta\hat{\mathsf{Q}}\|_2 = o_\p(1)$ and
$\E[|Y-\mathsf{Q}||S=1,D=d,X]$
and $f_{Y|SDX}$ are uniformly bounded.


In sum, to control the reminder term $\E_{n_\ell}\big[g(W, \hat \xi) - g(W, \xi)\big] = o_\p(1/\sqrt{nh})$, the sufficient rate conditions are $\|\Delta\hat\xi_\iota\|_2\|\Delta\hat\xi_j\|_2 = o_\p(1/\sqrt{nh})$ for $\iota \neq j$, $\|\Delta\hat\xi_\iota\|_2= o_\p(1)$ for $\iota=1,2,3,4,5$, and $\sqrt{nh}\|\hat{\mathsf{Q}} - \mathsf{Q}\|_2^2 = o_\p(1)$.

Assuming $\sqrt{nh}h^2 \rightarrow C \in [0, \infty)$ and
$\|\Delta\hat \xi_\iota\|_2 = o_p(1)$ yields $\sqrt{nh} h^2\E_\ell\big[\big|\Delta\hat{\mathsf{s}}_{d_j}\big|\big]= o_p(1)$.

We derive the sufficient rate conditions in Assumption~\ref{ADMLX}(iii)(iv) based on $\|\Delta\hat\xi\|_2$.
For $\hat{\mathsf{Q}} = \hat Q^d(1-\hat p_{dd_j}(X), X)$,
$\Delta\hat{\mathsf{Q}} =
\hat Q^d(1-\hat p_{dd_j}(X), X)
-  Q^d(1-\hat p_{dd_j}(X), X)
+ Q^d(1-\hat p_{dd_j}(X), X)
+ Q^d(1- p_{dd_j}(X), X)
= O_\p\Big(\sup_{p\in(0,1)}\Big|\hat Q^d(p,X) - Q^d(p,X)\Big|\Big)  + O_\p\big(|\hat s(d,X)- s(d,X)| + |\hat s(d_j,X)- s(d_j,X)|\big)$.

For $\hat\rho = \hat{\E}[Y|Y\geq \hat{\mathsf{Q}}, S=1, D=d, X]$, $\|\Delta\hat\rho\|_2 \leq  \|\hat{\E}[Y|Y\geq \hat{\mathsf{Q}}, S=1, D=d, X] -  {\E}[Y|Y\geq \hat{\mathsf{Q}}, S=1, D=d, X] \|_2 +  \|{\E}[Y|Y\geq \hat{\mathsf{Q}}, S=1, D=d, X] -  {\E}[Y|Y\geq {\mathsf{Q}}, S=1, D=d, X]\|_2
\leq \sup_{y\in \mathcal{Y}_0}\|\Delta\hat{\E}[Y|Y\geq y, S=1, D=d, X]\|_2 + O_\p(\sup_{p\in(0,1)}\| \Delta\hat{\mathsf{Q}} \|_2)
$.

\paragraph{Bias}
We use the law of iterated expectations,
 the dominated convergence theorem, and standard algebra of kernel in the following.
{\small \begin{align*}
&\E\bigg[ \mathsf{K}_d \lambda_d S\E[Y\mathsf{1}|S, D, X] + \mathsf{Q} \Big( \mathsf{K}_{d_j} \lambda_{d_j} (s(D,X)-\mathsf{s}_{d_j})
+ \mathsf{K}_d \lambda_{d} \big(\mathsf{s}_{d_j}  - S(1-F_{Y|SDX}(\mathsf{Q}|S,D,X))\big)\Big)
\\&\ \  +(1- \E[\mathsf{K}_d|X] \lambda_d) \rho \mathsf{s}_{d_j} \bigg]
\\&=
\E\bigg[ \mathsf{K}_d \lambda_d s(D,X)\E[Y|Y\geq \mathsf{Q},S=1, D, X] \big(1-F_{Y|SDX}(\mathsf{Q}|1,D,X)\big) + \mathsf{Q} \bigg( \lambda_{d_j} \Big\{ f_{D|X}(d_j,X) s(d_j,X) \\
&\ \ + \frac{h^2}{2}\kappa  \frac{\partial^2}{\partial d_j^2}\big(f_{D|X}(d_j,X) s(d_j,X) \big)
- f_{D|X}(d_j|X) \mathsf{s}_{d_j} -\frac{h^2}{2}\kappa  \mathsf{s}_{d_j}  \frac{\partial^2}{\partial d_j^2}f_{D|X}(d_j|X)
\Big\}
\\
&\ \ +  \lambda_{d} \Big\{f_{D|X}(d|X)\mathsf{s}_{d_j} + \frac{h^2}{2}\kappa  \mathsf{s}_{d_j}\frac{\partial^2}{\partial d^2}f_{D|X}(d,X)  - \mathsf{K}_d s(D,X)(1-F_{Y|SDX}(\mathsf{Q}|S=1,D,X))\Big\}\bigg)
\\
&\ \  +\big(1- f_{D|X}(d|X)\lambda_d - \frac{h^2}{2}\kappa   \lambda_d  \frac{\partial^2}{\partial d^2}f_{D|X}(d,X) \big) \rho \mathsf{s}_{d_j} \bigg]
+ o_\p(h^2) \\
&=
\E\bigg[\rho s(d_j,X) + \frac{h^2}{2}\kappa  \frac{\partial^2}{\partial d^2} \Big(f_{D|X}(d|X) s(d,x) \E[Y|Y\geq \mathsf{Q},S=1, D=d, X]  \big(1-F_{Y|SDX}(\mathsf{Q}|1,d,X)\big) \Big)
\\
&\ \  + \mathsf{Q} \bigg(  \mathsf{s}_{d_j}+ \frac{h^2}{2}\kappa  \lambda_{d_j} \frac{\partial^2}{\partial d_j^2}\big(f_{D|X}(d_j,X) s(d_j,X) \big)
- \mathsf{s}_{d_j} -\frac{h^2}{2}\kappa  \lambda_{d_j} \mathsf{s}_{d_j}  \frac{\partial^2}{\partial d_j^2}f_{D|X}(d_j|X)
+  \mathsf{s}_{d_j} \\
&\ \ + \frac{h^2}{2}\kappa  \lambda_{d} \mathsf{s}_{d_j}\frac{\partial^2}{\partial d^2}f_{D|X}(d,X)  -  \mathsf{s}_{d_j}
- \frac{h^2}{2}\kappa  \lambda_{d} \frac{\partial^2}{\partial d^2}\Big(f_{D|X}(d|X)s(d,X)\big(1-F_{Y|SDX}(\mathsf{Q}|S=1,d,X)\big)
\Big)\bigg)
\\
&\ \  - \frac{h^2}{2}\kappa   \lambda_d  \rho \mathsf{s}_{d_j}  \frac{\partial^2}{\partial d^2}f_{D|X}(d,X) \bigg]
 + o_\p(h^2) \\
 &= \E\big[\rho \mathsf{s}_{d_j}\big] + h^2 \mathsf{B}_{gU} + o_\p(h^2).
  \end{align*}
  }
The same arguments apply to $\mathsf{B}_{gL}$.

\paragraph{Variance.}
Note that $\E[S|D=d_j, X] = s(d_j, X) = \mathsf{s}_{d_j}$ and hense $var(S| D=d_j, X) = \mathsf{s}_{d_j}(1-\mathsf{s}_{d_j})$.
Theorem~3.1 in \cite{CL} gives $hvar(\psi(W_i,\xi)) \rightarrow R_k\E[ \E[(S-\mathsf{s}_{d_j})^2| D=d_j, X] \lambda_{d_j}] = R_k\E[ var(S| D=d_j, X) \lambda_{d_j}] = \mathsf{V}_\psi$.

First compute $h\E[\mathsf{K}_{d_j}^2(S - \mathsf{s}_{d_j})^2|X] =
h\int_{\mathcal{D}} h^{-2}k((v-d_j)/h)^2 \E[(S - \mathsf{s}_{d_j})^2|D = v,X] f_{D|X}(v) dv
=
\int k(u)^2 \E[(S - \mathsf{s}_{d_j})^2|D = d_j + hu,X] f_{D|X}(d_j+uh) du
=
R_k var(S|D=d_j, X) f_{D|X}(d_j) + o(1) = R_k \mathsf{s}_{d_j}(1-\mathsf{s}_{d_j}) \lambda_{d_j}^{-1} + o(1)$.

In $h\E[g(W,\xi) \psi(W_i,\xi)]$,
the product term with $\mathsf{K}_{d_j}\mathsf{K}_d$ results in a  convolution kernel $\int k((d-d_j)/h + u) k(u) du$ and hence is $o(1)$.
So $h\E[g(W,\xi) \psi(W_i,\xi)]= h\E[\mathsf{Q} \mathsf{K}_{d_j}^2\lambda_{d_j}^2(S - \mathsf{s}_{d_j})^2] + o(1)
\rightarrow \mathsf{V}_{g\psi}$.

Next we compute $h var(g(W_i, \xi)) = h \E[(g(W, \xi) -  \E[\rho \mathsf{s}_{d_j}])^2] + o(1)$ with
{\small \begin{align}
g(W, \xi) -  \E[\rho \mathsf{s}_{d_j}]
=\mathsf{K}_d \lambda_d (SY\mathsf{1}-\rho \mathsf{s}_{d_j}) + \mathsf{Q} \mathsf{K}_{d_j} \lambda_{d_j} (S-\mathsf{s}_{d_j})
+ \mathsf{Q}\mathsf{K}_d \lambda_{d} \big(\mathsf{s}_{d_j}  - S\mathsf{1}\big)
 + \big(\rho \mathsf{s}_{d_j} - \E[\rho \mathsf{s}_{d_j}]\big), 
\label{Eg}
\end{align}
}


For the first term in (\ref{Eg}),
$h \E[\mathsf{K}_d^2\lambda_d^2(SY\mathsf{1} - \rho \mathsf{s}_{d_j})^2]
=
h\E[\E[ \mathsf{K}_d^2(SY^2\mathsf{1} -2 SY\mathsf{1}\rho \mathsf{s}_{d_j} + \rho^2\mathsf{s}_{d_j}^2)|X]\lambda_d^2]
$
{\small \begin{align*}
&=
h\E[\E[ \mathsf{K}_d^2(
\E[Y^2|Y\geq \mathsf{Q}, D, X](1-F_{Y|S,D,X}(\mathsf{Q}|1,D,X)) s(D,X)
\\
&- 2 \E[Y|Y\geq \mathsf{Q}, D, X](1-F_{Y|S,D,X}(\mathsf{Q}|1,D,X)) s(D,X)\rho \mathsf{s}_{d_j}
+ \rho^2\mathsf{s}_{d_j}^2)|X]\lambda_d^2]
\\&=R_k\E[
\E[Y^2|Y\geq \mathsf{Q}, D=d, X] \mathsf{s}_{d_j}
- 2 \rho^2\mathsf{s}_{d_j}^2
+ \rho^2\mathsf{s}_{d_j}^2) f_{D|X}(d,X)\lambda_d^2] + o(1)
\\&=
R_k\E[
\E[Y^2|Y\geq \mathsf{Q}, D=d, X] \mathsf{s}_{d_j}
- \rho^2\mathsf{s}_{d_j}^2
+ \rho^2\mathsf{s}_{d_j} -\rho^2\mathsf{s}_{d_j}) \lambda_d] + o(1)
\\&=
R_k\E[(
var(Y^2|Y\geq \mathsf{Q}, D=d, X] \mathsf{s}_{d_j}
+ \rho^2\mathsf{s}_{d_j}(1- \mathsf{s}_{d_j}) )\lambda_d] + o(1).
\end{align*}
}
Similar arguments apply to the second term in (\ref{Eg}), so $h\E[\mathsf{Q}^2\mathsf{K}_{d_j}^2\lambda_{d_j}^2 (S-\mathsf{s}_{d_j})^2]
=R_k\E[\mathsf{Q}^2\lambda_{d_j}\mathsf{s}_{d_j}(1- \mathsf{s}_{d_j})] + o(1)$.
And for the third term,
$h\E[\mathsf{Q}^2\mathsf{K}_{d}^2\lambda_d^2 (\mathsf{s}_{d_j} - S\mathsf{1})^2]
= R_k\E[\mathsf{Q}^2\lambda_d \mathsf{s}_{d_j}(1- \mathsf{s}_{d_j})] + o(1)$.
 For the last term, $h\E[(\rho \mathsf{s}_{d_j} - \E[\rho \mathsf{s}_{d_j}])^2] = O(h) = o(1)$.

 For the product of the first and third terms, $h\E[
 \mathsf{Q}\mathsf{K}_d^2\lambda_d^2(SY\mathsf{1} - \rho \mathsf{s}_{d_j})
 (\mathsf{s}_{d_j} - S\mathsf{1})
 ] =
 -R_k\E[
 \mathsf{Q}\lambda_d \rho \mathsf{s}_{d_j}(1- \mathsf{s}_{d_j})] + o(1)$.
The other cross products are $o(1)$ by the law of iterated expectations.
Therefore, we obtain $h var(g(W_i, \xi)) = \mathsf{V}_g + o(1)$.

\paragraph{Asymptotic normality.}
Asymptotic normality follows from the Lyapunov central limit theorem with the third absolute moment.
%
Let $s_n^2 := \sum_{i=1}^n var\big(\sqrt{nh}n^{-1}\phi(W_i,\xi)\big) = hvar(\phi) = \mathsf{V}_d + o(1)$ as shown above.
If $\E\big[\big|\sqrt{nh}n^{-1}\phi(W, \xi)\big|^3\big] = O((n^3h)^{-1/2})$, then the Lyapunov condition holds: $\sum_{i=1}^n \E\big[\big|\sqrt{nh}n^{-1} \phi(W_i, \xi)\big|^3\big]/s_n^3 = O((nh)^{-1/2}) = o(1)$.
That is, it suffices to show that $\E[|\phi|^3] = O(h^{-2})$, which holds by
assuming that $\E[|Y|^3{\bf 1}|D=d, S=1,X]$ is continuous in $d$
uniformly over $\mathcal{X}$ and other assumed conditions.

\paragraph{Misclassification.}
Following the proof of Theorem~\ref{TEstLB},
let the true $ \tau_{j}(x) =  s(d_{j+1}, x) -  s(d_j, x)$ with the estimate $\hat \tau_{j}(x) = \hat s(d_{j+1}, x) - \hat s(d_j, x)$.
Let
$\mathcal{M}_{j} = \{x: \hat \tau_{j}(x) < 0 < \tau_{j}(x)\} \cup \{x: \hat \tau_{j}(x) > 0 > \tau_{j}(x)\}$.
As we have discussed in the proof of Theorem~\ref{TEstLB}, there is no mis-classified problem when $\tau_j(x) = 0$.  We define the correctly-classified set  $\mathcal{C}_{j} = \{x: \hat \tau_{j}(x) \times \tau_{j}(x) > 0 \} \cup \{x: \tau_j(x) = 0\}$.

Let $\1_{\mathcal{M}_J} = \1\{X \in \cup_{j \in\{1,..,J-1\}} \mathcal{M}_{j}\}$ for  the mis-classified set and
$\1_{\mathcal{C}_J} = \1\{X \in \cap_{j \in\{1,..,J-1\}}\mathcal{C}_{j}\}$ for the correctly classified set.
So $\1_{\mathcal{M}_J} + \1_{\mathcal{C}_J} =1$.
We have shown that $\E_n\big[ g(W, \hat \xi) \1_{\mathcal{C}_J}\big] - \E_n\big[ g(W,\xi)\big] = o_\p(1/\sqrt{nh})$.
The goal is to show $\E_n\big[ g(W,\hat\xi) \1_{\mathcal{M}_J}\big] = o_\p(1/\sqrt{nh})$.

Because $|\hat \tau_{j}(x) - \tau_{j}(x)| \leq |\hat s(d_{j+1}, x) - s(d_{j+1}, x)| + |\hat s(d_j, x) - s(d_j, x)| \leq 2\mathsf{s}_n$, $\mathcal{M}_{j} \subseteq \mathcal{M}_{j}^1:= \{x: 0 < |\tau_{j}(x)| \leq 2\mathsf{s}_n\}$.
Thus it suffices to show  $\E_n\big[ g(W,\hat\xi) \1_{\mathcal{M}^1_{j}}\big] = o_\p(1/\sqrt{nh})$ for all $j \in\{1,..,J-1\}$.

Let $\mathsf{D} = \overline{\mathcal{D}} - \underline{\mathcal{D}}$, so $d_{j+1} - d_j = \mathsf{D}/(J-1)$.
By the Taylor expansion and mean-value theorem,
for $d_j \in \mathcal{D}_{sx}$, for some $\bar d_j$ between $d_j$ and $d_{j+1}$, and for some generic constant $C$,
\begin{align*}
|\tau_j(x)| = \left|\sum_{m=1}^{\bar M_x-1} s^{(m)}(d_j,x) \mathsf{D}^m/(m! (J-1)^m) + s^{(\bar M_x)}(\bar d_j,x) \mathsf{D}^{\bar M_x}/({\bar M_x}! (J-1)^{\bar M_x})\right| \geq C/J^{\bar M}.
\end{align*}
So $\inf_{j\in\{1,..., J-1\}, x\in\mathcal{X}}
|\tau_j(x)|  \geq CJ^{-\bar M}$.  Assuming $2\mathsf{s}_n < CJ^{-\bar M}$ for $n$ large enough, $\1_{\mathcal{M}_j^1} = 0$.

The above calculation of variance $h\E[g(W,\hat\xi_\ell)^2|X, \mathcal{W}_\ell^c] = O_\p(1)$ uniformly in $X$.
By the condtional Markov inequality, for any $\epsilon>0$ and  for $n$ large enough,
$\p\big(\big|\sqrt{nh}\E_{n\ell}[g(W, \hat\xi_\ell)\1_{\mathcal{M}_{j}^1}]\big| > \epsilon\big|\mathcal{W}_\ell^c\big) <
 \E\big[h g(W, \hat\xi_\ell)^2 \1_{\mathcal{M}_{j}^1}\big|\mathcal{W}_\ell^c\big]/\epsilon^2
\leq  \E\big[h \E[g(W, \hat \xi_\ell)^2|X, \mathcal{W}_\ell^c]\1_{\mathcal{M}_{j}^1}\big|\mathcal{W}_\ell^c\big]/\epsilon^2
=0$.
\hfill$\square$

\paragraph{Acknowledgements}
We thank Adriana Lleras-Muney for providing the Civilian Conservation Corps (CCC) data.
We thank Vira Semenova, David McKenzie, and participants in the seminars in Academia Sinica, Emory University, UCI, UCSD, UCLA,  2024 California Econometrics Conference, and 2025 Econometric Society World Congress for helpful comments.

\singlespacing

\bibliographystyle{chicagoa}	
\bibliography{database}	

\begin{thebibliography}{}

\bibitem[\protect\citeauthoryear{Ahn and Powell}{Ahn and
  Powell}{1993}]{AhnPowell}
Ahn, H. and J.~L. Powell (1993).
\newblock Semiparametric estimation of censored selection models with a
  nonparametric selection mechanism.
\newblock {\em Journal of Econometrics\/}~{\em 58\/}(1), 3--29.


\bibitem[\protect\citeauthoryear{Aizer, Early, Eli, Imbens, Lee, Lleras-Muney,
  and Strand}{Aizer et~al.}{2024}]{Aizer}
Aizer, A., N.~Early, S.~Eli, G.~Imbens, K.~Lee, A.~Lleras-Muney, and A.~Strand
  (2024).
\newblock The lifetime impacts of the new deal's youth employment program.
\newblock {\em Quarterly Journal of Economics\/}~{\em 139\/}(4), 2579--2635.


\bibitem[\protect\citeauthoryear{Angrist, Imbens, and Rubin}{Angrist
  et~al.}{1996}]{AIR}
Angrist, J., G.~Imbens, and D.~Rubin (1996).
\newblock Identification of causal effects using instrumental variables.
\newblock {\em Journal of the American Statistical Association\/}~{\em 91},
  444–455.


\bibitem[\protect\citeauthoryear{Ao, Calonico, and Lee}{Ao et~al.}{2021}]{ACL}
Ao, W., S.~Calonico, and Y.-Y. Lee (2021).
\newblock Multivalued treatments and decomposition analysis: An application to
  the {WIA} program.
\newblock {\em Journal of Business \& Economic Statistics\/}~{\em 39},
  358--371.


\bibitem[\protect\citeauthoryear{Behaghel, Crépon, Gurgand, and
  Le~Barbanchon}{Behaghel et~al.}{2015}]{BCGL}
Behaghel, L., B.~Crépon, M.~Gurgand, and T.~Le~Barbanchon (2015).
\newblock Please call again: Correcting nonresponse bias in treatment effect
  models.
\newblock {\em The Review of Economics and Statistics\/}~{\em 97\/}(5),
  1070--1080.


\bibitem[\protect\citeauthoryear{Belloni, Chernozhukov, Fern\'{a}ndez-Val, and
  Hansen}{Belloni et~al.}{2017}]{BCFH17}
Belloni, A., V.~Chernozhukov, I.~Fern\'{a}ndez-Val, and C.~Hansen (2017).
\newblock Program evaluation and causal inference with high-dimensional data.
\newblock {\em Econometrica\/}~{\em 85\/}(1), 233--298.


\bibitem[\protect\citeauthoryear{Cesarini, Lindqvist, Notowidigdo, and
  \"Ostling}{Cesarini et~al.}{2017}]{swedish}
Cesarini, D., E.~Lindqvist, M.~J. Notowidigdo, and R.~\"Ostling (2017).
\newblock The effect of wealth on individual and household labor supply:
  Evidence from {Swedish} lotteries.
\newblock {\em American Economic Review\/}~{\em 107\/}(12), 3917--46.


\bibitem[\protect\citeauthoryear{Chen and Roth}{Chen and Roth}{2023}]{ChenRoth}
Chen, J. and J.~Roth (2023).
\newblock Logs with zeros? some problems and solutions.
\newblock {\em The Quarterly Journal of Economics\/}~{\em 139\/}(2), 891--936.


\bibitem[\protect\citeauthoryear{Chernozhukov, Chetverikov, Demirer, Duflo,
  Hansen, Newey, and Robins}{Chernozhukov et~al.}{2018}]{CCDDHNR}
Chernozhukov, V., D.~Chetverikov, M.~Demirer, E.~Duflo, C.~Hansen, W.~Newey,
  and J.~Robins (2018).
\newblock Double/debiased machine learning for treatment and structural
  parameters.
\newblock {\em The Econometrics Journal\/}~{\em 21\/}(1), C1--C68.


\bibitem[\protect\citeauthoryear{Chernozhukov, Newey, and Singh}{Chernozhukov
  et~al.}{2022a}]{CNS21ADML}
Chernozhukov, V., W.~Newey, and R.~Singh (2022a).
\newblock Automatic debiased machine learning of causal and structural effects.
\newblock {\em Econometrica\/}~{\em 90\/}(3), 967--1027.


\bibitem[\protect\citeauthoryear{Chernozhukov, Newey, and Singh}{Chernozhukov
  et~al.}{2022b}]{CNS21EJ}
Chernozhukov, V., W.~K. Newey, and R.~Singh (2022b).
\newblock {Debiased machine learning of global and local parameters using
  regularized {Riesz} representers}.
\newblock {\em The Econometrics Journal\/}~{\em 25\/}(3), 576--601.


\bibitem[\protect\citeauthoryear{Colangelo and Lee}{Colangelo and
  Lee}{2025}]{CL}
Colangelo, K. and Y.-Y. Lee (2025).
\newblock Double debiased machine learning nonparametric inference with
  continuous treatments.
\newblock {\em Journal of Business \& Economic Statistics\/}, 1--13.


\bibitem[\protect\citeauthoryear{Das, Newey, and Vella}{Das
  et~al.}{2003}]{DNV03}
Das, M., W.~K. Newey, and F.~Vella (2003).
\newblock Nonparametric estimation of sample selection models.
\newblock {\em Review of Economic Studies\/}~{\em 70\/}(1), 33--58.


\bibitem[\protect\citeauthoryear{DiNardo, Matsudaira, McCrary, and
  Sanbonmatsu}{DiNardo et~al.}{2021}]{DiNardo}
DiNardo, J., J.~Matsudaira, J.~McCrary, and L.~Sanbonmatsu (2021).
\newblock A practical proactive proposal for dealing with attrition:
  Alternative approaches and an empirical example.
\newblock {\em Journal of Labor Economics\/}~{\em 39\/}(S2).


\bibitem[\protect\citeauthoryear{Donald, Hsu, and Barrett}{Donald
  et~al.}{2012}]{DHB}
Donald, S.~G., Y.-C. Hsu, and G.~F. Barrett (2012).
\newblock Incorporating covariates in the measurement of welfare and
  inequality: methods and applications.
\newblock {\em The Econometrics Journal\/}~{\em 15\/}(1), C1--C30.


\bibitem[\protect\citeauthoryear{D’Amour, Ding, Feller, Lei, and
  Sekhon}{D’Amour et~al.}{2021}]{DAmour}
D’Amour, A., P.~Ding, A.~Feller, L.~Lei, and J.~Sekhon (2021).
\newblock Overlap in observational studies with high-dimensional covariates.
\newblock {\em Journal of Econometrics\/}~{\em 221\/}(2), 644--654.


\bibitem[\protect\citeauthoryear{Escanciano, Jacho-Ch\'{a}vez, and
  Lewbel}{Escanciano et~al.}{2016}]{EJL16}
Escanciano, J.~C., D.~T. Jacho-Ch\'{a}vez, and A.~Lewbel (2016).
\newblock Identification and estimation of semiparametric two-step models.
\newblock {\em Quantitative Economics\/}~{\em 7\/}(2), 561--589.


\bibitem[\protect\citeauthoryear{Estrada}{Estrada}{2024}]{Estrada}
Estrada, P. (2024).
\newblock Spillover effects with nonrandom sample selection.
\newblock Working paper, {Emory University}.

\bibitem[\protect\citeauthoryear{Fan and Park}{Fan and Park}{2010}]{FanPark}
Fan, Y. and S.~S. Park (2010).
\newblock Sharp bounds on the distribution of treatment effects and their
  statistical inference.
\newblock {\em Econometric Theory\/}~{\em 26\/}(3), 931–51.


\bibitem[\protect\citeauthoryear{Farrell}{Farrell}{2015}]{Farrell15}
Farrell, M.~H. (2015).
\newblock Robust inference on average treatment effects with possibly more
  covariates than observations.
\newblock {\em Journal of Econometrics\/}~{\em 189\/}(1), 1--23.


\bibitem[\protect\citeauthoryear{Flores, Flores-Lagunes, Gonzalez, and
  Neumann}{Flores et~al.}{2012}]{FFGN12ReStat}
Flores, C.~A., A.~Flores-Lagunes, A.~Gonzalez, and T.~C. Neumann (2012).
\newblock Estimating the effects of length of exposure to instruction in a
  training program: The case of {Job Corps}.
\newblock {\em The Review of Economics and Statistics\/}~{\em 94\/}(1),
  153--171.


\bibitem[\protect\citeauthoryear{Garlick and Hyman}{Garlick and
  Hyman}{2022}]{Garlick}
Garlick, R. and J.~Hyman (2022).
\newblock Quasi-experimental evaluation of alternative sample selection
  corrections.
\newblock {\em Journal of Business \& Economic Statistics\/}~{\em 40\/}(3),
  950--964.


\bibitem[\protect\citeauthoryear{Gerard, Rokkanen, and Rothe}{Gerard
  et~al.}{2020}]{GRR}
Gerard, F., M.~Rokkanen, and C.~Rothe (2020).
\newblock Bounds on treatment effects in regression discontinuity designs with
  a manipulated running variable.
\newblock {\em Quantitative Economics\/}~{\em 11\/}(3), 839--870.


\bibitem[\protect\citeauthoryear{Hansen}{Hansen}{2022a}]{HansenBook}
Hansen, B.~E. (2022a).
\newblock {\em Econometrics}.
\newblock Princeton University Press.


\bibitem[\protect\citeauthoryear{Hansen}{Hansen}{2022b}]{HansenProb}
Hansen, B.~E. (2022b).
\newblock {\em Probability and Statistics for Economists}.
\newblock Princeton University Press.


\bibitem[\protect\citeauthoryear{Heckman}{Heckman}{1976}]{RePEc:nbr:nberch:10491}
Heckman, J. (1976).
\newblock The common structure of statistical models of truncation, sample
  selection and limited dependent variables and a simple estimator for such
  models.
\newblock In {\em Annals of Economic and Social Measurement, Volume 5, number
  4}, pp.\  475--492. National Bureau of Economic Research, Inc.

\bibitem[\protect\citeauthoryear{Heckman, Smith, and Clements}{Heckman
  et~al.}{1997}]{HSC}
Heckman, J., J.~Smith, and N.~Clements (1997).
\newblock Making the most out of program evaluations and social experiments:
  accounting for heterogeneity in program impacts.
\newblock {\em Review of Economic Studies\/}~{\em 64}, 487--535.


\bibitem[\protect\citeauthoryear{Heckman}{Heckman}{1979}]{Heckman}
Heckman, J.~J. (1979).
\newblock Sample selection bias as a specification error.
\newblock {\em Econometrica\/}~{\em 47\/}(1), 153--161.


\bibitem[\protect\citeauthoryear{Heiler}{Heiler}{2024}]{HEILER24}
Heiler, P. (2024).
\newblock Heterogeneous treatment effect bounds under sample selection with an
  application to the effects of social media on political polarization.
\newblock {\em Journal of Econometrics\/}~{\em 244\/}(1), 105856.


\bibitem[\protect\citeauthoryear{Ho and Rosen}{Ho and Rosen}{2017}]{HoRosen}
Ho, K. and A.~M. Rosen (2017).
\newblock {\em Partial Identification in Applied Research: Benefits and
  Challenges}, pp.\  307–359.
\newblock Econometric Society Monographs. Cambridge University Press.

\bibitem[\protect\citeauthoryear{Honoré and Hu}{Honoré and
  Hu}{2020}]{HonoreHu}
Honoré, B.~E. and L.~Hu (2020).
\newblock Selection without exclusion.
\newblock {\em Econometrica\/}~{\em 88\/}(3), 1007--1029.


\bibitem[\protect\citeauthoryear{Honoré and Hu}{Honoré and
  Hu}{2022}]{HonoreHuJoE}
Honoré, B.~E. and L.~Hu (2022).
\newblock Sample selection models without exclusion restrictions: Parameter
  heterogeneity and partial identification.
\newblock {\em Journal of Econometrics\/}, 105360.


\bibitem[\protect\citeauthoryear{Horowitz and Manski}{Horowitz and
  Manski}{1995}]{HM95}
Horowitz, J.~L. and C.~F. Manski (1995).
\newblock Identification and robustness with contaminated and corrupted data.
\newblock {\em Econometrica\/}~{\em 63}, 281--302.


\bibitem[\protect\citeauthoryear{Hsu, Huber, Lee, and Lettry}{Hsu
  et~al.}{2020}]{HHLP}
Hsu, Y.-C., M.~Huber, Y.-Y. Lee, and L.~Lettry (2020).
\newblock Direct and indirect effects of continuous treatments based on
  generalized propensity score weighting.
\newblock {\em Journal of Applied Econometrics\/}~{\em 35\/}(7), 814--840.


\bibitem[\protect\citeauthoryear{Hsu, Huber, Lee, and Liu}{Hsu
  et~al.}{2023}]{HHLL}
Hsu, Y.-C., M.~Huber, Y.-Y. Lee, and C.-A. Liu (2023).
\newblock Testing monotonicity of mean potential outcomes in a continuous
  treatment with high-dimensional data.
\newblock {\em The Review of Economics and Statistics\/}, forthcoming.


\bibitem[\protect\citeauthoryear{Hsu, Huber, Liu, and Lee}{Hsu
  et~al.}{2023}]{HHLLdata}
Hsu, Y.-C., M.~Huber, C.-A. Liu, and Y.-Y. Lee (2023).
\newblock {Replication data for: Testing Monotonicity of Mean Potential
  Outcomes in a Continuous Treatment with High-Dimensional Data}.

\bibitem[\protect\citeauthoryear{Imbens}{Imbens}{2004}]{Imbens04REStat}
Imbens, G. (2004).
\newblock Nonparametric estimation of average treatment effects under
  exogeneity: A review.
\newblock {\em Review of Economics and Statistics\/}~{\em 86\/}(1), 4–29.


\bibitem[\protect\citeauthoryear{Imbens and Angrist}{Imbens and
  Angrist}{1994}]{ImbensAngrist}
Imbens, G.~W. and J.~D. Angrist (1994).
\newblock Identification and estimation of local average treatment effects.
\newblock {\em Econometrica\/}~{\em 62\/}(2), 467--475.


\bibitem[\protect\citeauthoryear{Imbens and Newey}{Imbens and
  Newey}{2009}]{IN09ETA}
Imbens, G.~W. and W.~K. Newey (2009).
\newblock Identification and estimation of triangular simultaneous equations
  models without additivity.
\newblock {\em Econometrica\/}~{\em 77\/}(5), 1481--1512.


\bibitem[\protect\citeauthoryear{Kennedy, Ma, McHugh, and Small}{Kennedy
  et~al.}{2017}]{Kennedy}
Kennedy, E.~H., Z.~Ma, M.~D. McHugh, and D.~S. Small (2017).
\newblock Nonparametric methods for doubly robust estimation of continuous
  treatment effects.
\newblock {\em Journal of the Royal Statistical Society: Series B\/}~{\em
  79\/}(4), 1229--1245.


\bibitem[\protect\citeauthoryear{Kline and Tamer}{Kline and
  Tamer}{2023}]{KlineTamer}
Kline, B. and E.~Tamer (2023).
\newblock Recent developments in partial identification.
\newblock {\em Annual Review of Economics\/}~{\em 15}, 125--150.


\bibitem[\protect\citeauthoryear{Kline and Santos}{Kline and
  Santos}{2013}]{KlineSantos}
Kline, P. and A.~Santos (2013).
\newblock Sensitivity to missing data assumptions: Theory and an evaluation of
  the {U.S.} wage structure.
\newblock {\em Quantitative Economics\/}~{\em 4\/}(2), 231--267.


\bibitem[\protect\citeauthoryear{Kroft, Mourifi\'{e}, and Vayalinkal}{Kroft
  et~al.}{2024}]{KMV}
Kroft, K., I.~Mourifi\'{e}, and A.~Vayalinkal (2024).
\newblock Lee bounds with multilayered sample selection.
\newblock {NBER} working paper 32952.

\bibitem[\protect\citeauthoryear{Lee}{Lee}{2009}]{LeeBound}
Lee, D. (2009).
\newblock Training, wages, and sample selection: Estimating sharp bounds on
  treatment effects.
\newblock {\em Review of Economic Studies\/}~{\em 76\/}(3), 1071--1102.


\bibitem[\protect\citeauthoryear{Lee}{Lee}{2015}]{Lee15}
Lee, Y.-Y. (2015).
\newblock Partial mean processes with generated regressors: Continuous
  treatment effects and nonseparable models.
\newblock Working paper.

\bibitem[\protect\citeauthoryear{Molinari}{Molinari}{2020}]{MolinariReview}
Molinari, F. (2020).
\newblock Microeconometrics with partial identification.
\newblock In S.~N. Durlauf, L.~P. Hansen, J.~J. Heckman, and R.~L. Matzkin
  (Eds.), {\em Handbook of Econometrics, Volume 7A}, pp.\  355--486. Elsevier.

\bibitem[\protect\citeauthoryear{Olma}{Olma}{2021}]{Olma}
Olma, T. (2021).
\newblock Nonparametric estimation of truncated conditional expectation
  functions.
\newblock arxiv:2109.06150.

\bibitem[\protect\citeauthoryear{Powell, Stock, and Stoker}{Powell
  et~al.}{1989}]{PSS89ETA}
Powell, J.~L., J.~H. Stock, and T.~M. Stoker (1989).
\newblock Semiparametric estimation of index coefficients.
\newblock {\em Econometrica\/}~{\em 57\/}(6), 1403--30.


\bibitem[\protect\citeauthoryear{Powell and Stoker}{Powell and
  Stoker}{1996}]{PS96}
Powell, J.~L. and T.~M. Stoker (1996).
\newblock Optimal bandwidth choice for density-weighted averages.
\newblock {\em Journal of Econometrics\/}~{\em 75\/}(2), 291--316.


\bibitem[\protect\citeauthoryear{Rubin}{Rubin}{1976}]{Rubin}
Rubin, D.~B. (1976).
\newblock Inference and missing data.
\newblock {\em Biometrika\/}~{\em 63}, 581--592.


\bibitem[\protect\citeauthoryear{Schochet, Burghardt, and McConnell}{Schochet
  et~al.}{2008}]{SBM}
Schochet, P.~Z., J.~Burghardt, and S.~McConnell (2008).
\newblock Does {Job Corps} work? impact findings from the national {Job Corps}
  study.
\newblock {\em American Economic Review\/}~{\em 98\/}(5).


\bibitem[\protect\citeauthoryear{Semenova}{Semenova}{2024}]{SGLee}
Semenova, V. (2024).
\newblock Generalized {Lee} bounds.
\newblock {\em Journal of Econometrics\/}, forthcoming.


\bibitem[\protect\citeauthoryear{Su, Ura, and Zhang}{Su et~al.}{2019}]{SUZ}
Su, L., T.~Ura, and Y.~Zhang (2019).
\newblock Non-separable models with high-dimensional data.
\newblock {\em Journal of Econometrics\/}~{\em 212\/}(2), 646--677.


\bibitem[\protect\citeauthoryear{Velez}{Velez}{2025}]{Velez}
Velez, A. (2025).
\newblock On the asymptotic properties of debiased machine learning estimators.
\newblock Working paper.

\bibitem[\protect\citeauthoryear{Vytlacil}{Vytlacil}{2002}]{Vytlacil02}
Vytlacil, E. (2002).
\newblock Independence, monotonicity, and latent index models: An equivalence
  result.
\newblock {\em Econometrica\/}~{\em 70\/}(1), 331--341.


\bibitem[\protect\citeauthoryear{Zhang and Rubin}{Zhang and Rubin}{2003}]{ZR03}
Zhang, J.~L. and D.~B. Rubin (2003).
\newblock Estimation of causal effects via principal stratification when some
  outcomes are truncated by “death”.
\newblock {\em Journal of Educational and Behavioral Statistics\/}~{\em
  28\/}(4), 353--368.


\bibitem[\protect\citeauthoryear{Zhang, Rubin, and Mealli}{Zhang
  et~al.}{2009}]{ZRM}
Zhang, J.~L., D.~B. Rubin, and F.~Mealli (2009).
\newblock Likelihood-based analysis of causal effects of job-training programs
  using principal stratification.
\newblock {\em Journal of the American Statistical Association\/}~{\em
  104\/}(485), 166–176.


\end{thebibliography}

\newpage

\renewcommand{\theequation}{S.\arabic{equation}}
\setcounter{equation}{0} \setcounter{page}{1}

\begin{center}
{\Large Online Supplementary Appendix for \\[0pt]
Lee Bounds with a Continuous Treatment in Sample Selection}\\[5pt]
Ying-Ying Lee$^{\dagger }$ \let\thefootnote\relax%
\footnotetext{$^{\dagger }$Department of economics, University of California Irvine, Irvine, CA 92697, U.S.A.  \\
E-mail:\ \href{yingying.lee@uci.edu}{yingying.lee@uci.edu}.
\href{https://sites.google.com/site/yyleelilian}{https://sites.google.com/site/yyleelilian}.
Tel: +1 9498244834. Fax: +1 9498242492.
}
\hspace{1cm}
Chu-An Liu$^{\ddagger }$ \let\thefootnote\relax%
\footnotetext{$^{\ddagger }$Institute of Economics, Academia Sinica, Taipei City 115, Taiwan. \\
E-mail:\ \href{caliu@econ.sinica.edu.tw}{caliu@econ.sinica.edu.tw}.
\href{https://chuanliu.weebly.com/}{https://chuanliu.weebly.com/}.
}
\end{center}

\medskip

Section~\ref{AsecClaim} presents the proofs of Claim-Step1, 2, 3 in Proof of Theorem~\ref{TEstLB} and Corollaries~\ref{CATE} and ~\ref{CATEX}.
Section~\ref{ASecEM} presents the details of the first-step Lasso estimation in Section~\ref{Sec1stEst}
and supplementary material for the empirical applications.

\section{Additional proofs}
\label{AsecClaim}

\subsection{Proofs of Claim-Step1, 2, 3 in Proof of Theorem~\ref{TEstLB}:}
\paragraph{Proof of Claim-Step1:}
(i) The kernel regression estimator
$\hat s_\ell(d) = {\E_{n_\ell}\left[SK_h(D-d)\right]}/{\hat f_{D\ell}(d)}$, where the denominator
$\hat f_{D\ell}(d) = \E_{n_\ell}\left[K_h(D-d)\right]$,
is well studied.
For example, Theorems 19.1 and 19.2 \cite{HansenBook} provide that
$\hat s_\ell(d) - s(d) = O_{\p}(1/\sqrt{nh} + h^2)$.
By a linearization as in~(\ref{Elinear}),
\begin{align*}
\hat s_\ell(d)
&= \frac{\E_{n_\ell}\left[SK_h(D-d)\right] - s(d)f_D(d)}{f_D(d)} - \frac{s(d)}{f_D(d)}\left(\hat f_{D\ell}(d) - f_D(d)\right)\\
&\ \ \ + O_{\p}\left(\|\hat f_{D\ell} - f_D\|^2 + \|\hat f_{D\ell} - f_D\| \|
\E_{n_\ell}\left[SK_h(D-d)\right]-s(d)f_D(d)\|\right)
\\
&= \frac{1}{f_D(d)}\left( \E_{n_\ell}\left[SK_h(D-d)\right] - s(d) \E_{n_\ell}\left[K_h(D-d)\right]
\right) + o_{\p}(1/\sqrt{nh}) \\
&=
\E_{n_\ell}\left[\phi_2(d)\right]  + o_{\p}(1/\sqrt{nh}).
\end{align*}

(ii) A standard algebra (e.g., Theorem 19.2 in \cite{HansenBook}) yields $V_{s(d)}$.
Specifically, $h\E[\phi_{s(d)}^2] = h\E[(S - s(d))^2 K_h(D - d)^2/f_D(d)^2]
=
h \int \E[(S - s(d))^2|D=v] K_h(v-d)^2 f_D(v) dv/f_D(d)^2
=
\int \E[(S - s(d))^2|D= d+uh] k(u)^2 f_D(d+uh) du/f_D(d)^2
= \E[(S - s(d))^2|D= d]R_k/f_D(d) + O(h) \rightarrow V_{s(d)}R_k/s(d)$,
under the condition that $s(d)f_D(d)$ and its first derivative are bounded uniformly.

(iii) Let the convolution kernel $\bar k(x) = \int k(u)k(u - x) du$.
$h\E[\phi_{s(d')} \phi_{s(d)}] = h \E\big[(S-s(d'))(S-s(d)) K_h(D-d') K_h(D-d)\big]/(f_D(d') f_D(d)) =
 \int \E[(S-s(d'))(S-s(d))|D = d+uh] k(u) k\big(\frac{d-d'}{h} + u\big) f_D(d+uh) du/(f_D(d') f_D(d))
 =
 \E[(S-s(d'))(S-s(d))|D = d] \bar k\big(\frac{d'-d}{h}\big) /f_D(d') + o(1)
 =
 s(d)(1-s(d)) \bar k\left(\frac{d'-d}{h}\right) /f_D(d') + o(1) = o(1)$,
as $h\rightarrow 0$.

(iv) Similar arguments and a linearization as in (\ref{Elinear}) give the result for $\hat p_\ell$.

\paragraph{Proof of Claim-Step3:}
The kernel regression estimator of $\E[Y\1|D=d, S=1, \mathcal{W}_\ell^c]$ is standard as in Claim-Step1.

\paragraph{Proof of Claim-SE:}
Let $W_i := Y_i(\hat \1_i^\ell - \1_i)K_h(D_i-d)S_i$.
\begin{align*}
\E_{\ell}[W]
&=
\E\left[\int_{\hat Q_\ell^d(1-\hat p_{\ell})}^{Q^d(1-p)} y f_{Y|DS}(y|D, S) K_h(D-d)S \bigg|\mathcal{W}_\ell^c\right]
\\
&= - \left(\hat Q_\ell^d(1-\hat p_{\ell}) - Q^d(1-p)\right)Q^d(1-p) f_{Y|DS}(Q^d(1-p)|d,1) s(d) f_D(d) + O_{\p}(h^2)
\end{align*}
by a Taylor series expansion and Leibniz rule.
By Claim-Step1 and Claim-Step2, we show
\begin{align*}
\hat Q_\ell^d(1-\hat p_{\ell}) - Q^d(1-p)
&=
\hat Q_\ell^d(1-\hat p_{\ell}) -  Q_\ell^d(1-\hat p_{\ell}) +  Q_\ell^d(1- \hat p_{\ell}) - Q^d(1-p)
\\
&=
\E_{n\ell}^c[\phi_2(1-p)]
  - (\hat p_{\ell} - p) \partial Q^d(\tau)/\partial\tau|_{\tau = 1-p}
 + o_\p(1/\sqrt{nh}) +  O_{\p}\big(\|\hat p_\ell - p\|^2\big),
\end{align*}
where the second inequality $\hat Q_\ell^d(1-\hat p_{\ell}) -  Q_\ell^d(1-\hat p_{\ell}) - \big( \hat Q_\ell^d(1- p) - Q^d(1-p)\big) = o_\p(1/\sqrt{nh})$ that we show below.

Claim-Step2, Theorem 4.1 in \cite{DHB}, and the functional delta method imply that
$\sqrt{nh}\big(\hat Q_\ell^d(p_1) -  Q_\ell^d(p_{1})
- \big( \hat Q_\ell^d(p_0) - Q^d(p_0)\big)\big)
= \sqrt{nh}\E_{n\ell}^c[\phi_2(p_1) - \phi_2(p_0)] + o_\p(1)$ weakly converges to a Gaussian process indexed by $(p_0, p_1)\in [0,1]^2$, which has mean zero and variance
$\lim_{n\rightarrow \infty}\E\big[h\big(\phi_2(p_1) - \phi_2(p_0)\big)^2\big] =
\lim_{n\rightarrow \infty}\E\Big[hK_h(D-d)^2 S\Big(\frac{p_1 - \1\{Y\leq Q^d(p_1)\}}{f_{Y|DS}(Q^d(p_1)|d,1)} - \frac{p_0 - \1\{Y\leq Q^d(p_0)\}}{f_{Y|DS}(Q^d(p_0)|d,1)}\Big)^2
\Big]$\\$\times(s(d)^2 f_D(d)^2)^{-1}
= O\big(\|p_1 - p_0\|\big)$.
So the condition in Theorem 18.5 in \cite{HansenProb} holds, i.e.,
for all $\delta >0$ and $(p_0, p_1) \in [0,1]^2$, $\Big(\E\big[ \sup_{\|p_1 - p_0\| \leq \delta} \| \sqrt{h} (\phi_2(p_1) - \phi_2(p_0))\|^2\big]\Big)^{1/2} \leq C \delta^\psi$
for some $C < \infty$ and $0 < \psi < \infty$.
It follows that $\sqrt{nh} \E_{n\ell}^c\left[ \phi_2(p)\right]$ is stochastic equicontinuous, i.e., $\forall \eta, \epsilon >0$, there exists some $\delta > 0$ such that
$\lim\sup_{n\rightarrow \infty} \p\Big(
\sup_{\|p_1 - p_0\| \leq \delta} \big\| \sqrt{nh} \E_{n\ell}^c\big[ \phi_2(p_1) - \phi_2(p_0)\big]\big\|
> \eta\Big) \leq \epsilon$.
We obtain
$\sqrt{nh}\E_{n\ell}^c[\phi_2(p_1) - \phi_2(p_0)] = o_\p(1)$ uniformly over $\|p_1 - p_0\| \leq \delta$,
and hence $
\hat Q_\ell^d(1-\hat p_{\ell}) -  Q_\ell^d(1-\hat p_{\ell}) - \big( \hat Q_\ell^d(1- p) - Q^d(1-p)\big) =
\E_{n\ell}^c[\phi_2(1-\hat p_\ell) - \phi_2(1-p)] + o_\p(1/\sqrt{nh}) = o_\p(1/\sqrt{nh})$ as $\|\hat p_\ell - p\| = o_\p(1)$ by Claim-Step1.

Further assuming $\sqrt{nh} h^2 = o(1)$ and $\sqrt{nh} \|\hat p_\ell - p\|^2= o_{\p}(1)$, we obtain $\E_{\ell}[W] = num_{12}^\ell \times den + o_{\p}(1/\sqrt{nh})$.

\begin{align*}
h \E_\ell\left[W^2\right]
&=
h \E\left[ Y^2(\hat \1^\ell - \1)^2 K^2_h(D-d)S \bigg|\mathcal{W}_\ell^c\right]
\\
&= \E[Y^2(\hat \1^\ell - \1)^2|D=d, S=1, \mathcal{W}_\ell^c] s(d) f_D(d)R_k + O_{\p}(h) \\
&= o_{\p}(1)
\end{align*}
by the consistency of Step 1 and Step 2.
By the conditional Markov inequality, $\sqrt{n_\ell h}(\E_{n_\ell}[W] -\E_\ell[W]) =
\sqrt{n_\ell h}(\E_{n_\ell}[W] -num_{12}^\ell \times den) + o_{\p}(1) =  o_{\p}(1)$.

By a linearization as (\ref{Elinear}),
\begin{align*}
\widehat{num}_{12}^\ell  = num_{12}^\ell
+
\frac{\E_{n_\ell}[W] - num_{12}^\ell \times den}{den}
-
\frac{num_{12}^\ell}{den}  (\widehat{den}- den)
+ o_{\p}(1/\sqrt{nh}).
\end{align*}
By the consistency of Step 1 and Step 2, $num_{12}^\ell = o_{\p}(1)$, and by Claim-Step3,
 the above third term $\frac{num_{12}^\ell}{den}  (\widehat{den}- den) = o_{\p}(1/\sqrt{nh} + h^2)$.
Note that $ \partial Q^d(\tau)/\partial\tau|_{\tau = 1-p} \times f_{Y|DS}(Q^d(1-p)|d,1) = 1$.


For $\rho_{LB}(\tau)$, $\hat\1 = \1\{Y \leq \hat Q^d(\hat p_\ell)\}$.
\begin{align*}
\E_{\ell}[W]
&=
\E\left[\int_{Q^d(p)}^{\hat Q^d(\hat p_\ell)} y f_{Y|DS}(y|D, S) K_h(D-d)S \bigg|\mathcal{W}_\ell^c\right]
\\
&= \left(\hat Q_\ell^d(\hat p_{\ell}) - Q^d(p)\right)Q^d(p) f_{Y|DS}(Q^d(p)|d,1) s(d) f_D(d) + O_{\p}(h^2), \mbox{ where
}
\end{align*}
\begin{align*}
\hat Q_\ell^d(\hat p_{\ell}) - Q^d(p)
&=
\hat Q_\ell^d(\hat p_{\ell}) -  Q_\ell^d(\hat p_{\ell}) + Q_\ell^d(\hat p_{\ell}) - Q^d(p)
\\
&=
 \hat Q_\ell^d(p) - Q^d(p)+ (\hat p_{\ell} - p) \partial Q^d(\tau)/\partial\tau|_{\tau = p}
 + O_{\p}(\|\hat Q_\ell^d - Q^d\|^2 + \|\hat p_\ell - p\|^2).
\end{align*}

\subsection{Proofs of Corrollaries}

\paragraph{Proof of Corollary~\ref{CATE}:}
By (\ref{IF}) in the proof of Theorem~\ref{TEstLB}, $\hat{\bar\Delta}_{d_1d_2} - \bar\Delta_{d_1d_2} = n^{-1}\sum_{i=1}^n
\big(\phi_{d_2Ui} - \phi_{d_1Li}\big) + o_\p((nh)^{-1/2})$, where $h = \min\{h_{d_2U}, h_{d_1L}\}$.
We next verify the third-absolute-moment condition for the Lyapunov CLT.
By the variance calculation in the proof of Theorem~\ref{TEstLB}, let $s_n^2 = \sum_{i=1}^n \E\big[\big(\phi_{d_2Ui} - \phi_{d_1Li}\big)\big/\sqrt{n}\big)^2\big]=\mathsf{V}_{Un} = O(h^{-1})$
and $\E[|\phi|^3] = O(h^{-2})$.
So $\sum_{i=1}^n \E\big[\big|\big(\phi_{d_2Ui} - \phi_{d_1Li}\big)\big/\sqrt{n} \big|^3\big]/s_n^3 = O(n h^{-2}n^{-3/2}/h^{-3/2}) = O((nh)^{-1/2}) = o(1)$.
Then by the Lyapunov CLT,
$s_n^{-1}\sum_{i=1}^n \big(\phi_{d_2Ui} - \phi_{d_1Li}
- (h_{d_2U}^2{B}_{d_2U} -h_{d_1L}^2{B}_{d_1L})
\big)/\sqrt{n}
= \mathsf{V}_{Un}^{-1/2} \sqrt{n} n^{-1}\sum_{i=1}^n \big(\phi_{d_2Ui} - \phi_{d_1Li}
 - (h_{d_2U}^2{B}_{d_2U} -h_{d_1L}^2{B}_{d_1L})
\big)
\stackrel{d}{\rightarrow} \mathcal{N}(0,1)$.

The remainder term $\mathsf{V}_{Un}^{-1/2} \sqrt{n} o_\p((nh)^{-1/2}) = o_\p(1)$, so we obtain $\mathsf{V}_{Un}^{-1/2} \sqrt{n}\big(\hat{\bar\Delta}_{d_1d_2} - \bar\Delta_{d_1d_2} - (h_{d_2U}^2{B}_{d_2U} -h_{d_1L}^2{B}_{d_1L})\big)
\stackrel{d}{\rightarrow} \mathcal{N}(0,1)$.
The same arguments apply to $\hat{\underline{\Delta}}_{d_1d_2}$.

We show that as $n\rightarrow \infty, h\rightarrow 0$, $hV_{Un} = h\E[\phi_{d_2U}^2 + \phi_{d_1L}^2 - 2\phi_{d_2U}\phi_{d_1L}] = V_{d_2U} + V_{d_1L} -2 C_{d_1d_2U}+ o(1)$.
In $h\E[\phi_{d_2U} \phi_{d_1L}]$,
the cross-product terms with $K_h(D-d_1)K_h(D-d_2)$ result in a  convolution kernel $\int k((d_2-d_1)/h + u) k(u) du$ and hence is $o(1)$.
From the proof of Theorem~\ref{TEstLB}, $h\E[\phi_\pi^2] = R_k V_{\pi}/f_D(d_{\text{AT}}) + o(1)$.
Then
$h\E[\phi_{d_2U} \phi_{d_1L}] = h\E[\phi_\pi^2(
Q^{d_2}(1-p_{d_2}) - \rho_{d_2U}(\pi)
)(
Q^{d_1}(p_{d_1}) - \rho_{d_1L}(\pi)
)/(s(d_1)s(d_2)p_{d_1}p_{d_2})] + o(1) = C_{d_1d_2U} + o(1)$.
The same arguments apply to $C_{d_1d_2L}$ for the lower bound $\underline{\Delta}_{d_1d_2}$.\hfill $\square$

\paragraph{Proof of Corollary~\ref{CATEX}:}
By Theorem~\ref{TDMLX}, $\hat{\bar\Delta}_{d_1d_2} - \bar\Delta_{d_1d_2} = n^{-1}\sum_{i=1}^n
\big(\phi_{d_2U}(W_i, \xi) - \phi_{d_1L}(W_i, \xi)\big) + o_\p((nh)^{-1/2})$, where $h = \min\{h_{d_2U}, h_{d_1L}\}$.
We next verify the third-absolute-moment condition for the Lyapunov CLT.
By the variance calculation in the proof of Theorem~\ref{TDMLX}, let $s_n^2 = \sum_{i=1}^n \E\big[\big(\phi_{d_2U}(W_i, \xi) - \phi_{d_1L}(W_i, \xi)\big)\big/\sqrt{n}\big)^2\big]=\mathsf{V}_{Un} = O(h^{-1})$
and $\E[|\phi|^3] = O(h^{-2})$.
Therefore $\sum_{i=1}^n \E\big[\big|\big(\phi_{d_2U}(W_i, \xi) - \phi_{d_1L}(W_i, \xi)\big)\big/\sqrt{n} \big|^3\big]/s_n^3 = O(n h^{-2}n^{-3/2}/h^{-3/2}) = O((nh)^{-1/2}) = o(1)$.
By the Lyapunov CLT,
$s_n^{-1}\sum_{i=1}^n \big(\phi_{d_2U}(W_i, \xi) - \phi_{d_1L}(W_i, \xi)
- (h_{d_2U}^2\mathsf{B}_{d_2U} -h_{d_1L}^2\mathsf{B}_{d_1L})
\big)/\sqrt{n}
= \mathsf{V}_{Un}^{-1/2} \sqrt{n} n^{-1}\sum_{i=1}^n \big(\phi_{d_2U}(W_i, \xi) - \phi_{d_1L}(W_i, \xi)
 - (h_{d_2U}^2\mathsf{B}_{d_2U} -h_{d_1L}^2\mathsf{B}_{d_1L})
\big)
\stackrel{d}{\rightarrow} \mathcal{N}(0,1)$.

The remainder term $\mathsf{V}_{Un}^{-1/2} \sqrt{n} o_\p((nh)^{-1/2}) = o_\p(1)$, so we obtain $\mathsf{V}_{Un}^{-1/2} \sqrt{n}\big(\hat{\bar\Delta}_{d_1d_2} - \bar\Delta_{d_1d_2} - (h_{d_2U}^2\mathsf{B}_{d_2U} -h_{d_1L}^2\mathsf{B}_{d_1L})\big)
\stackrel{d}{\rightarrow} \mathcal{N}(0,1)$.
The same arguments apply to $\hat{\underline{\Delta}}_{d_1d_2}$.
\hfill $\square$

\section{Supplements and details for empirical applications}
\label{ASecEM}

\subsection{Step 0 strict overlap sub-sample}
\label{ASecStep0}
We prepare a sub-sample that satisfies Assumption~\ref{ADMLX}(i) for estimating the bounds.
In Step 0, we use the full sample and bandwidth $h_1$ to estimate the GPS by $\tilde\mu_d(X_i)$ and the selection probability by $\tilde s(d, X_i)$.

We choose the fixed trimming parameter $trim_{GPS}$ by extending the idea of Imbens (2004) for a binary treatment to a continuous treatment.  We limit the ``importance weight" $\bar k/(\mu_d(X_i) n_\ell  h) \leq 5\%$, where
 $n_\ell $ is the floor of $n/L$ (the largest integer less than or equal to $n/L$) and
the kernel function is bounded by $\bar k := \max_{u \in \mathcal{R}} k(u)$ (for the Epanechnikov kernel, $\bar k =  0.75/\sqrt{5}$).
 So the trimming rule is to drop the observation $i$ if
 \[
\tilde\mu_d(X_i)  < \bar k/(5\%n_\ell h_1) =: trim_{GPS}
 \]
 for {\it some} $d \in \mathcal{D}_J$.  So we obtain an overlap sample, denoted as $Sample_{GPS}$, where $\tilde\mu_d(X_i) \geq trim_{GPS}$ for {\it all} $i \in Sample_{GPS}$ and for {\it all} $d \in\mathcal{D}_J$.
For JC, $trim_{GPS} = 0.000055$.  For CCC, $trim_{GPS} = 0.01205$.

For the selection probability and sufficient always-takers,
we drop observation $i$ in the full sample if $\min_{d' \in \mathcal{D}_J} \tilde s(d', X_i) < 5\%$.
We obtain a sample $Sample_{S}$ where $\tilde s(d, X_i) \geq 5\%$  for {\it all} $i \in Sample_{S}$  and for {\it all} $d \in \mathcal{D}_J$.


Finally, we obtain a sub-sample that is an intersection of $Sample_{GPS}$ and $Sample_{S}$ for estimating the bounds.

But in the cross-fitting sub-samples in Step~1, it is possible to obtain a GPS estimate below $trim_{GPS}$ and a small proportion of always-takers.
So we set the GPS estimate below $trim_{GPS}$ to $trim_{GPS}$, following \cite{HHLL} to obtain a more stable estimator.
That is, replace $\hat \mu_{d\ell}(X_i)$ with $\max\{ \hat \mu_{d\ell}(X_i), trim_{GPS}\}$.

To further address possible small proportion of always-takers $\hat p_{dd_i\ell}(X_i)$ in estimating the conditional bounds, we restrict the conditional bounds estimates by the maximum and minimum of the outcome variable $Y$ in the sample.
That is, we estimate $\rho_{dU}(\pi_{\text{AT}}(X_i), X_i)$ by $ \min\big\{ \widehat{\E}_\ell[Y|Y \geq \hat Q_\ell^d(1-\hat p_{dd_j \ell}(X_i),X_i)_\ell, D=d, S=1, X=X_i], \max\{Y_i, i=1,...,n\}\big\}$, for $i \in I_\ell$.
Similarly, estimate
 $\rho_{dL}(\pi_{\text{AT}}(X_i), X_i)$ by $ \max\big\{ \widehat{\E}_\ell[Y|Y \leq \hat Q_\ell^d(\hat p_{dd_j \ell}(X_i),X_i)_\ell, D=d, S=1, X=X_i], \min\{Y_i, i=1,...,n\}\big\}$.

\subsection{First-step Lasso estimation in Section~\ref{Sec1stEst}}
\label{ASec1stEst}
Let the logistic likelihood $M(y, x; g) = -\big(y\log(\Lambda(b(x)'g)) + (1-y)\log(1-\Lambda(b(x)'g))\big)$, where $\Lambda$ is the logistic CDF.
Penalty loading matrix $\hat\Psi_{d\ell}$, $\hat\Psi_{dy\ell}$, $\hat\Xi_{d\ell}$ are computed by Algorithm~\ref{ASUZ3.2} below from the iterative Algorithms 3.1 and 3.2 in SUZ.
Let the final penalty loading matrix $\hat \Psi_{d\ell}$ be $\hat\Psi_{d\ell}^m$ from Algorithm~\ref{ASUZ3.2} for some fixed positive integer $M$.

For $\ell\in\{1,...,L\}$,
\begin{itemize}
\item
$\hat F_{D|X_\ell}(D|X) = \Lambda(b(X)'\hat\beta_{d\ell})$, where
\begin{align}
\hat\beta_{d\ell} = \arg\min_\beta \frac{1}{N_\ell}\sum_{i\notin I_\ell} M({\bf 1}\{D_i \leq d\}, X_i; \beta) + \frac{\tilde \lambda}{N_\ell}\|\hat\Psi_{d\ell}\beta\|_1,
\label{SUZ3.6}
\end{align}
$N_\ell = n -n_\ell$,
the penalty $\tilde\lambda = 1.1\Phi^{-1}(1-r/\{p\vee nh_1\}) n^{1/2}$,
for some $r\rightarrow 0$ and $h_1\rightarrow 0$, with the standard normal CDF  $\Phi$. We follow SUZ and set $r=1/\log(n)$.

Compute $\hat F_{D|X_\ell}(d|x) = \Lambda(b(X)'\hat\beta_{d\ell})$ from (\ref{SUZ3.6}).
Then the conditional density estimator
\[
\hat \mu_{d\ell}(x) = \frac{\hat F_{D|X_\ell}(d+h_1|x) - \hat F_{D|X_\ell}(d-h_1|x)}{2h_1}.
\]

\item
$\hat Q^d_\ell(p,x) = \inf\{y: \hat F_{Y|SDX_\ell}(y|1,d,x) \geq p\}$, where
$\hat F_{Y|SDX_\ell}(y|1,d,x) = \Lambda(x'\hat \alpha_{dy\ell})$ with \begin{align}
\hat \alpha_{dy\ell}:= \arg\min_\alpha \frac{1}{N_{\ell}}\sum_{i\notin I_\ell}
M(\1\{Y_i\leq y\}, X_i, \alpha)S_ik\left(\frac{D_i - d}{h_1}\right)  + \frac{\lambda}{N_{\ell}}\|\hat\Psi_{dy\ell} \alpha\|_1,
\label{EMLEalpha}
\end{align}
 $N_{\ell} = \sum_{i\notin I_\ell} S_i$, and
the penalty $\lambda = \ell_n(\log(p\vee nh_1)nh_1)^{1/2}$ and $\ell_n=\sqrt{\log(\log(nh_1))}$.

\item $\hat s_\ell(d,x) = \Lambda(b(x)'\hat\theta_{d\ell})$, where
\begin{align}
\hat\theta_{d\ell}:= \arg\min_\theta \frac{1}{N_\ell}\sum_{i\notin I_\ell}
 M(S_i, X_i; \theta)k\left(\frac{D_i - d}{h_1}\right) + \frac{\lambda}{N_\ell}\|\hat\Upsilon_{d\ell}\theta\|_1,
\label{EMLEtheta}
 \end{align}
 $N_\ell = n-n_\ell$.

\item $\hat \rho_{dU_\ell}(\pi, x) = \hat{\E}[Y| Y\geq \hat Q^d_\ell(1-\hat\pi_\ell/\hat s_\ell(d,x),x),S=1, D=d, X=x] = b(x)'\hat\gamma_{d\ell}$, where
\begin{align}
\hat\gamma_{d\ell} := \arg\min_{\gamma} &\frac{1}{2N_{\ell}}\sum_{i\notin I_\ell} (Y_i - b(X_i)'\gamma)^2 k\left(\frac{D_i - d}{h_1}\right)S_i\1\{Y_i\geq \hat Q^d_\ell(1-\hat\pi_\ell/\hat s_\ell(d,X_i),X_i)\} \notag\\
& + \frac{\lambda}{N_{\ell}}\|\hat{\Xi}_{d\ell} \gamma\|_1,
\label{EMLEgamma}
\end{align}
with $N_{\ell} = \sum_{i\notin I_\ell} S_i \1\{Y_i\geq \hat Q^d_\ell(1-\hat\pi_\ell/\hat s_\ell(d,X_i),X_i)\}$.

\end{itemize}

\begin{algorithm}[SUZ Algorithm 3.1 and 3.2]
For $\ell\in\{1,...,L\}$,
\begin{itemize}
\item For $\hat \mu_{d\ell}(x)$,
\begin{enumerate}
\item
Let $\hat \Psi_{d\ell}^0 = diag(l_{d\ell,1}^0,...,l_{d\ell,p}^0)$, where $l_{d\ell,j}^0 = \|{\bf 1}\{D\leq d\} b_j(X)\|_{{\p}_{N_\ell},2}$.
Compute $\hat\beta_{d\ell}^0$  by (\ref{SUZ3.6})
with $\hat \Psi_{d\ell}^0$ in place of  $\hat \Psi_{d\ell}$.
Let $\hat F_{D|X_\ell}^0(D|X) = \Lambda(b(x)'\hat\beta_{d\ell}^0)$.

\item Compute $\hat\Psi_{d\ell}^m = diag(l_{d\ell,1}^m,...,l_{d\ell,p}^m)$, where
$l_{d\ell,j}^m = \Big\|\Big({\bf 1}\{D\leq d\}  - \hat F_{D|X_\ell}^{m-1}(d|X)\Big)b_j(X)\Big\|_{{\p}_{N_\ell},2}$,
for $m=1,...,M$.
Compute $\hat\beta_{d\ell}^m$  by (\ref{SUZ3.6})
with $\hat \Psi_{d\ell}^m$ in place of  $\hat \Psi_{d\ell}$.
Let $\hat F_{D|X_\ell}^m(d|x) = \Lambda(b(x)'\hat\beta_{d\ell}^m)$.
\end{enumerate}

\item For $\hat Q^d_\ell(p,x)$,
\begin{enumerate}
\item
Let $\hat \Psi_{dy\ell}^0 = diag(l_{dy\ell,1}^0,...,l_{dy\ell,p}^0)$, where $l_{dy\ell,j}^0 = \big\|{\bf 1}\{Y\leq y\} S b_j(X)k((D-d)/h_1)h_1^{-1/2}\big\|_{{\p}_{N_\ell},2}$.
Compute $\hat\alpha_{dy\ell}^0$  by (\ref{EMLEalpha})
with $\hat \Psi_{dy\ell}^0$ in place of  $\hat \Psi_{dy\ell}$.
Let $\hat F_{Y|SDX_\ell}^0(y|1,d,x) = \Lambda(b(x)'\hat\alpha_{dy\ell}^0)$.

\item Compute $\hat\Psi_{dy\ell}^m = diag(l_{dy\ell,1}^m,...,l_{dy\ell,p}^m)$, where
$l_{dy\ell,j}^m = \Big\|\Big({\bf 1}\{Y\leq y\}  - \hat F_{Y|SDX_\ell}^{m-1}(y|1,d,X)\Big) S$\\
$\times b_j(X)K((D-d)/h_1) h_1^{-1/2}\Big\|_{{\p}_{N_{\ell}},2}$,
for $m=1,...,M$.
Compute $\hat\alpha_{dy\ell}^m$  by (\ref{EMLEalpha})
with $\hat \Psi_{dy\ell}^m$ in place of  $\hat \Psi_{dy\ell}$.
Let $\hat F_{Y|SDX_\ell}^m(y|1,d,x) = \Lambda(b(x)'\hat\alpha_{dy\ell}^m)$.
\end{enumerate}

\item For $\hat s_\ell(d,x)$,
\begin{enumerate}
\item
Let $\hat \Upsilon_{d\ell}^0 = diag(l_{d\ell,1}^0,...,l_{d\ell,p}^0)$, where $l_{d\ell,j}^0 = \|S b_j(X)k((D-d)/h_1)h_1^{-1/2}\|_{{\p}_{N_\ell},2}$.
Compute $\hat\theta_{d\ell}^0$  by (\ref{EMLEtheta})
with $\hat \Upsilon_{d\ell}^0$ in place of  $\hat \Upsilon_{d\ell}$.
Let $\hat s_\ell^0(d,x) = \Lambda(b(x)'\hat\theta_{d\ell}^0)$.

\item Compute $\hat\Upsilon_{d\ell}^m = diag(l_{d\ell,1}^m,...,l_{d\ell,p}^m)$, where
$l_{d\ell,j}^m = \Big\|\Big(S - \hat s_\ell^{m-1}(d,X) \Big) b_j(X)K((D-d)/h_1) h_1^{-1/2}\Big\|_{{\p}_{N_\ell},2}$,
for $m=1,...,M$.
Compute $\hat\theta_{d\ell}^m$  by (\ref{EMLEtheta})
with $\hat \Upsilon_{d\ell}^m$ in place of  $\hat \Upsilon_{d\ell}$.
Let $\hat s^m_\ell(d,x) = \Lambda(b(x)'\hat\theta_{d\ell}^m)$.
\end{enumerate}

\item For $\hat \rho_{dU_\ell}(\pi,x)$,
\begin{enumerate}
\item
Let $\hat \Xi_{d\ell}^0 = diag(l_{d\ell,1}^0,...,l_{d\ell,p}^0)$, where $l_{d\ell,j}^0 = \|Y b_j(X) S \1\{Y\geq \hat Q^d_\ell(1-\hat\pi_\ell/\hat s_\ell(d,X),X)\} k((D-d)/h_1)h_1^{-1/2}\|_{{\p}_{N_{\ell}},2}$.
Compute $\hat\gamma_{d\ell}^0$  by (\ref{EMLEgamma})
with $\hat \Xi_{d\ell}^0$ in place of  $\hat \Xi_{d\ell}$.
Let $\hat \rho_{dUU_\ell}^0(\pi,x) = b(x)'\hat\gamma_{d\ell}^0$.

\item Compute $\hat\Xi_{d\ell}^m = diag(l_{d\ell,1}^m,...,l_{d\ell,p}^m)$, where
$l_{d\ell,j}^m = \Big\|\Big(Y - \hat \rho_{dU_\ell}^{m-1}(\pi,X) \Big) S \1\{Y\geq \hat Q^d_\ell(1-\hat\pi_\ell/\hat s_\ell(d,X),X)\} b_j(X)K((D-d)/h_1) h_1^{-1/2}\Big\|_{{\p}_{N_{\ell}},2}$,
for $m=1,...,M$.
Compute $\hat\gamma_{d\ell}^m$  by (\ref{EMLEgamma})
with $\hat \Xi_{d\ell}^m$ in place of  $\hat \Xi_{d\ell}$.
Let $\hat \rho^m_{dU\ell}(\pi,x) = b(x)'\hat\gamma_{d\ell}^m$.
\end{enumerate}

\end{itemize}
\label{ASUZ3.2}
\end{algorithm}
Following SUZ, we choose $M = 5$.

\subsection{Supplements for empirical applications: Job Corps}
\label{ASecJC}

\begin{figure}[!htp]
\centering
\caption{(Job Corps) Estimated bounds and 95\% confidence intervals with quadratic $b(X)$
}
\includegraphics[width=0.4\textwidth]{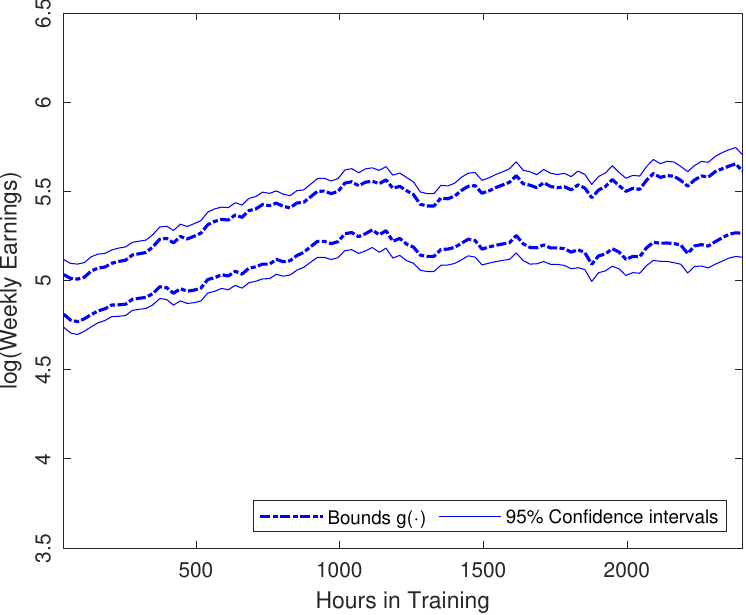}
\label{FJCXquad}
\end{figure}

Let the covariates $X = (X_c, X_b, X_{ca})$, with continuous $X_c$, binary $X_b$, and categorical $X_{ca}$.
Figure~\ref{FJCXquad} presents the estimated bounds with a quadratic basis function
$b(X) = (X_c, X_c^2, X_b, X_{ca}, X_{ca}^2)$ for the Job Corps.
The bounds estimates for the largest effect of switching training hours from 87.677 to 1112.727 are bounded by $[0.276, 0.787]$ with the 95\% confidence interval $[0.172, 0.928]$.\footnote{
The undersmoothing bandwidths range from 160.298 to 274.602. $\nu=0.01$ and $c_1=1$.}
These results are similar to the estimates using the linear basis function in Figure~\ref{FbetadXJC}.

  \begin{table}[htp]
	\begin{center}
		\caption{(Job Corps) Descriptive statistics}\label{TSS_JC}
        {\small		
          \begin{tabular}{lcccccc}
			\hline\hline
			Variable &Mean&Median&Std Dev&Min&Max&Nonmissing \\
			\hline
            weekly earnings in fourth year ($Y$)&215.52&194.31&202.62&0.00&1879.17&4024\\
            hours in training ($D$)&1195.32&966.43&965.93&0.86&5142.86&4024\\
            selection status ($S$)&0.83&1.00&0.37&0.00&1.00&4024\\
            female&0.43&0.00&0.50&0.00&1.00&4024\\
            age&18.33&18.00&2.14&16.00&24.00&4024\\
            White&0.25&0.00&0.43&0.00&1.00&4024\\
            Black&0.50&1.00&0.50&0.00&1.00&4024\\
            Hispanic&0.17&0.00&0.38&0.00&1.00&4024\\
            years of education&9.91&10.00&1.93&0.00&20.00&3968\\
            native English&0.85&1.00&0.36&0.00&1.00&4024\\
            has children&0.18&0.00&0.38&0.00&1.00&4024\\
            ever worked&0.14&0.00&0.35&0.00&1.00&4024\\
            mean gross weekly earnings&19.59&0.00&98.67&0.00&2000.00&4024\\
            household size&3.48&3.00&2.03&0.00&15.00&3968\\
            household gross income brackets&2.21&1.00&2.44&0.00&7.00&2529\\
            personal gross income brackets&0.49&0.00&0.63&0.00&7.00&1789\\
            mother's years of education&9.40&12.00&5.03&0.00&20.00&3288\\
            father's years of education&7.17&10.00&6.00&0.00&20.00&2519\\
            welfare receipt during childhood&1.93&1.00&1.26&0.00&4.00&3753\\
            poor or fair general health&0.12&0.00&0.33&0.00&1.00&4024\\
            physical or emotional problems&0.04&0.00&0.20&0.00&1.00&4024\\
            ever arrested&0.24&0.00&0.43&0.00&1.00&4024\\
            extent of recruiter support&1.56&1.00&1.07&0.00&5.00&3934\\
            idea about the desired training&0.84&1.00&0.37&0.00&1.00&4024\\
            expected hourly wage after Job Corps&4.50&0.00&6.73&0.00&96.00&1808\\
            expected to be training for a job&1.03&1.00&0.27&0.00&3.00&3945\\
            expected stay in Job Corps&6.60&0.00&9.78&0.00&36.00&4024\\
			\hline
		\end{tabular}}
	\end{center}
		{\small Note: We use missing dummies for missing observations in covariates.}
\end{table}

\begin{figure}[!htp]
\centering
\caption{(Job Corps) Histogram of hours of training}
\includegraphics[width=0.4\textwidth]{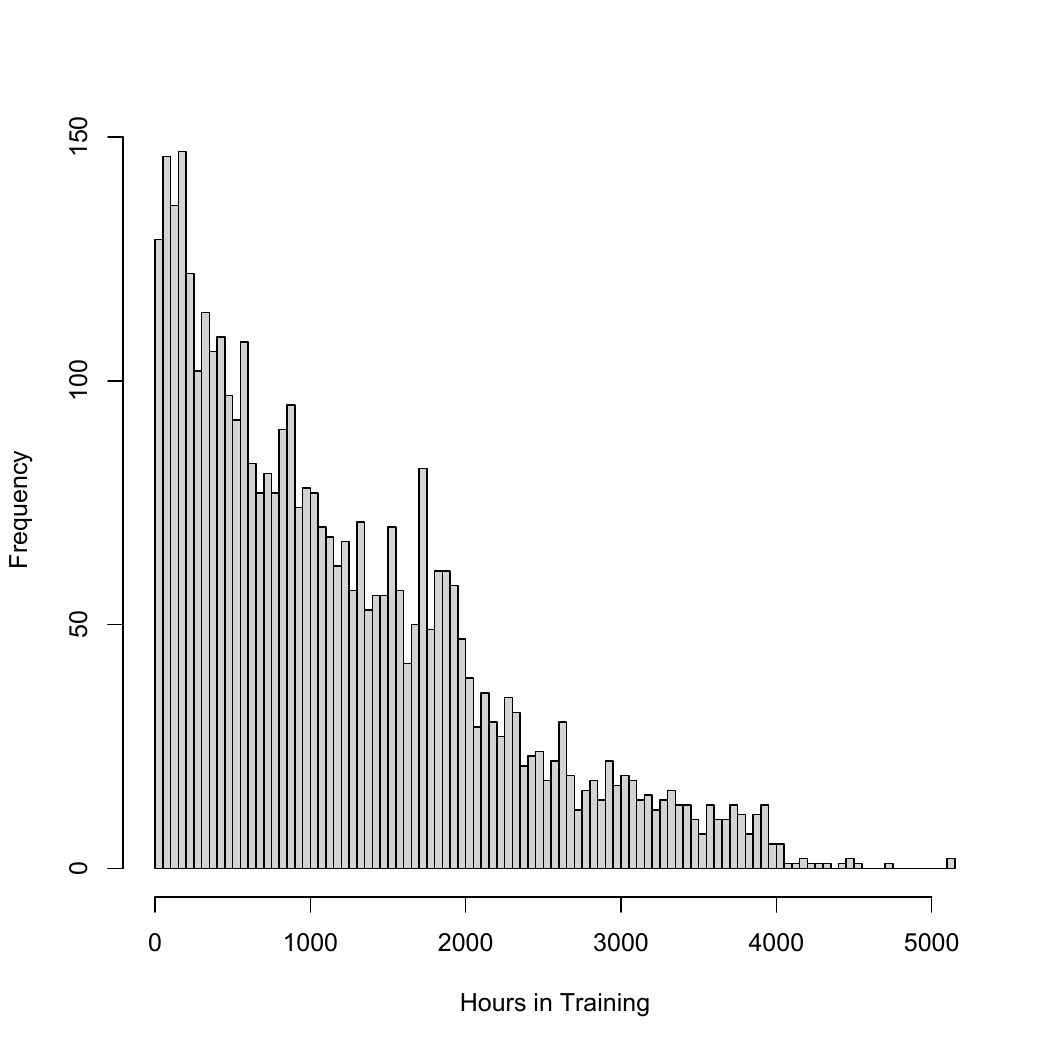}
\label{HisT}
\end{figure}

 \begin{figure}[!htp]
\centering
\caption{\small (Job Corps) Histograms of $Y$ and $\log(Y)$ in $\{Y_i > 0, i=1,...,n\}$}
\includegraphics[width=0.45\textwidth]{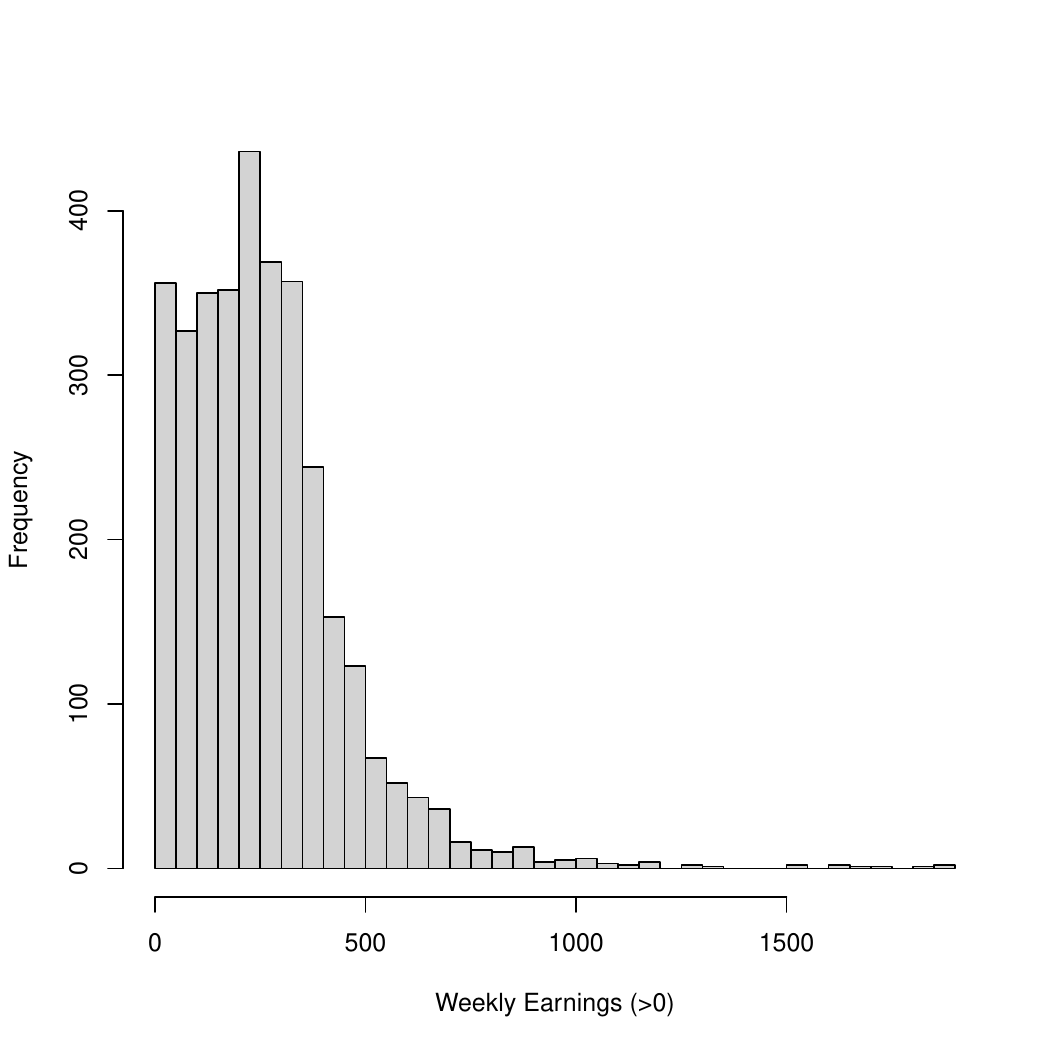}
\includegraphics[width=0.45\textwidth]{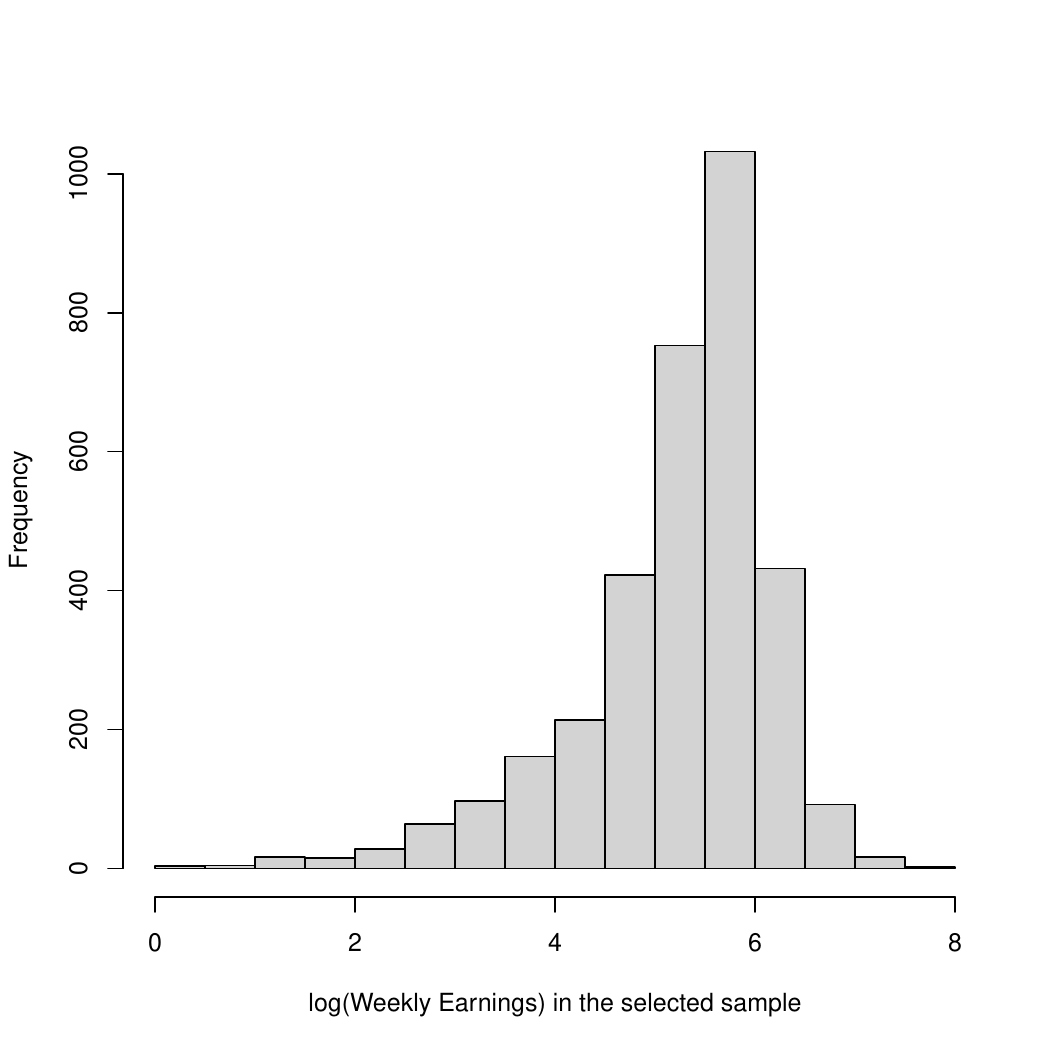}
\label{FJCHY}
\end{figure}

\begin{figure}[!htp]
\centering
\caption{{\small (Job Corps) Estimated bounds and 95\% confidence intervals for Weekly Earnings without $X$}
}
\includegraphics[width=0.4\textwidth]{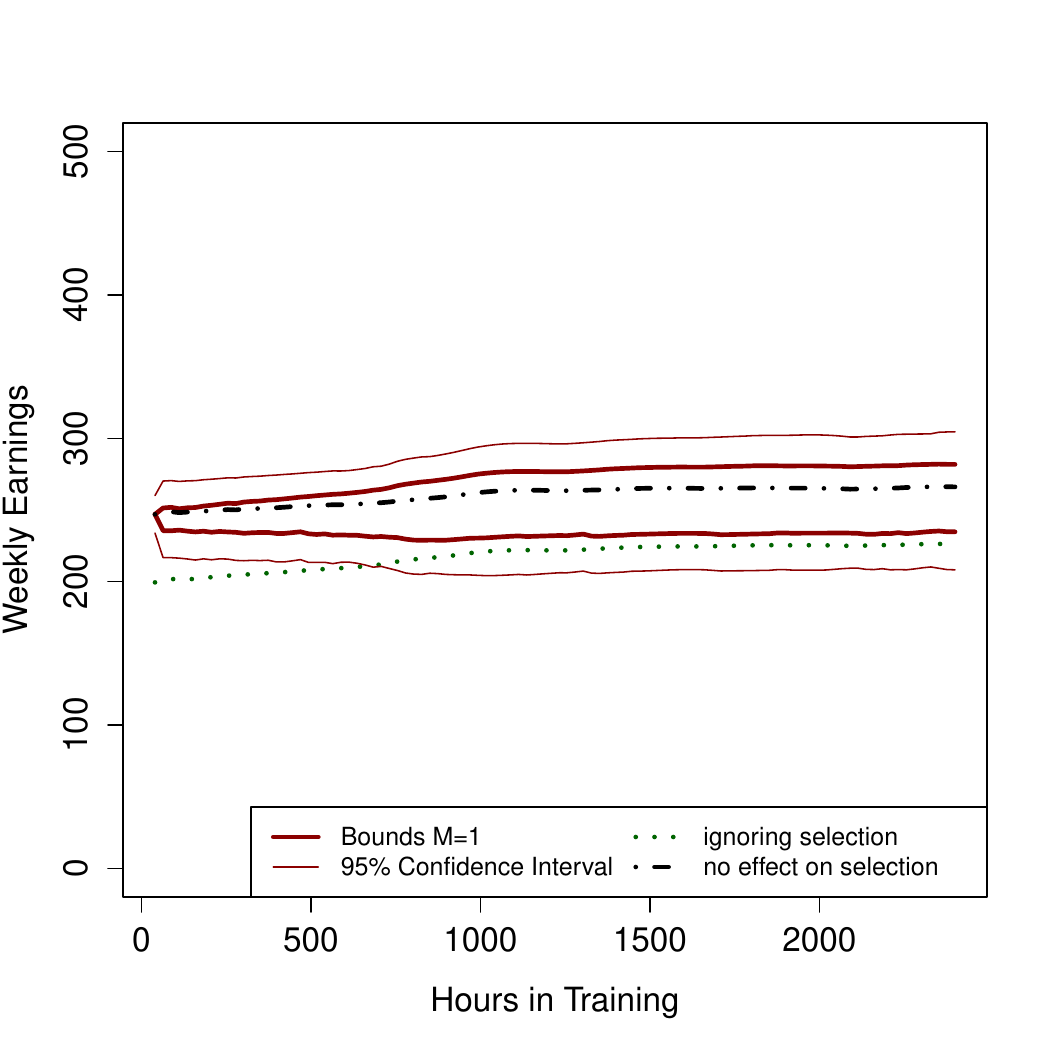}
\label{FbetaY}
\end{figure}

\subsection{Supplements for empirical applications: CCC}
\label{ASecCCC}

\begin{figure}[h!]
\centering
\caption{(CCC) Histogram of duration of service}
\includegraphics[width=0.45\textwidth]{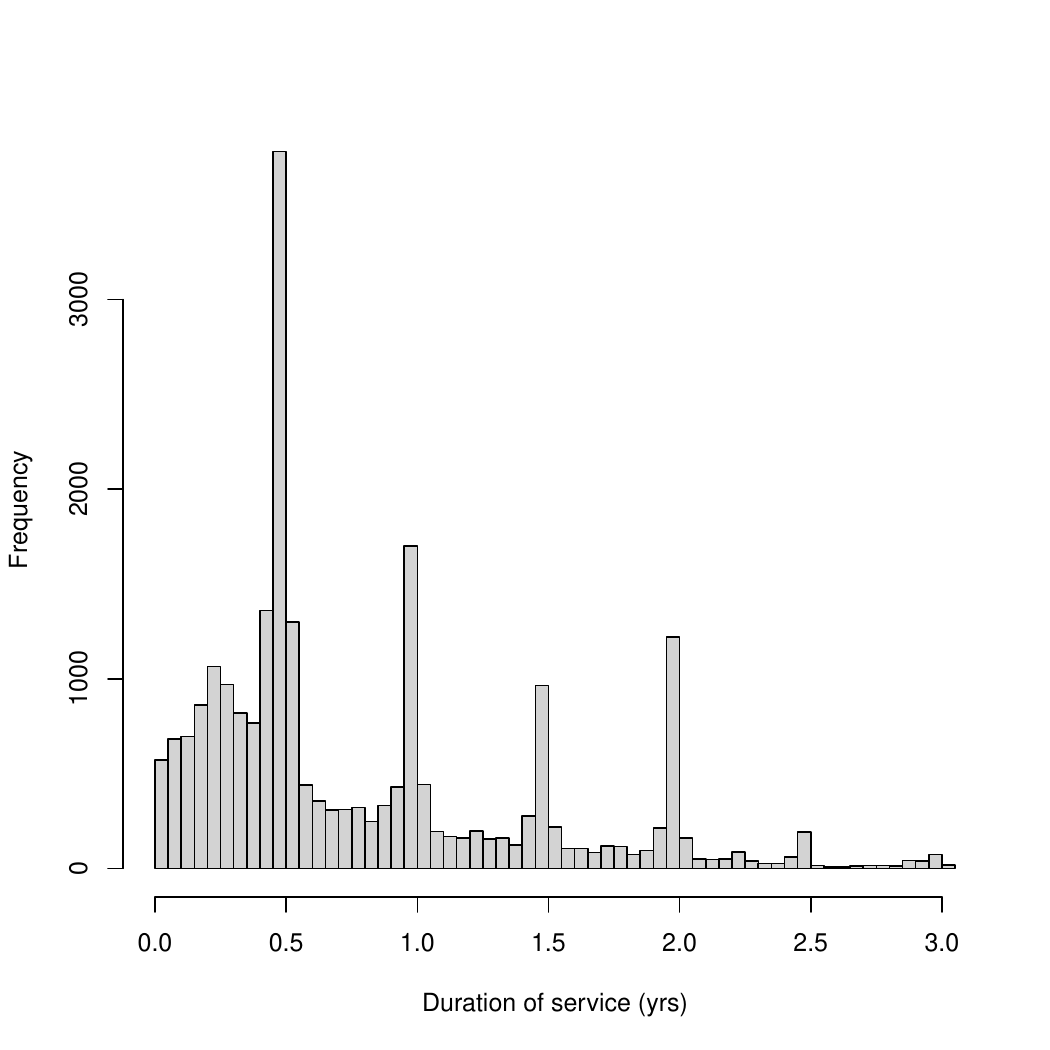}
\label{FhistdCCC}
\end{figure}

 \begin{figure}[!htp]
\centering
\caption{\small (CCC) Histograms of $Y$ and $\log(Y)$ in $\{Y_i > 0, i=1,...,n\}$}
\includegraphics[width=0.45\textwidth]{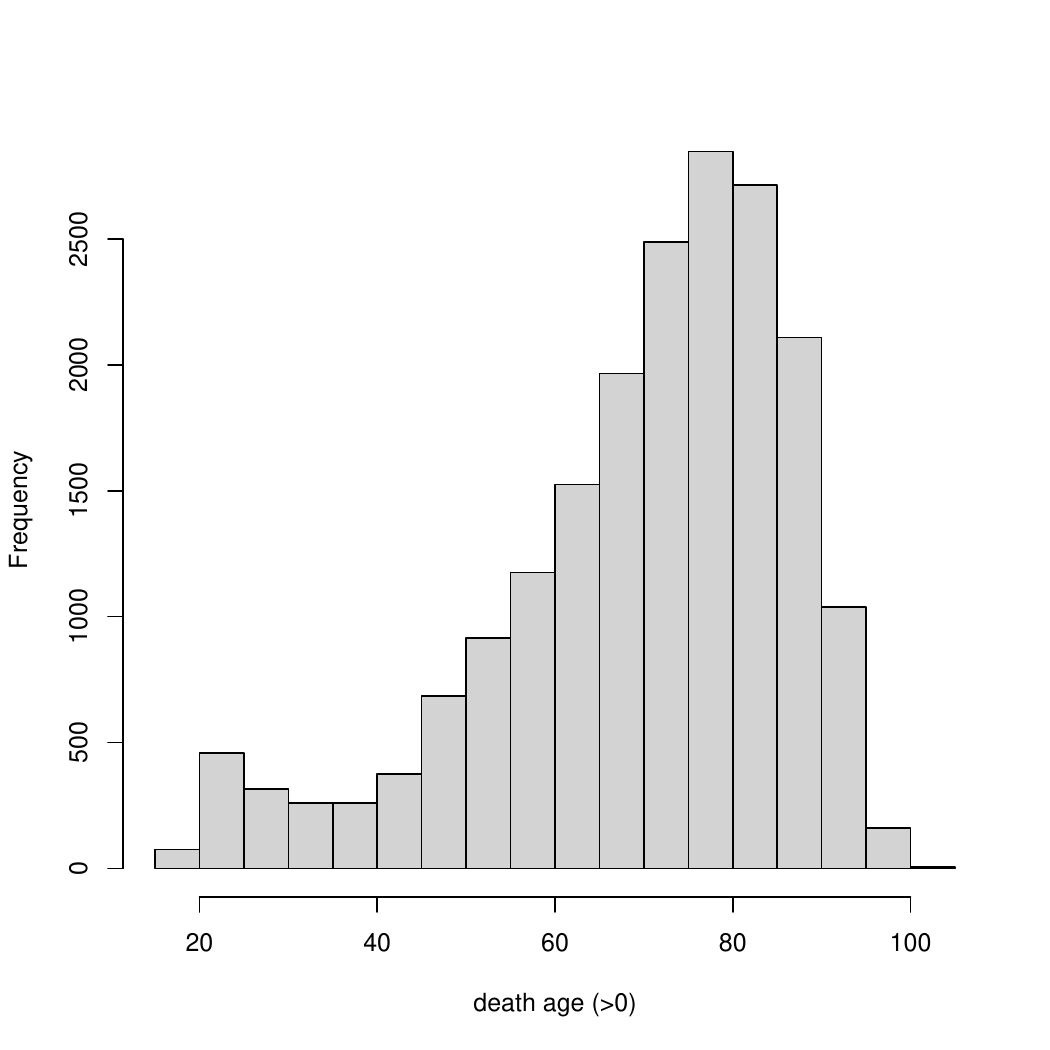}
\includegraphics[width=0.45\textwidth]{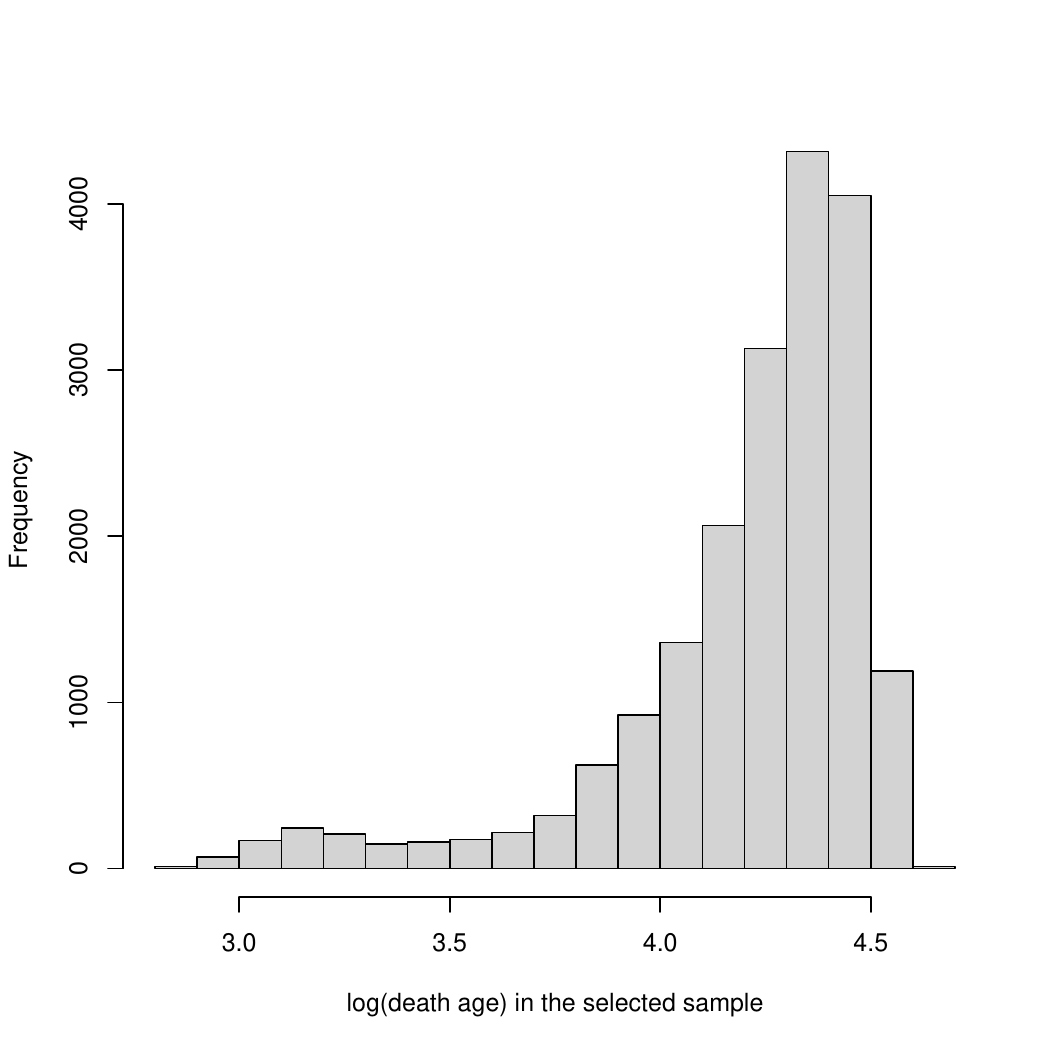}
\label{FCCCHY}
\end{figure}

\begin{table}[htp]
\begin{center}
\caption{(CCC) Descriptive statistics}\label{SSCCC}
{\small		
\begin{tabular}{lccccccc}
\hline\hline
Variable&Mean&Median&Std Dev&Minimum&Maximum&Non-imputing\\
\hline
death age (Y)&57.03&68.79&31.00&0.00&102.80&23722\\
duration of service (D)&0.82&0.51&0.71&0.00&8.37&23722\\
selection status (S)&0.82&1.00&0.39&0.00&1.00&23722\\
birth yerar&1919.78&1920.00&3.71&1870.00&1929.00&23722\\
ever rejected&0.02&0.00&0.14&0.00&1.00&23722\\
disabled&0.01&0.00&0.09&0.00&1.00&23722\\
non-junior&0.01&0.00&0.08&0.00&1.00&23722\\
reported age younger than DMF&0.09&0.00&0.28&0.00&1.00&23722\\
reported age older than DMF&0.17&0.00&0.37&0.00&1.00&23722\\
not eligible&0.02&0.00&0.12&0.00&1.00&23722\\
age is 17 or 18&0.56&1.00&0.49&0.00&1.00&23722\\
first allottee amount&21.63&22.00&3.71&0.00&30.00&22970\\
allottee is father&0.33&0.00&0.47&0.00&1.00&23722\\
allottee is mother&0.47&0.00&0.50&0.00&1.00&23722\\
gap in service&0.16&0.00&0.37&0.00&1.00&23722\\
log distance from home to camp&4.25&4.40&1.66&-9.40&8.04&23722\\
hispanic&0.48&0.00&0.50&0.00&1.00&23722\\
highest grade completed&8.60&8.61&1.65&0.00&17.00&14506\\
household size excluding applicant&4.74&4.74&1.50&0.00&22.00&7870\\
live on farm&0.25&0.25&0.25&0.00&1.00&8101\\
height&67.79&67.79&1.81&49.00&91.00&8141\\
weight&1.38&1.38&0.10&0.68&2.90&8234\\
father living&0.80&0.80&0.23&0.00&1.00&7943\\
mother living&0.85&0.85&0.21&0.00&1.00&8006\\
tenure in county&12.66&12.66&3.10&0.00&35.00&5432\\
\hline
\end{tabular}}
\end{center}
{\small Note: We use imputation dummies for imputed observations in covariates.}
\end{table}

We include the following covariates as in column (3) of \cite{Aizer}.
\begin{itemize}
\item X4-X7: X4 and X6 are family and individual characteristics, and X5 and X7 are imputation indicators. There are 8 continuous variables and 24 dummy variables.
\item Year dummies:  The birth year is varied from 1870 to 1929. After dropping the birth year dummies with the observations less than 10, we have 23 year dummies.
\end{itemize}

\begin{figure}[!htp]
\centering
\caption{(CCC) Estimated bounds and 95\% confidence intervals with quadratic $b(X)$
}
\includegraphics[width=0.4\textwidth]{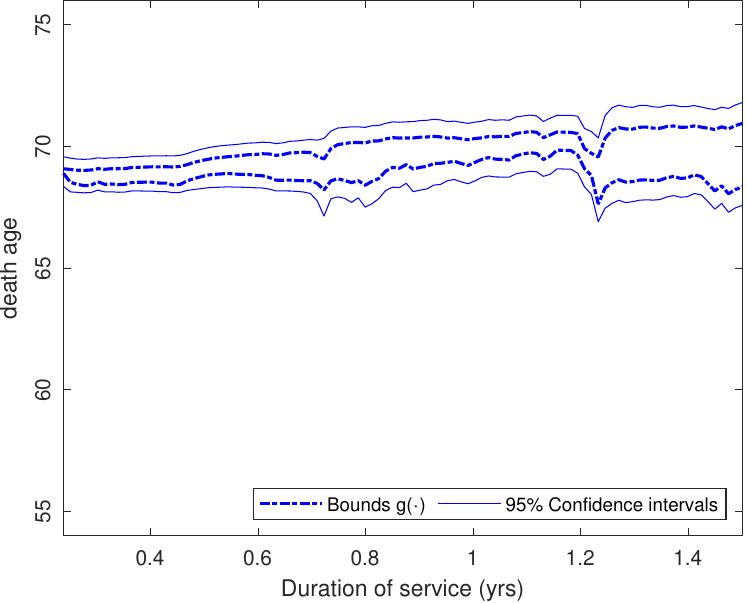}
\label{FCCCXquad}
\end{figure}

Figure~\ref{FCCCXquad} presents the estimated bounds with a quadratic basis function $b(X)$ for the CCC. For the estimates with quadratic $b(X)$ on $\mathcal{D}_{100}$, the largest ATE is when increasing
the duration from 0.276 to 1.156 years with the bounds $[0.836, 2.205]$ and 95\% confidence
interval $[0.119, 3.230]$.\footnote{The undersmoothing bandwidths range from 0.166 to 0.356.  $c_1 = 1.5, \nu = 0.01$.}

\end{document}